\documentclass[fleqn,10pt]{article}

\usepackage[english]{babel}
\usepackage[utf8]{inputenc}
\usepackage[T1]{fontenc}

\usepackage{latexsym, graphicx, epsfig, amsmath,  amssymb,amsfonts, mathtools}
\usepackage{natbib,amsthm,version}
\usepackage{amsbsy,bm,multirow}
\usepackage[titletoc,page]{appendix}
\usepackage[mathscr]{eucal}
\usepackage{titlesec}
\usepackage{enumerate}
\usepackage{float}
\restylefloat{table}
\usepackage{subfigure}

\usepackage{amsmath}
\usepackage{graphicx}

\usepackage[colorinlistoftodos]{todonotes}
\usepackage[colorlinks=true, allcolors=blue]{hyperref}

\newtheorem{remark}{Remark}

\theoremstyle{definition}

\title{
  A Three-Dimensional Hybrid Spectral Element-Fourier Spectral Method for Wall-Bounded Two-Phase Flows
} 
\author{
  S.H. Challa$^1$, \ S. Dong$^2$\thanks{Author of correspondence. Email: sdong@purdue.edu},
  \ L.D. Zhu$^3$ \\ [0.1in]
  $^1$School of Aeronautics and Astronautics \\
  $^2$Department of Mathematics \\
  Purdue University \\
  West Lafayette, Indiana, USA \\[0.1in]
  $^3$Department of Mathematics \\
  Indiana University-Purdue University Indianapolis \\
  Indianapolis, Indiana, USA
 } 
 
 \date{} 
\begin{document}
\maketitle


\begin{abstract}

We present a hybrid spectral element-Fourier spectral method
for solving the coupled system of Navier-Stokes and Cahn-Hilliard equations to simulate wall-bounded two-phase flows in a three-dimensional domain which is homogeneous in at least one direction. 
Fourier spectral expansions are employed along
the homogeneous direction and $C^0$ high-order spectral
element expansions are employed in the other directions.
A critical component of the method is a strategy we
developed in a previous work for dealing with the variable
density/viscosity of the two-phase mixture,
which makes the efficient use of Fourier expansions
in the current work possible for two-phase flows with different densities 
and viscosities for the two fluids.
The attractive feature of the presented method lies in that 
the two-phase computations in the three-dimensional space
are transformed into a set of de-coupled
two-dimensional computations
in the planes of the non-homogeneous directions.
The overall scheme consists of solving a set of
de-coupled two-dimensional equations for the flow and phase-field
variables in these planes. The linear algebraic systems
for these two-dimensional equations have constant
coefficient matrices that need to be computed only once
and can be pre-computed.
We present ample numerical simulations for different cases
to demonstrate the accuracy and capability of the presented method
in simulating the class of two-phase problems involving 
solid walls and moving contact lines. 

\end{abstract}

\noindent Key words: {\em hybrid discretization; two-phase flow; contact angle;
Fourier spectral method; spectral element method; wall-bounded flow}

\section{Introduction}



This work focuses on three-dimensional (3D) incompressible 
two-phase flows, comprised of two immiscible incompressible
fluids with different densities and viscosities, 
on flow domains which are homogeneous in at least one
direction and can be arbitrarily complex in the other
directions.
%
By ``homogeneous in one direction'' we mean that
the 3D domain is an extrusion of a two-dimensional (2D) domain
in the third (homogeneous) direction and it is 
infinite along that direction.
While seemingly a little restrictive with the homogeneity requirement, 
this class of problems encompasses a wide
range of important two-phase flows  such as those in the pipes,
channels, boundary layers, shear layers, and the two-phase
flows past long bluff bodies (e.g.~cylinders with cross sections
of various shapes), to name just a few.
It should be noted that flows in such domains are in general three-dimensional,
except at low Reynolds numbers with certain
flow configurations. For example, for the single-phase flow
past an infinitely long circular cylinder, the cylinder wake is two-dimensional
at Reynolds numbers below about $185$ and it transitions to
a three-dimensional flow for Reynolds numbers beyond about $260$~\cite{Williamson1996}.
In the current work
we assume that in the non-homogeneous directions 
the domain may be bounded by solid walls
with certain wetting properties. 
It is further assumed that the two-phase system consists of two immiscible incompressible fluid components, which can be
liquid/liquid or liquid/gas, and no solid phase (e.g.~solid
particles) is involved except solid-wall boundaries.


This class of 3D two-phase problems can potentially
allow the use of fast Fourier transforms (FFT) along
the homogeneous direction. One can then transform 
the 3D problem into a set of weakly-coupled or un-coupled
2D problems about the Fourier modes
on the discrete level.
The 3D computations can be reformulated into the solution
of a series of 2D problems, which can be performed largely in
parallel. This will enable very efficient simulations of  
 3D two-phase flows. This  idea motivates
 the current work.

Applying this idea to two-phase flows with generally
different densities and viscosities for the two fluids,
one immediately encounters difficulties.
The predominant difficulty lies in
the field-dependent
variable fluid properties (e.g.~mixture density and viscosity),
which would give rise to differential equations in space
with variable coefficients upon temporal discretization.
Performing FFT thus induces convolutions with the coefficient
functions in the frequency space, which couples together
all the Fourier modes of the unknown flow variable to be solved
for and completely defeats the purpose for using FFTs.
This situation is very different from single-phase 
flows (see e.g.~\cite{MaKK2000,DongKER2006,Dong2007,DongZ2011})
and two-phase problems with matched densities and viscosities
for the two fluids (see e.g.~\cite{LiuS2003,YueFLS2004}), in which
the fluid properties are all constant and cause no difficulty
when Fourier transforms and expansions are used.
Note that the energy-stable type schemes
(see e.g.~\cite{ShenY2010,ShenYY2015,AlandSebastian2016Aeae,GongYuezheng2018SOFD}, among others) are 
not suitable with FFT because they
give rise to differential equations in space
with variable coefficients even with matched fluid properties for the two fluids 
and with constant fluid properties for the two-phase mixture.

This difficulty can be circumvented. In \cite{DongS2012} we
have developed a strategy to deal with the variable density
and variable viscosity for two-phase problems in two dimensions.
The main idea lies in reformulations of the 
pressure and viscous terms in the momentum equation as follows,
\begin{equation}
\left\{
\begin{split}
&
\frac{1}{\rho}\nabla p \approx \frac{1}{\rho_0}\nabla p
+ \left(\frac{1}{\rho} - \frac{1}{\rho_0} \right)\nabla p^{*} \\
&
\frac{\mu}{\rho}\nabla^2\mathbf{u} \approx 
\nu_m\nabla^2\mathbf{u} - \left(\frac{\mu}{\rho} - \nu_m \right)
\nabla\times\nabla\times\mathbf{u}^{*},
\end{split}
\right.
\label{equ:reform_idea}
\end{equation}
where $\rho$ and $\mu$ are the variable density and dynamic
viscosity of the mixture (field functions), 
$\rho_0$ and $\nu_m$ are two appropriate constants,
$p$ and $\mathbf{u}$ are the pressure and velocity to be
computed,
and $p^{*}$ and $\mathbf{u}^{*}$ are the pressure and
velocity that are approximated explicitly with a prescribed
order of accuracy in time. The second approximation in the above
has employed the velocity divergence-free property ($\nabla\cdot\mathbf{u}=0$).
These approximations essentially incorporate certain zero terms,
such as $\frac{1}{\rho_0}\nabla(p-p^*)$, which
are equivalent to zero to the prescribed order of accuracy.
They can lead to differential equations
in space with constant coefficients upon temporal
discretization, and make the efficient use of FFTs possible
for two-phase flows with different densities and
viscosities for the two fluids.
This strategy is critical to 
the  developments in the current work.

The approach taken in this work for dealing
with two phases and fluid interfaces falls into
the phase field framework. Phase field is one of
the often-used approaches for two-phase 
problems~\cite{AndersonD.M.1998DMIF,LowengrubT1998,Jacqmin1999,LiuS2003,YueFLS2004}. We refer the reader to the 
reviews~\cite{OsherS1988,SethianJ.A.2003LSMF,TryggvasonG.2001AFMf,ScardovelliZ1999} and the references
therein for related approaches such as level set,
volume of fluids, and front tracking.
%
%
With the phase field approach, the fluid interface is
treated as a diffuse interface (thin  transition layer),
and regions of different fluids are marked by the phase
field function. The phase field function varies smoothly
over the diffuse interfacial layer and is mostly uniform
in the bulk of the fluids. The evolution of the system
is described by the system of the Navier-Stokes equations
and the Cahn-Hilliard equation, which stems from the
mass balance relation of the system~\cite{AbelsGG2012,Dong2014}.

The increasing popularity of phase field method for two-phase
flows in recent years 
is due to the contributions
of many researchers (see e.g.~\cite{AndersonD.M.1998DMIF,LowengrubT1998,Jacqmin2000,LiuS2003,BadalassiCB2003,YueFLS2004,DingSS2007,AbelsGG2012,DongS2012,Kim2012,Dong2014obc}, among others). 
Phase field is a physics-based approach~\cite{Dong2018}.
The effect of
surface tension in the two-phase
system is characterized by the free energy density function.
Certain component terms in the free energy density
promote the mixing of the two fluids, while the other 
terms tend to segregate different fluids. The interplay
of these two tendencies determines the profile of 
the interface. This method can deal with  contact
lines and contact angles relatively easily due to the diffuse interface~\cite{Jacqmin2000,QianWS2006,YueZF2010,Dong2012},
which is a prominent feature when compared with other
related methods. The contact angle on the solid wall
can be enforced by imposing appropriate 
boundary conditions. Since the phase field Cahn-Hilliard
equation is of fourth order in space, two independent
boundary conditions are required on the wall boundary.
One boundary condition can be used to enforce
the mass conservation, and the other condition can be
used to impose the contact angle at the 
wall~\cite{Jacqmin2000,YueZF2010,Dong2012,Dong2017}. 
Only the static (equilibrium) contact angle will be considered
in the current paper. We refer to e.g.~\cite{Dong2012,CarlsonDA2011} 
for discussions of the effect of dynamic contact
angles in the literature.

In this paper we present an efficient hybrid spectral element-Fourier
spectral method for wall-bounded two-phase flows on 3D domains
with at least one homogeneous direction. Fourier expansions of
the field variables are performed along the homogeneous direction,
and the difficulty caused by the variable
density/viscosity of the two-phase mixture is overcome by
exploiting the 
strategy represented by equation \eqref{equ:reform_idea}.
This allows
the efficient simulations of 3D two-phase flows with different
densities and dynamics viscosities for the two fluids.
In the plane of the non-homogeneous directions
high-order $C^0$ spectral elements are employed,
which can allow us to deal with
domains with complex  geometry 
in those directions.
The presented method has the advantage that the two-phase
computations in 3D space are decomposed into a
set of problems for the Fourier modes 
in the 2D planes, which are completely de-coupled from
one another. 
The method provides an efficient technique 
for simulating 3D two-phase flows.

The rest of this paper is organized as follows. 
In Section \ref{sec:method}, we discuss how to discretize
the governing equations and boundary conditions
for incompressible two-phase flows with the hybrid
Fourier spectral and spectral element method. 
In Section \ref{sec:tests} we present extensive
numerical examples of 3D two-phase problems to test the
convergence rates, accuracy and performance of the method developed
herein. Section \ref{sec:summary} concludes the
discussions with some closing remarks.

\section{Hybrid Spectral Element-Fourier Spectral Method}
\label{sec:method}

\subsection{Flow Domain, Governing Equations, and Boundary Conditions}

We consider 
the dynamics of 
two incompressible immiscible fluids contained in a flow domain in
the three-dimensional (3D) space. 
In the present work we concentrate on the
class of problems with the flow domain having the following property:
\begin{itemize}
\item The domain is homogeneous in at least one direction (denoted as $z$ direction), while in the other two ($x$ and $y$) directions
it can be arbitrarily complex.
\end{itemize}
As such, a Fourier spectral expansion of the field variables can be
performed along the homogeneous direction
to take advantage of the fast Fourier transforms (FFT),
and spectral element expansions of the field variables
can be performed within the $x$-$y$ planes to deal with the complex geometry.


To simulate this class of problems,
we choose a computational domain with dimension $L_z$ along
the homogeneous direction ($0\leqslant z\leqslant L_z$), and assume that all the field variables
and the domain are periodic at $z=0$ and $z=L_z$. 
Let $\Omega$ denote the 3D computational domain, and $\partial\Omega$ denote the 3D domain
boundary. Let $\Omega_{2D}$ denote the projection of the 3D domain onto the $x$-$y$ plane,
and $\partial\Omega_{2D}$ denote its boundary.
Then we have
\begin{equation}
\Omega = \Omega_{2D} \otimes [0,L_z], \quad
\partial\Omega = \partial\Omega_{2D} \otimes [0, L_z].
\end{equation}
Let $\mathbf{n}$ denote the outward-pointing unit vector normal to $\partial\Omega$,
and $\mathbf{n}_{2D}$ denote the outward-pointing unit vector normal to
$\partial\Omega_{2D}$ within the 2D $x$-$y$ plane. It should be noted that
$\mathbf{n} = (\mathbf{n}_{2D},0)$, namely, $\mathbf{n}$ has zero
component in the homogeneous direction.

Based on the phase-field approach, the dynamics of the flow is described 
by the following system of coupled Navier-Stokes and Cahn-Hilliard 
equations (see~\cite{LiuS2003,YueFLS2004,DongS2012}),
\begin{subequations}
\begin{equation}
\label{e4.1a}
 \rho\left(\frac{\partial\textbf{u}}{\partial t}+\textbf{u}\cdot\nabla\textbf{u}\right)=-\nabla p+\nabla\cdot[\mu\textbf{D}(\textbf{u})]-\nabla\cdot(\lambda\nabla\phi \nabla\phi) + \textbf{f}, 
\end{equation}    
\begin{equation}
\nabla\cdot\textbf{u}=0  \label{e4.1b}
\end{equation}
\begin{equation}
\label{e4.1c}
 \frac{\partial\phi}{\partial t}+\textbf{u}\cdot\nabla\phi=-\lambda\gamma_1\nabla{^2}\left[\nabla{^2}\phi-h(\phi)\right]+g(\textbf{x,}t).
\end{equation}
\end{subequations}
In the above equations, $\textbf{x}$ and $t$ are the spatial
coordinate and time, respectively.
$\textbf{u}(\textbf{x,}t)$ is the velocity, and $p(\textbf{x,}t)$ is the pressure.
$\textbf{D}(\textbf{u})=\nabla\textbf{u}+(\nabla\textbf{u}){^T}$, where ($\cdot$)${^T}$ denotes the transpose of ($\cdot$).
$\textbf{f}(\textbf{x,}t)$ is an external body force. $\phi(\textbf{x,}t)$ is the phase field function, $-1\le\phi\le 1.$ The flow regions containing the first and second fluids are denoted by $\phi=1$ and $\phi=-1$ respectively, and the iso-surface $\phi(\textbf{x,}t)=0$ denotes the fluid interface at  time $t$. The function $h(\phi)$ in \eqref{e4.1c} is given by, $h(\phi)=\frac{1}{\eta{}^2}\phi(\phi{^2}-1)$, where $\eta$ is the characteristic length scale of the interface thickness. $\lambda$ is the mixing energy density coefficient, given by $\lambda=\frac{3}{2\space \sqrt[]{2}}\sigma\eta$~\cite{YueFLS2004}, where $\sigma$ is the surface tension (assumed to be constant). $\gamma_1$ denotes the mobility of the interface, and it is also assumed to be constant. The mixture density $\rho$ and dynamic viscosity $\mu$ are given by,
\begin{equation}
\label{e4.2}
\rho(\phi)=\frac{\rho_1+\rho_2}{2}+\frac{\rho_1-\rho_2}{2}\phi, \quad
\space \mu(\phi)=\frac{\mu_1+\mu_2}{2}+\frac{\mu_1-\mu_2}{2}\phi
\end{equation}
where $\rho_1$ and $\rho_2$ are the constant densities of the two fluids, 
and $\mu_1$ and $\mu_2$ are the constant dynamic viscosities of the two fluids, respectively.
 Since $\phi$ is time-dependent, both $\rho(\phi)$ and $\mu(\phi)$ are field
 functions and vary in time. $g(\textbf{x,}t)$ in \eqref{e4.1c} is a prescribed source term for the purpose of numerical testing only, and will be set to zero in practical simulations.


 We assume that the flow domain is bounded in the non-homogeneous directions
 by solid walls 
with certain wetting properties. Correspondingly, the governing equations
\eqref{e4.1a}--\eqref{e4.1c} are supplemented by the
following boundary conditions:
\begin{subequations}
\begin{equation}
\label{e4.3}
\textbf{u}=\textbf{w}(\textbf{x,}t), \ \text{on} \ \partial\Omega 
\end{equation} 
\begin{equation}
\label{e4.4a}
\textbf{n}\cdot\nabla[\nabla{^2}\phi-h(\phi)]=\textbf{g}_c(\textbf{x,}t), \ \text{on} \ \partial\Omega
\end{equation}
\begin{equation}
\label{e4.4b}
\textbf{n}\cdot\nabla\phi +\frac{1}{\lambda} \textbf{f}'_w(\phi)+\textbf{g}_b(\textbf{x,}t)=0, \ \text{on} \ \partial\Omega
\end{equation}
\end{subequations}
 where $\mathbf{w}$ is the boundary velocity, and
 $\textbf{g}_b(\textbf{x,}t)$ and $\textbf{g}_c(\textbf{x,}t)$ are
 prescribed source terms for numerical
 testing only and will be set to $g_b=0$ and $g_c=0$ in actual simulations.
 The boundary condition \eqref{e4.4a} with $g_c=0$
 enforces the mass conservation and the condition \eqref{e4.4b} with $g_b=0$
 enforces the static contact angle on the wall~\cite{Dong2012}.
 The function $f_w(\phi)$ represents the wall energy density, and
 $\textbf{f}'_w(\phi)$ is given by (see~\cite{Dong2012}):
 \begin{equation}
  \textbf{f}'_w(\phi)=-\frac{3}{4} \sigma(1-\phi^2) \cos\theta_s.
  \label{equ:static_fw}
 \end{equation}
 where $\theta_s$ is the static contact angle
 formed between the solid wall and the fluid interface
 measured on the side of the first fluid.
 In the present work, only the effect of the
 static contact angle will be considered. One
     can refer to e.g.~\cite{Dong2012} for a study of the effect of dynamic
     contact angles.
     
 Finally, the governing equations are supplemented by the initial conditions for
 the velocity and the phase field function as follows:
 \begin{equation}
 \textbf{u}(\textbf{x},0)=\textbf{u}_{in}(\textbf{x}),
 \label{e4.5}
 \end{equation}
 \begin{equation}
 \phi(\textbf{x},0)=\phi_{in}(\textbf{x}),
 \label{e4.6}
 \end{equation}
 where $\mathbf{u}_{in}$ and $\phi_{in}$ are the initial velocity and
 phase field distributions.

\subsection{Variable Density/Viscosity and the  Difficulty Caused to Fourier Transform}

Using Fourier spectral expansions to numerically solve the
governing equations \eqref{e4.1a}-\eqref{e4.1c}
together with the boundary conditions \eqref{e4.3}--\eqref{e4.4b},
one confronts an immediate difficulty.
The difficulty is associated with 
the variable density $\rho(\phi)$ and variable viscosity $\mu(\phi)$
of the two-phase mixture, because these are field functions.
A naive treatment of those terms involving these variables
would lead to semi-discretized equations
about the flow variables whose coefficients are field functions (not constants).
Upon Fourier expansion, this will give rise to a convolution of
the Fourier modes for the flow variable to be solved for and
the Fourier modes of the coefficient function.
Therefore, all the Fourier modes of the flow variable will be
coupled together, which defeats the purpose for using Fourier transforms.

This difficulty associated with the 
variable density and variable viscosity can be circumvented.
The key lies in a strategy we developed in \cite{DongS2012} for two dimensions,
by a reformulation of the pressure term and the viscous term as given
in equation \eqref{equ:reform_idea}.
We will exploit this strategy in the current work.
This will enable the use of Fourier spectral expansions and efficient computations
based on FFT. 

We employ the following scheme for time stepping of the 3D two-phase
governing equations, which is
modified from that of \cite{DongS2012} and incorporates
the reformulations given in equation \eqref{equ:reform_idea}.
By introducing an auxiliary pressure, $P=p+\frac{\lambda}{2}\nabla\phi\cdot\nabla\phi$,
equation \eqref{e4.1a} can be transformed into
\begin{equation}
\label{equ:nse_trans}
 \frac{\partial\textbf{u}}{\partial t}+\textbf{u}\cdot\nabla\textbf{u}=-\frac{1}{\rho}\nabla P+\frac{\mu}{\rho}\nabla{^2}\textbf{u}
 +\frac{1}{\rho}\nabla\mu\cdot\mathbf{D}(\mathbf{u})
 -\frac{\lambda}{\rho}\left(\nabla{^2}\phi\right)\nabla\phi+ \frac{1}{\rho}\mathbf{f}. 
\end{equation}  
Let $n\geqslant 0$ denote the time step index and $(\cdot)^n$
denote the variable $(\cdot)$ at time step $n$.
Let $J$ ($J=1$ or $2$) denote the temporal order of accuracy.
Given $(\phi^n,P^n,\mathbf{u}^n)$, we solve $\phi^{n+1}$, $P^{n+1}$
and $\mathbf{u}^{n+1}$ successively as follows. \\
  \noindent \underline{For $\phi{^{n+1}}$:}	
  \begin{subequations}
   \begin{equation}
   \label{e4.8a}
    \frac{\gamma_0\phi^{n+1}-\hat{\phi}}{\Delta t} +\textbf{u}^{*,n+1}\cdot\nabla\phi^{*,n+1}=-\lambda\gamma_1\nabla^2\left[\nabla^2\phi^{n+1}-\frac{S}{\eta^2}\left(\phi^{n+1}-\phi^{*,n+1}\right)-h(\phi^{*,n+1})\right]+g^{n+1},
   \end{equation}
      \begin{equation}
   \label{e4.8b}
 \textbf{n}\cdot\nabla\left[\nabla^2\phi^{n+1}-\frac{S}{\eta^2}(\phi^{n+1}-\phi^{*,n+1})-h(\phi^{*,n+1})\right]=g_c^{n+1}, \quad \text{on} \ \partial\Omega,    
   \end{equation}
      \begin{equation}
   \label{e4.8c}
   \textbf{n}\cdot\nabla\phi^{n+1}=-\frac{1}{\lambda}f'_w(\phi^{*,n+1})-g_b^{n+1},
   \quad \text{on} \ \partial\Omega.   
   \end{equation}
  \end{subequations}
     \noindent \underline{For  $P^{n+1}$:}
\begin{subequations}
    \begin{multline}
	\frac{\gamma_0 \tilde{\textbf{u}}^{{n+1}}-\hat{\textbf{u}}}{\Delta t}+\frac{1}{\rho_0}\nabla P^{n+1}=-\textbf{u}^{*,n+1}\cdot\nabla\textbf{u}^{*,n+1}+\left(\frac{1}{\rho_0}-\frac{1}{\rho^{n+1}}\right)\nabla P^{*,n+1}-\frac{\mu^{n+1}}{\rho^{n+1}}\nabla\times\nabla\times\textbf{u}^{*,n+1}\\+\frac{1}{\rho^{n+1}}\nabla\mu^{n+1}\cdot\textbf{D}(\textbf{u}^{*,n+1})-\frac{\lambda}{\rho^{n+1}}\nabla^2\phi^{n+1}\nabla\phi^{n+1}+\frac{1}{\rho^{n+1}}\mathbf{f}^{n+1},
    \label{e4.9a}
	\end{multline}
\begin{equation}
    \label{e4.9b}
    \nabla\cdot\tilde{\textbf{u}}^{n+1}=0,
    \end{equation}
    \begin{equation}
    \label{e4.9c}
    \textbf{n}\cdot\tilde{\textbf{u}}^{n+1}=\textbf{n}\cdot\textbf{w}^{n+1}, \quad \text{on} \ \partial \Omega
    \end{equation}
\end{subequations}
\noindent \underline{For  $\textbf{u}^{n+1}$:}
\begin{subequations}
  %
  \begin{equation}
    \label{e4.10a}
    \begin{split}
    \frac{\gamma_0 \textbf{u}^{{n+1}}-\hat{\textbf{u}}}{\Delta t}&+\frac{1}{\rho_0}\nabla P^{n+1} - \nu_m\nabla^2\mathbf{u}^{n+1}
    =-\textbf{u}^{*,n+1}\cdot\nabla\textbf{u}^{*,n+1}+\left(\frac{1}{\rho_0}-\frac{1}{\rho^{n+1}}\right)\nabla P^{*,n+1} \\
    &+ \left(\nu_m-\frac{\mu^{n+1}}{\rho^{n+1}}\right)\nabla\times\nabla\times\textbf{u}^{*,n+1}
    +\frac{1}{\rho^{n+1}}\nabla\mu^{n+1}\cdot\textbf{D}(\textbf{u}^{*,n+1})\\
    &-\frac{\lambda}{\rho^{n+1}}\nabla^2\phi^{n+1}\nabla\phi^{n+1}+\frac{1}{\rho^{n+1}}\mathbf{f}^{n+1},
    \end{split}
  \end{equation}
\begin{equation}
\label{e4.10b}
\textbf{u}^{n+1}=\textbf{w}^{n+1}, \quad \text{on} \  \partial\Omega.
\end{equation}
\end{subequations}
   
In the above equations  $\Delta t$ is the time step size,
$\tilde{\textbf{u}}^{n+1}$ is an auxiliary velocity
approximating $\textbf{u}^{n+1}$,
and $\textbf{D}(\textbf{u})=\nabla\textbf{u}+(\nabla\textbf{u})^T$.
Let $\zeta$ denote a generic variable. Then $\zeta^{*,n+1}$ in the above equations
   is a $J$-th order explicit approximation of $\zeta^{n+1}$, given by: 
\begin{equation}
\zeta^{*,n+1}=\begin{cases}
\zeta^n, & \text{if}\ J=1,\\
2\zeta^n-\zeta^{n-1}, & \text{if}\ J=2. 
\end{cases}
\end{equation}
$\hat{\zeta}$ and $\gamma_0$ in the above equations are given by
   \begin{equation}
   \hat{\zeta}=\begin{cases}
   \zeta^n, & \text{if} \ J=1, \\
   2\zeta^n-\frac{1}{2}\zeta^{n-1}, & \text{if} \ J=2,
   \end{cases}
   \hspace{0.03\linewidth} 
   \gamma_0=\begin{cases}
    1, & \text{if} \ J=1,\\
    \frac{3}{2}, & \text{if} \ J=2.
   \end{cases}
   \end{equation}
   So $\frac{1}{\Delta t}\left(\gamma_0\zeta^{n+1}-\hat{\zeta}\right)$
in the above equations is an approximation of
$\left.\frac{\partial\zeta}{\partial t}\right|^{n+1}$ based on the $J$-th order
backward differentation formula (BDF).
$\rho_0$ is a  constant given by
$ 
 \rho_0= \min(\rho_1,\rho_2).
$ 
  $\nu_m$ is a chosen constant that is sufficiently large, with a
  reasonable condition given by
$ 
    \nu_m\geqslant\frac{1}{2} \left(\frac{\nu_1}{\rho_1}+\frac{\nu_2}{\rho_2}\right).
$ 
 $S$ is a chosen constant that satisfies the condition
 \begin{equation}
 S\geqslant \eta^2\sqrt[]{\frac{4\gamma_0}{\lambda\gamma_1\Delta t}}.
 \label{equ:S_cond}
 \end{equation}

In the scheme given by equations \eqref{e4.8a}--\eqref{e4.10b},
the computations for the phase field function $\phi^{n+1}$,
the pressure $P^{n+1}$ and the velocity $\mathbf{u}^{n+1}$
are de-coupled. Note that the auxiliary velocity $\tilde{\mathbf{u}}^{n+1}$
is not computed explicitly.
When solving for the phase field function, it should be noted that
a stabilization term $\frac{S}{\eta^2}(\phi^{n+1}-\phi^{*,n+1})$
has been added to both the discrete equation \eqref{e4.8a} and the discretized
boundary condition \eqref{e4.8b}. In the pressure sub-step,
the discrete momentum equation \eqref{e4.9a} incorporates the reformulation
of the pressure term given in \eqref{equ:reform_idea}, and because
of the explicit treatment of the viscous term the second reformulation
in \eqref{equ:reform_idea} has no contribution to this sub-step.
In the velocity sub-step, the discrete equation \eqref{e4.10a}
incorporates the reformulations of both the pressure term and the 
viscous term given in \eqref{equ:reform_idea}.

\subsection{Hybrid Spectral Element-Fourier Spectral Discretization
and Solution Algorithm}

Let us now focus on how to compute the velocity, pressure and 
the phase field function based on 
 the scheme given by 
 \eqref{e4.8a}--\eqref{e4.10b} on a 3D domain $\Omega$
 that is homogeneous in the $z$ direction.
The field variables will be represented by a Fourier spectral expansion
along the $z$ direction and a $C^0$ 
spectral element expansion within the $x-y$ planes.

\paragraph{Essential Integral Relations}
Since the spectral element bases are real-valued functions
in the $x$-$y$ plane and the Fourier bases involve complex functions in $z$,
the 3D basis and test functions will be complex-valued.
Because weak forms for the governing equations are required
for the spectral element discretizations, it is
important and convenient to first spell out several 
integral relations involving the 3D basis functions.
These relations will be extensively used in subsequent 
discussions.

Let $N_z$ denote the number of modes in the Fourier spectral expansion in 
$z$ direction. Let $\Phi_k(z)$ ($-\frac{N_z}{2}\leqslant k\leqslant \frac{N_z}{2}-1$)
denote the $k$-th Fourier basis function in $z$, given by
\begin{equation}
\Phi_k(z) = e^{i\beta_k z}, \quad \beta_k=\frac{2\pi k}{L_z},
\quad -\frac{N_z}{2}\leqslant k\leqslant \frac{N_z}{2}-1.
\label{equ:F_basis}
\end{equation}
The following property holds,
\begin{equation}
\int_{0}^{L_z} \bar{\Phi}_k(z)\Phi_m(z) dz = L_z\delta_{km},
\quad -\frac{N_z}{2}\leqslant k, m\leqslant \frac{N_z}{2}-1,
\label{equ:F_ortho}
\end{equation}
where the over-bar in $\bar{\Phi}_k$ denotes the complex conjugate of $\Phi_k$, and $\delta_{km}$ is the Kronecker delta.

Let $f(z)$ denote a generic scalar field function on $\Omega$ that is periodic in $z$
with a period $L_z$. Note that its dependence on $x$ and $y$
has been suppressed for brevity. The Fourier expansion is
\begin{equation}
f(z) = \sum_{k=-\frac{N_z}{2}}^{\frac{N_z}{2}-1} \hat{f}_k\Phi_k(z)
\label{equ:F_expansion}
\end{equation}
where $\hat{f}_k$ denotes the $k$-th Fourier mode of $f(z)$.
The following relations can be readily verified based on
\eqref{equ:F_basis}--\eqref{equ:F_expansion}:
\begin{equation}
\left\{
\begin{split}
&
\int_{0}^{L_z} f(z) \bar{\Phi}_k(z) dz = L_z \hat{f}_k, \\
&
\int_{0}^{L_z} f(z)\frac{\partial\bar{\Phi}_k}{\partial z} dz = -i\beta_k\hat{f}_kL_z, \\
&
\int_{0}^{L_z} \frac{\partial f}{\partial z} \bar{\Phi}_k(z) dz = i\beta_k\hat{f}_kL_z,
 \\
&
\int_{0}^{L_z} \frac{\partial f}{\partial z} \frac{\partial\bar{\Phi}_k}{\partial z} dz
= \beta_k^2\hat{f}_kL_z,
\quad -\frac{N_z}{2}\leqslant k\leqslant \frac{N_z}{2}-1.
\end{split}
\right.
\label{equ:F_relations}
\end{equation}


Let $\varphi(\vec{x})$, where $\vec{x}=(x,y)$, 
denote an arbitrary real test (basis) function in
the 2D $x-y$ plane. Then the 3D basis functions and test functions are respectively 
\begin{equation}
\left\{
\begin{split}
&
Q_k(\vec{x},z)=\varphi(\vec{x})\Phi_k(z), 
\quad -\frac{N_z}{2}\leqslant k\leqslant \frac{N_z}{2}-1;
\quad\text{(basis function)} \\
&
\bar{Q}_k(\vec{x},z) = \varphi(\vec{x})\bar{\Phi}_k(z), 
\quad -\frac{N_z}{2}\leqslant k\leqslant \frac{N_z}{2}-1.
\quad\text{(test function)}
\end{split}
\right.
\label{equ:3d_test_func}
\end{equation}
Define the gradient in the 2D $x-y$ plane,
$\nabla_{2D} = \frac{\partial}{\partial\vec{x}} 
= \left(\frac{\partial}{\partial x},\frac{\partial}{\partial y} \right)$.
Then the 3D gradient
$\nabla = \left(\nabla_{2D},\frac{\partial}{\partial z} \right)$.
The following relations can be verified based on equations
\eqref{equ:F_expansion}--\eqref{equ:F_relations}:
\begin{equation}
\left\{
\begin{split}
&
\int_{\Omega} f(\vec{x},z)\bar{Q}_{k}(\vec{x},z)d\Omega = 
L_z\int_{\Omega_{2D}}\hat{f}_k(\vec{x})\varphi(\vec{x})d\Omega_{2D}, \\
&
\int_{\Omega}\nabla f\cdot\nabla\bar{Q}_k d\Omega = L_z\int_{\Omega_{2D}}\left(\nabla_{2D}\hat{f}_{k}\cdot\nabla_{2D}\varphi
+\beta_k^2\hat{f}_k\varphi
\right) d\Omega_{2D}, \\
&
\int_{\partial\Omega} f(\vec{x},z)\bar{Q}_{k}(\vec{x},z)dA = 
L_z\int_{\partial\Omega_{2D}}\hat{f}_k(\vec{x})\varphi(\vec{x})dA,
\quad -\frac{N_z}{2}\leqslant k\leqslant \frac{N_z}{2}-1,
\end{split}
\right.
\label{equ:F_relations_1}
\end{equation}
where $d\Omega = d\Omega_{2D}dz = d\vec{x}dz=dxdydz$.

Let $\bm{\chi}(\vec{x},z) = (\bm{\chi}_{2D},\chi_z)=(\chi_x,\chi_y,\chi_z)$
denote a generic vector-valued field function on $\Omega$ that is periodic
in $z$ with a period $L_z$, and $\bm{\chi}_{2D}$ denote its component
vector in the 2D $x-y$ plane. 
Then the following relations hold based on equations
\eqref{equ:F_expansion}--\eqref{equ:F_relations}:
\begin{equation}
\left\{
\begin{split}
&
\int_{\Omega}\bm{\chi}\cdot\nabla\bar{Q}_k d\Omega
= L_z\left(\int_{\Omega_{2D}} \hat{\bm{\chi}}_{2D,k}\cdot\nabla_{2D}\varphi d\Omega_{2D}
-i\beta_k\int_{\Omega_{2D}} \hat{\chi}_{z,k}\varphi d\Omega_{2D} \right),
 \\
&
\int_{\partial\Omega}(\mathbf{n}\cdot\bm{\chi})\bar{Q}_k dA
= L_z\int_{\partial\Omega_{2D}}\mathbf{n}_{2D}\cdot\hat{\bm{\chi}}_{2D,k}\varphi dA,
\quad -\frac{N_z}{2}\leqslant k\leqslant \frac{N_z}{2}-1
\end{split}
\right.
\label{equ:F_relations_2}
\end{equation}
where $\mathbf{n}=(\mathbf{n}_{2D},0)$, and
$\hat{\bm{\chi}}_{2D,k}$ and $\hat{\chi}_{z,k}$ are
the Fourier expansion coefficients of $\bm{\chi}_{2D}$ and
$\chi_z$, respectively.

With the above notation and preparation, let us now consider the
discretization and solution for the phase field function,
the pressure and the velocity.

\paragraph{Phase Field Function}

Equation \eqref{e4.8a} can be written as
  \begin{equation}
 \nabla^2\left[\nabla^2\phi^{n+1}-\frac{S}{\eta^2}\phi^{n+1}\right]
 +\frac{\gamma_0}{\lambda\gamma_1\Delta t}\phi^{n+1}=R_1+\nabla^2 R_{2}=R
 \label{equ:CH_trans_1}
 \end{equation}
 where  
     \begin{equation}
     \left\{
     \begin{split}
     &
     R_{1}=\frac{1}{\lambda\gamma_1}\left[g^{n+1}-                        \textbf{u}^{*,n+1}\cdot\nabla\phi^{*,n+1}+\frac{\hat{\phi}}{\Delta t}\right]\hspace{0.03\linewidth} \\
     &
     R_{2}=h(\phi^{*,n+1})-\frac{S}{\eta^2}\phi^{*,n+1}.
     \end{split}
     \right.
     \label{equ:def_R1_R2}
     \end{equation}
Under the condition \eqref{equ:S_cond} for the constant $S$,
equation \eqref{equ:CH_trans_1} can be reformulated into two
de-coupled Helmholtz type equations given by
(see \cite{YueFLS2004,DongS2012} for details),
\begin{subequations}
\begin{equation}
 \nabla^{2}\psi^{n+1} -\left(\alpha+\frac{S}{\eta^2}\right)\psi^{n+1} =R,
 \label{equ:CH_helm_1}
 \end{equation}
\begin{equation}
\nabla^2\phi^{n+1}+\alpha\phi^{n+1}=\psi^{n+1},
\label{equ:CH_helm_2}
\end{equation}
\end{subequations}
where $\psi^{n+1}$ is an auxiliary variable defined by \eqref{equ:CH_helm_2}
and the constant $\alpha$ is given by
$
\alpha = -\frac{S}{2\eta^2} \left( 1+\sqrt[]{1-\frac{4\gamma_0}{\lambda\gamma_1\Delta t}\frac{\eta^4}{S^2}} \right ).
$
The boundary condition \eqref{e4.8b} can be transformed into
\begin{equation}
\textbf{n}\cdot\nabla\psi^{n+1}=\textbf{n}\cdot\nabla R_{2}-\left( \alpha+\frac{S}{\eta^2}\right)\left[\frac{1}{\lambda}f'_w(\phi^{*,n+1})+g_b^{n+1}\right]+g_c^{n+1}, \hspace{0.03\linewidth}on\hspace{0.03\linewidth}\partial\Omega
\label{equ:psi_bc}
\end{equation}
where equations \eqref{e4.8c} and \eqref{equ:CH_helm_2} have been used.


We now take the $L^2$ inner product between the 3D test function,
$\bar{Q}_k(\vec{x},z)=\varphi(\vec{x})\bar{\Phi}_k(z)$ ($\varphi(\vec{x})$
denoting an arbitrary 2D test function), 
and equation \eqref{equ:CH_helm_1}, and integrate by part.
This leads to
\begin{multline}
\int_{\Omega}\nabla\psi^{n+1}\cdot\nabla\bar{Q}_k
+ \left(\alpha+\frac{S}{\eta^2} \right)\int_{\Omega}\psi^{n+1}\bar{Q}_k
=-\int_{\Omega}R_1\bar{Q}_k + \int_{\Omega}\nabla R_2\cdot\nabla\bar{Q}_k \\
+\int_{\partial\Omega} \left[
-\left(\alpha+\frac{S}{\eta^2} \right)\left(\frac{1}{\lambda}f_w^{\prime}(\phi^{*,n+1})+g_b^{n+1} \right)
+ g_c^{n+1}
\right] \bar{Q}_k,
\quad -\frac{N_z}{2}\leqslant k\leqslant \frac{N_z}{2}-1, \quad \forall \varphi
\label{equ:CH_psi_1}
\end{multline}
where equation \eqref{equ:psi_bc} has been used.
Perform the Fourier spectral expansion of $\psi^{n+1}$ in $z$,
$
\psi^{n+1} = \sum_{k=-N_z/2}^{N_z/2-1} \hat{\psi}_k^{n+1}(\vec{x}) \Phi_k(z).
$
In light of the relations \eqref{equ:F_relations}--\eqref{equ:F_relations_1},
equation \eqref{equ:CH_psi_1} can be reduced into
\begin{multline}
\int_{\Omega_{2D}}\nabla_{2D}\hat{\psi}_k^{n+1}\cdot\nabla_{2D}\varphi
+\left(\alpha + \frac{S}{\eta^2}+\beta^2_k \right)\int_{\Omega_{2D}}\hat{\psi}_k^{n+1}\varphi
=\int_{\Omega_{2D}}\left(\beta_k^2\hat{R}_{2,k}-\hat{R}_{1,k}\right)\varphi \\
+\int_{\Omega_{2D}}\nabla_{2D}\hat{R}_{2,k}\cdot\nabla_{2D}\varphi 
+\int_{\partial\Omega_{2D}} \hat{\mathcal{T}}_k\varphi,
\quad -\frac{N_z}{2}\leqslant k\leqslant \frac{N_z}{2}-1, \quad \forall \varphi
\label{equ:psi_weakform}
\end{multline}
where $\hat{R}_{1,k}$ and $\hat{R}_{2,k}$ are the Fourier coefficients of
$R_1$ and $R_2$, respectively, and $\hat{\mathcal{T}}_k$ is the Fourier coefficient
of the variable $\mathcal{T}$ given by
\begin{equation}
\mathcal{T} = 
-\left(\alpha+\frac{S}{\eta^2} \right)\left(\frac{1}{\lambda}f_w^{\prime}(\phi^{*,n+1})+g_b^{n+1} \right)
+ g_c^{n+1}.
\end{equation}
Equation \eqref{equ:psi_weakform} is the weak form of a Helmholtz type equation
about the $N_z$ Fourier modes $\hat{\psi}_k^{n+1}$ in the 2D plane, and the different
Fourier modes are not coupled in the computations. The terms on the right hand side (RHS) of 
the equation can be computed explicitly.


Taking the $L^2$ inner product between the 3D
test function $\bar{Q}_k$ and equation \eqref{equ:CH_helm_2}
and integrating by part, we have
\begin{multline}
\int_{\Omega} \nabla\phi^{n+1}\cdot\nabla\bar{Q}_k
-\alpha\int_{\Omega}\phi^{n+1}\bar{Q}_k
=-\int_{\Omega}\psi^{n+1}\bar{Q}_k
-\int_{\partial\Omega}\left[\frac{1}{\lambda}f_w^{\prime}(\phi^{*,n+1})+g_b^{n+1} \right]\bar{Q}_k, \\
-\frac{N_z}{2}\leqslant k\leqslant \frac{N_z}{2}-1, \quad \forall \varphi
\end{multline}
where  the boundary condition \eqref{e4.8c} has been used.
In light of the relations \eqref{equ:F_relations}--\eqref{equ:F_relations_1},
this equation is reduced to
\begin{multline}
\int_{\Omega_{2D}}\nabla_{2D}\hat{\phi}_k^{n+1}\cdot\nabla_{2D}\varphi
+\left(-\alpha+\beta_k^2\right)\int_{\Omega_{2D}}\hat{\phi}_k^{n+1}\varphi
=-\int_{\Omega_{2D}}\hat{\psi}_k^{n+1}\varphi
-\int_{\partial\Omega_{2D}} \hat{\mathcal{M}}_k\varphi, \\
-\frac{N_z}{2}\leqslant k\leqslant \frac{N_z}{2}-1, \quad \forall \varphi
\label{equ:phi_weakform}
\end{multline}
where $\hat{\phi}_k^{n+1}$ are the Fourier expansion coefficients of $\phi^{n+1}$,
and $\hat{\mathcal{M}}_k$ are the Fourier coefficients of the variable $\mathcal{M}$
defined by 
$
\mathcal{M} = \frac{1}{\lambda}f_w^{\prime}(\phi^{*,n+1})+g_b^{n+1}.
$
Equation \eqref{equ:phi_weakform} is the weak form about the $N_z$ Fourier modes
$\hat{\phi}_k^{n+1}$ in the 2D $x-y$ plane. Once $\hat{\psi}_k^{n+1}$ are known,
these equations can be solved for $\hat{\phi}_k^{n+1}$. It is noted that
the different Fourier modes are de-coupled from one another.

To compute the phase field function, we therefore take two steps:
(i) solve equation \eqref{equ:psi_weakform} for the $\hat{\psi}_k^{n+1}$,
$-N_z/2\leqslant k\leqslant N_z/2-1$;
(ii) solve equation \eqref{equ:phi_weakform} for the $\hat{\phi}_k^{n+1}$,
$-N_z/2\leqslant k\leqslant N_z/2-1$.

\paragraph{Pressure}

Take the $L^2$ inner product between equation \eqref{e4.9a} and
$\nabla\bar{Q}_k$, where $\bar{Q}_k(\vec{x},z)$
denotes the 3D test function given by \eqref{equ:3d_test_func},
and we get
\begin{multline}
\int_\Omega \nabla P^{n+1}\cdot\nabla \bar{Q}_k =\rho_0\int_\Omega\textbf{T}\cdot\nabla \bar{Q}_k
   -\rho_0\int_{\partial \Omega } \frac{\mu^{n+1}}{\rho^{n+1}}\textbf{n}\times\bm{\omega}^{*,n+1}\cdot\nabla \bar{Q}_k -\frac{\rho_0 \gamma_0}{\Delta t}\int_{\partial\Omega} 
   \textbf{n}\cdot\textbf{w}^{n+1}\bar{Q}_k, \\
   -\frac{N_z}{2}\leqslant k\leqslant \frac{N_z}{2}, 
   \quad \forall\varphi
   \label{equ:p_trans_1}
\end{multline}
where $\bm{\omega}=\nabla\times\mathbf{u}$ is the
vorticity, and
we have used integration by part, equation \eqref{e4.9c} and the identity
$
\frac{\mu}{\rho}\nabla\times\bm{\omega}\cdot\nabla \bar{Q}_k=\nabla \cdot\left(\frac{\mu}{\rho}\bm{\omega} \times\nabla \bar{Q}_k\right)-\nabla\left(\frac{\mu}{\rho}\right)\times\bm{\omega}\cdot \bar{Q}_k.
$
In this equation,
\begin{multline}
\textbf{T}=\frac{1}{\rho^{n+1}} \left[\textbf{f}^{n+1}-\lambda(\psi^{n+1}-\alpha\phi^{n+1})\nabla\phi^{n+1}+\nabla\mu^{n+1}\cdot\textbf{D}(\textbf{u}^{*,n+1})\right]
+\frac{\hat{\textbf{u}}}{\Delta t}-\textbf{u}^{*,n+1}\cdot\nabla\textbf{u}^{*,n+1} \\
+\left(\frac{1}{\rho_0}-\frac{1}{\rho^{n+1}}\right)\nabla P^{*,n+1}
+ \nabla\left(\frac{\mu^{n+1}}{\rho^{n+1}}\right)\times\bm{\omega}^{*,n+1}.
\label{equ:def_T}
\end{multline}
where we have used equation \eqref{equ:CH_helm_2}.

Let 
$
\mathbf{J}^{n+1} = (\mathbf{J}_{2D},J_z)=(J_x,J_y,J_z)
= \frac{\mu^{n+1}}{\rho^{n+1}}\textbf{n}\times\bm{\omega}^{*,n+1}
$
defined on $\partial\Omega$.
In light of the relations \eqref{equ:F_relations}--\eqref{equ:F_relations_1},
equation \eqref{equ:p_trans_1} is reduced to
\begin{multline}
\int_{\Omega_{2D}}\nabla_{2D}\hat{P}_k^{n+1}\cdot\nabla_{2D}\varphi
+\beta_k^2\int_{\Omega_{2D}}\hat{P}_k^{n+1}\varphi
=\rho_0\int_{\Omega_{2D}}\hat{\mathbf{T}}_{2D,k}\cdot\nabla_{2D}\varphi
-i\beta_k\rho_0\int_{\Omega_{2D}}\hat{T}_{z,k}\varphi \\
-\rho_0\int_{\partial\Omega_{2D}}\hat{\mathbf{J}}_{2D,k}^{n+1}\cdot\nabla_{2D}\varphi
+i\beta_k\rho_0\int_{\partial\Omega_{2D}}\hat{J}_{z,k}^{n+1}\varphi
-\frac{\gamma_0\rho_0}{\Delta t}\int_{\partial\Omega_{2D}}
\mathbf{n}_{2D}\cdot\hat{\mathbf{w}}_{2D,k}^{n+1}\varphi, \\
\quad -\frac{N_z}{2}\leqslant k\leqslant \frac{N_z}{2}-1,
\quad \forall\varphi
\label{equ:p_weakform}
\end{multline}
where $\hat{P}_k^{n+1}$ are the Fourier coefficients of $P^{n+1}$,
$(\hat{\mathbf{T}}_{2D,k},\hat{T}_{z,k})$ are the Fourier coefficients
of $\mathbf{T}^{n+1}$, 
and $(\hat{\mathbf{J}}_{2D,k}^{n+1},J_{z,k}^{n+1})$ are 
the Fourier coefficients
of $\mathbf{J}^{n+1}$.
$\hat{\mathbf{w}}_{2D,k}^{n+1}$ are the Fourier expansion coefficients
of the component vector of $\mathbf{w}^{n+1}$ in the $x$-$y$ plane.
This equation is the weak form about the $N_z$ Fourier modes of
the pressure $\hat{P}_k^{n+1}$ in the 2D $x-y$ plane.
The terms on RHS can be computed explicitly once $\phi^{n+1}$ and
$\psi^{n+1}$ are computed. The terms $\mathbf{T}$ and $\mathbf{J}^{n+1}$
are first computed in the physical space, which can then be 
transformed into the Fourier space for computing the RHS of
this equation.

\begin{remark}

Since the Fourier modes are complex-valued, due to the $i\beta_k$
terms on the RHS of \eqref{equ:p_weakform},
the real part (resp. imaginary part) of $\hat{P}_k^{n+1}$
will be affected by the imaginary part (resp. real part)
of $\hat{T}_{z,k}$ and $\hat{J}_{z,k}^{n+1}$. 
These are prone to errors in the implementation.

\end{remark}

\paragraph{Velocity}

Equation  \eqref{e4.10a} can be written as
\begin{equation}
 \frac{\gamma_0 }{\nu_m\Delta t}\textbf{u}^{n+1} -\nabla^2\textbf{u}^{n+1}=
 \frac{1}{\nu_m}\left[\textbf{Y} -\nabla\left(\frac{\mu^{n+1}}{\rho^{n+1}} \right)\times\bm{\omega}^{*,n+1}
 \right] 
 -\frac{1}{\nu_m}\left(\frac{\mu^{n+1}}{\rho^{n+1}}-\nu_m\right)\nabla\times\bm{\omega}^{*,n+1}
\label{equ:vel_trans_1}
\end{equation}
where 
\begin{equation}
\mathbf{Y} = \mathbf{T} -\frac{1}{\rho_0}\nabla P^{n+1}
\label{equ:def_Y}
\end{equation}
and $\mathbf{T}$ is given by \eqref{equ:def_T}.

Let $\varphi^{(0)}(\vec{x})$ denote an arbitrary test function in
the 2D $x-y$ plane that vanishes on $\partial\Omega_{2D}$,
i.e.~$\left.\varphi^{(0)}\right|_{\partial\Omega_{2D}}=0$.
Define the test function for 3D
\begin{equation}
\bar{Q}^{(0)}_k(\vec{x},z) = \varphi^{(0)}(\vec{x})\bar{\Phi}_k(z),
\quad -\frac{N_z}{2}\leqslant k\leqslant \frac{N_z}{2}-1
\label{equ:def_testfunc_3d_vel}
\end{equation}
which satisfies $\left.\bar{Q}_k^{(0)}\right|_{\partial\Omega}=0$.
Take the $L^2$ inner product between equation \eqref{equ:vel_trans_1}
and the test function $\bar{Q}_k^{(0)}$, and we have
\begin{multline}
\frac{\gamma_0}{\nu_m\Delta t}\int_{\Omega}\mathbf{u}^{n+1}\bar{Q}_k^{(0)}
+\int_{\Omega}\nabla\bar{Q}_k^{(0)}\cdot\nabla\mathbf{u}^{n+1}
=\frac{1}{\nu_m}\int_{\Omega}\mathbf{Y}\bar{Q}_k^{(0)}
-\frac{1}{\nu_m}\int_{\Omega}\left( 
\frac{\mu^{n+1}}{\rho^{n+1}}-\nu_m
\right)\bm{\omega}^{*,n+1}\times\nabla\bar{Q}_k^{(0)}, \\
-\frac{N_z}{2}\leqslant k\leqslant \frac{N_z}{2}-1,
\quad \forall\varphi^{(0)}|_{\partial\Omega_{2D}}=0
\label{equ:vel_trans_2}
\end{multline}
where we have used integration by part, the property 
$\left.\bar{Q}_k^{(0)} \right|_{\partial\Omega}=0$, and
the identity ($\kappa$ and $\xi$ denoting two scalar functions),
$
\int_{\Omega} (\nabla\times\bm{\omega})\kappa\xi
=\int_{\partial\Omega}(\mathbf{n}\times\bm{\omega})\kappa\xi
+\int_{\Omega}(\bm{\omega}\times\nabla\kappa)\xi
+\int_{\Omega}(\bm{\omega}\times\nabla\xi)\kappa.
$

Let 
\begin{equation}
\mathbf{K}^{n+1} = \left(\frac{\mu^{n+1}}{\rho^{n+1}}-\nu_m \right)
\bm{\omega}^{*,n+1}
\label{equ:def_K}
\end{equation}
and define ($f(\vec{x})$ denoting a generic function in the $x$-$y$ plane)
\begin{equation}
\nabla_3 = (\nabla_{2D},-i\beta_k), \quad
\nabla_3 f = (\nabla_{2D}f, -i\beta_k f).
\label{equ:def_grad_3}
\end{equation}
The last term on the RHS of equation \eqref{equ:vel_trans_2}
can be transformed into
\begin{equation}
\begin{split}
\int_{\Omega}\mathbf{K}^{n+1}\times\nabla\bar{Q}_k^{(0)}
&=\sum_{m=-N_z/2}^{N_z/2-1}\left(\int_{\Omega_{2D}}\hat{\mathbf{K}}^{n+1}_m\times
\nabla_3\varphi^{(0)}\right)\left(\int_{0}^{L_z}\Phi_m(z)\bar{\Phi}_k(z)
\right) \\
&=L_z\int_{\Omega_{2D}}\hat{\mathbf{K}}^{n+1}_k\times
\nabla_3\varphi^{(0)}
\end{split}
\label{equ:vel_term}
\end{equation}
where $\hat{\mathbf{K}}_k^{n+1}$ is the Fourier expansion
coefficients of $\mathbf{K}^{n+1}$, and we have used the equation
\eqref{equ:F_ortho}.

In light of the relations \eqref{equ:F_relations}--\eqref{equ:F_relations_1}
and \eqref{equ:vel_term}, we can transform equation
\eqref{equ:vel_trans_2} into
\begin{multline}
\left(\frac{\gamma_0}{\nu_m\Delta t}+\beta_k^2 \right)
\int_{\Omega_{2D}}\hat{\mathbf{u}}_k^{n+1}\varphi^{(0)}
+\int_{\Omega_{2D}}\nabla_{2D}\varphi^{(0)}\cdot\nabla\hat{\mathbf{u}}_k^{n+1} =
\frac{1}{\nu_m}\int_{\Omega_{2D}}\hat{\mathbf{Y}}_k\varphi^{(0)} \\
-\frac{1}{\nu_m}\int_{\Omega_{2D}}
\hat{\mathbf{K}}_k^{n+1}\times\nabla_3\varphi^{(0)},
\quad -\frac{N_z}{2}\leqslant k\leqslant \frac{N_z}{2}-1,
\quad \forall\varphi^{(0)}|_{\partial\Omega_{2D}}=0,
\label{equ:vel_weakform}
\end{multline}
where $\hat{\mathbf{u}}_k^{n+1}$ and $\hat{\mathbf{Y}}_k$
are the Fourier expansion coefficients of $\mathbf{u}^{n+1}$
and $\mathbf{Y}$, respectively.
This is the weak form about the velocity Fourier modes
$\hat{\mathbf{u}}_k^{n+1}$ in the 2D $x$-$y$ plane.
It is noted that the three velocity components 
are not coupled, and the different Fourier modes for any
velocity component are also de-coupled.
Furthermore, the computations for the real and imaginary
parts of each Fourier mode of any of the flow variables
($\psi^{n+1}$, $\phi^{n+1}$, $P^{n+1}$ and $\mathbf{u}^{n+1}$)
are un-coupled.
When computing the RHS of this equation,
$\mathbf{Y}$ and $\mathbf{K}^{n+1}$ will be first computed
in physical space based on equations \eqref{equ:def_Y}
and \eqref{equ:def_K}, and then their
Fourier coefficients can be calculated.

\begin{remark}

Because $\nabla_3\varphi^{(0)}$ involves an $i\beta_k$ term,
the real parts (resp.~imaginary parts) of $\hat{\mathbf{K}}_k^{n+1}$
will contribute to the imaginary parts (resp.~real parts)
of $\hat{\mathbf{u}}_k^{n+1}$. In addition,
the cross product in the last term of equation \eqref{equ:vel_weakform}
will blend the contributions of this term to different
velocity components. These points are error-prone in
the implementation.

\end{remark}

\paragraph{Spectral Element Discretization in 2D $x$-$y$ Plane}


Let us now  discuss the discretization
of the equations \eqref{equ:psi_weakform}, \eqref{equ:phi_weakform},
\eqref{equ:p_weakform} and \eqref{equ:vel_weakform} using $C^0$ spectral
elements~\cite{SherwinK1995,KarniadakisS2005,ZhengD2011} in the 2D domain $\Omega_{2D}$.
We partition  $\Omega_{2D}$ with a spectral element mesh.
Let $\Omega_{2D,h}$ denote 
the discretized $\Omega_{2D}$, 
$
\Omega_{2D,h} = \cup_{e=1}^{N_{el}}\Omega_{2D,h}^e,
$
where $N_{el}$ is the number of elements in the mesh
and $\Omega_{2D,h}^e$ ($1\leqslant e\leqslant N_{el}$) denotes
the element $e$.
Let $\partial\Omega_{2D,h}$ denote
the boundary of $\Omega_{2D,h}$.
We use $\prod_K(\Omega_{2D,h}^e)$ to denote the linear space of
polynomials of degree characterized by $K$,
which will be referred to as the element order hereafter.
Let $H^1(\Omega_{2D,h})$ denote the set of globally continuous
and square-integrable functions defined on $\Omega_{2D,h}$.
Define function spaces
\begin{equation}
\left\{
\begin{split}
&
X_h=\left\{\ v\in H^{1}(\Omega_{2D,h})\ :\  v|_{\Omega_{2D,h}^e}\in\Pi_K(\Omega_{2D,h}^e), \ 1\leqslant e\leqslant N_{el}   \  \right\}, \\
&
X_{h0} = \left\{\ v\in X_h \ :\ v|_{\partial\Omega_{2D,h}}=0
\  \right\}.
\end{split}
\right.
\label{equ:def_func_space}
\end{equation}

In what follows, we use $(\cdot)_h$ to denote the discretized
version of variable $(\cdot)$, and
use $Re(\cdot)$ and $Im(\cdot)$ to denote the real and imaginary
parts of a complex-valued variable $(\cdot)$.
The fully discretized version of equation \eqref{equ:psi_weakform}
reads: find $\hat{\psi}_{k,h}^{n+1}$ such that
$Re(\hat{\psi}_{k,h}^{n+1})\in X_h$ and 
$Im(\hat{\psi}_{k,h}^{n+1})\in X_h$ and
\begin{multline}
\int_{\Omega_{2D,h}}\nabla_{2D}\hat{\psi}_{kh}^{n+1}\cdot\nabla_{2D}\varphi_h
+\left(\alpha + \frac{S}{\eta^2}+\beta^2_k \right)\int_{\Omega_{2D,h}}\hat{\psi}_{kh}^{n+1}\varphi_h
=\int_{\Omega_{2D,h}}\left(\beta_k^2\hat{R}_{2,kh}-\hat{R}_{1,kh}\right)\varphi_h \\
+\int_{\Omega_{2D,h}}\nabla_{2D,h}\hat{R}_{2,kh}\cdot\nabla_{2D}\varphi_h 
+\int_{\partial\Omega_{2D,h}} \hat{\mathcal{T}}_{kh}\varphi_h,
\quad -\frac{N_z}{2}\leqslant k\leqslant \frac{N_z}{2}-1, \quad \forall \varphi_h\in X_h.
\label{equ:psi_weakform_disc}
\end{multline}
The fully discretized version of equation
\eqref{equ:phi_weakform} reads: find $\hat{\phi}_{kh}^{n+1}$
such that $Re(\hat{\phi}_{kh}^{n+1})\in X_h$ and
$Im(\hat{\phi}_{kh}^{n+1})\in X_h$ and
\begin{multline}
\int_{\Omega_{2D,h}}\nabla_{2D}\hat{\phi}_{kh}^{n+1}\cdot\nabla_{2D}\varphi_h
+\left(-\alpha+\beta_k^2\right)\int_{\Omega_{2D,h}}\hat{\phi}_{kh}^{n+1}\varphi_h
=-\int_{\Omega_{2D,h}}\hat{\psi}_{kh}^{n+1}\varphi_h
-\int_{\partial\Omega_{2D,h}} \hat{\mathcal{M}}_{kh}\varphi_h, \\
-\frac{N_z}{2}\leqslant k\leqslant \frac{N_z}{2}-1, 
\quad \forall \varphi_h\in X_h.
\label{equ:phi_weakform_disc}
\end{multline}
The fully discretized version of equation \eqref{equ:p_weakform}
reads: find $\hat{P}_{kh}^{n+1}$ such that
$Re(\hat{P}_{kh}^{n+1})\in X_h$ and
$Im(\hat{P}_{kh}^{n+1})\in X_h$ and
\begin{multline}
\int_{\Omega_{2D,h}}\nabla_{2D}\hat{P}_{kh}^{n+1}\cdot\nabla_{2D}\varphi_h
+\beta_k^2\int_{\Omega_{2D,h}}\hat{P}_{kh}^{n+1}\varphi_h
=\rho_0\int_{\Omega_{2D,h}}\hat{\mathbf{T}}_{2D,kh}\cdot\nabla_{2D}\varphi_h
-i\beta_k\rho_0\int_{\Omega_{2D,h}}\hat{T}_{z,kh}\varphi_h \\
-\rho_0\int_{\partial\Omega_{2D,h}}\hat{\mathbf{J}}_{2D,kh}^{n+1}\cdot\nabla_{2D}\varphi_h
+i\beta_k\rho_0\int_{\partial\Omega_{2D,h}}\hat{J}_{z,kh}^{n+1}\varphi_h
-\frac{\gamma_0\rho_0}{\Delta t}\int_{\partial\Omega_{2D,h}}
\mathbf{n}_{2D,h}\cdot\hat{\mathbf{w}}_{2D,kh}^{n+1}\varphi_h, \\
\quad -\frac{N_z}{2}\leqslant k\leqslant \frac{N_z}{2}-1,
\quad \forall\varphi_h\in X_h.
\label{equ:p_weakform_disc}
\end{multline}
The fully discretized version of equation \eqref{equ:vel_weakform}
reads: find $\hat{\mathbf{u}}_k^{n+1}$ such that
$Re(\hat{\mathbf{u}}_k^{n+1})\in [X_h]^3$ and
$Im(\hat{\mathbf{u}}_k^{n+1})\in [X_h]^3$ and
\begin{multline}
\left(\frac{\gamma_0}{\nu_m\Delta t}+\beta_k^2 \right)
\int_{\Omega_{2D,h}}\hat{\mathbf{u}}_{kh}^{n+1}\varphi_h
+\int_{\Omega_{2D,h}}\nabla_{2D}\varphi_h\cdot\nabla\hat{\mathbf{u}}_{kh}^{n+1} =
\frac{1}{\nu_m}\int_{\Omega_{2D,h}}\hat{\mathbf{Y}}_{kh}\varphi_h \\
-\frac{1}{\nu_m}\int_{\Omega_{2D,h}}
\hat{\mathbf{K}}_{kh}^{n+1}\times\nabla_3\varphi_h,
\quad -\frac{N_z}{2}\leqslant k\leqslant \frac{N_z}{2}-1,
\quad \forall\varphi_h\in X_{h0}.
\label{equ:vel_weakform_disc}
\end{multline}

The velocity Dirichlet boundary condition \eqref{e4.10b}
also needs to be discretized. Upon Fourier expansion in
$z$ direction, the fully discretized version of equation
\eqref{e4.10b} becomes
\begin{equation}
\hat{\mathbf{u}}_{kh}^{n+1} = \hat{\mathbf{w}}_{kh}^{n+1},
\quad -\frac{N_z}{2}\leqslant k\leqslant \frac{N_z}{2}-1,
\quad \text{on} \ \partial\Omega_{2D,h}
\label{equ:dbc_disc}
\end{equation}
where $\hat{\mathbf{w}}_{kh}^{n+1}$ are the Fourier expansion coefficients
of the discretized boundary velocity $\mathbf{w}_h^{n+1}$.

\paragraph{Solution Algorithm}

Within a time step, with ($\phi_h^n$, $\psi^n_h$, $P_h^n$, $\mathbf{u}_h^n$) known,
we compute the physical variables at the new time step through the 
following procedure:
\begin{itemize}

\item
Solve equations \eqref{equ:psi_weakform_disc} for $\hat{\psi}_{kh}^{n+1}$
($-N_z/2\leqslant k\leqslant N_z/2-1$);

\item
Solve equations \eqref{equ:phi_weakform_disc} for $\hat{\phi}_{kh}^{n+1}$
($-N_z/2\leqslant k\leqslant N_z/2-1$);

\item 
Solve equations \eqref{equ:p_weakform_disc} for $\hat{P}_{kh}^{n+1}$
($-N_z/2\leqslant k\leqslant N_z/2-1$);

\item
Solve equations \eqref{equ:vel_weakform_disc}, together with the boundary
condition \eqref{equ:dbc_disc},
for $\hat{\mathbf{u}}_{kh}^{n+1}$ ($-N_z/2\leqslant k\leqslant N_z/2-1$).

\end{itemize}


It should be emphasized that all the equations to solve involved in this
algorithm are 2D equations in the $x$-$y$ plane, and that
all the equations are de-coupled thanks
to the Fourier expansions along the homogeneous direction.
These characteristics provide extensive opportunities 
for efficient parallel processing.
Furthermore, the linear algebraic systems resulting from these
equations all involve 
coefficient matrices that are time-independent. Therefore
these coefficient matrices only need to be computed once and
can be pre-computed.

\section{Representative Numerical Tests}
\label{sec:tests}

In this section we use several two-phase flow problems in three
dimensions to test the performance of the method developed
in the previous section and to assess its accuracy by
comparing simulation results with analytic solutions or with
theory. We will also investigate the effects of contact angles
on the  dynamics of
two-phase flows.

\begin{table}[tb]
\begin{center}
\begin{tabular}{ll | ll}
\hline
Variables/parameters & Normalization constant &
Variables/parameters & Normalization constant \\
$\mathbf{x}$, $\vec{x}$, $x$, $y$, $z$, $\eta$ & $L$
& $t$, $\Delta t$, $g(\mathbf{x},t)$ & $L/U_0$ \\
$\mathbf{u}$, $\mathbf{w}$ & $U_0$
& $\rho$, $\rho_1$, $\rho_2$, $\rho_0$ & $\varrho_d$ \\
$\mu$, $\mu_1$, $\mu_2$ & $\varrho_d U_0 L$
& $P$, $p$ & $\varrho_d U_0^2$\\
$\psi$ & $1/L^2$ 
& $\phi$, $\theta_s$, $S$, $\gamma_0$ & $1$ \\
$\sigma$ & $\varrho_d U_0^2 L$
& $\lambda$ & $\varrho_d U_0^2 L^2$ \\
$\gamma_1$ & $L/(\varrho_d U_0)$ 
& $\mathbf{f}$ & $\varrho_d U_0^2/L$ \\
$g_c$ & $L^3$ 
& $g_b$ & $1/L$ \\
\hline
\end{tabular}
\end{center}
\caption{Normalization constants for physical variables and
parameters. $L$, $U_0$ and $\varrho_d$ denote  characteristic
scales for the length, velocity and density, respectively.
}
\label{tab:general_normalization}
\end{table}

We first briefly discuss the normalization of the governing
equations and the physical
variables and parameters.
As expounded in previous works (see e.g.~\cite{Dong2014} for 
details), if the physical variables and parameters are normalized
consistently, the resultant non-dimensionalized
governing equations and boundary/initial
conditions for two-phase flows
 will retain the same 
forms as their dimensional counterparts. This will greatly simplify
the discussions and presentations. 
Let $L$ denote a characteristic length scale, $U_0$ denote a
characteristic velocity scale, and $\varrho_d$ denote
a characteristic density scale. The normalization constants
for consistently non-dimensionalizing different physical variables
and parameters are listed in Table \ref{tab:general_normalization}.
For example, the non-dimensional surface tension is
$\frac{\sigma}{\varrho_dU_0^2L}$ (inverse of the Weber number)
according to this table. In the following sections  all the
physical variables and parameters have been
consistently normalized based on Table~\ref{tab:general_normalization}, unless otherwise specified.

\subsection{Convergence Tests}

In this subsection
we employ a manufactured analytic solution to the system of 3D governing
equations for two-phase flows to demonstrate the convergence
rates in space and time for the method developed in
Section \ref{sec:method}.

\begin{figure}[tb]
  \centerline{
  \subfigure[Test Domain]
{\label{Test_Domain}\includegraphics[width=0.4\textwidth]{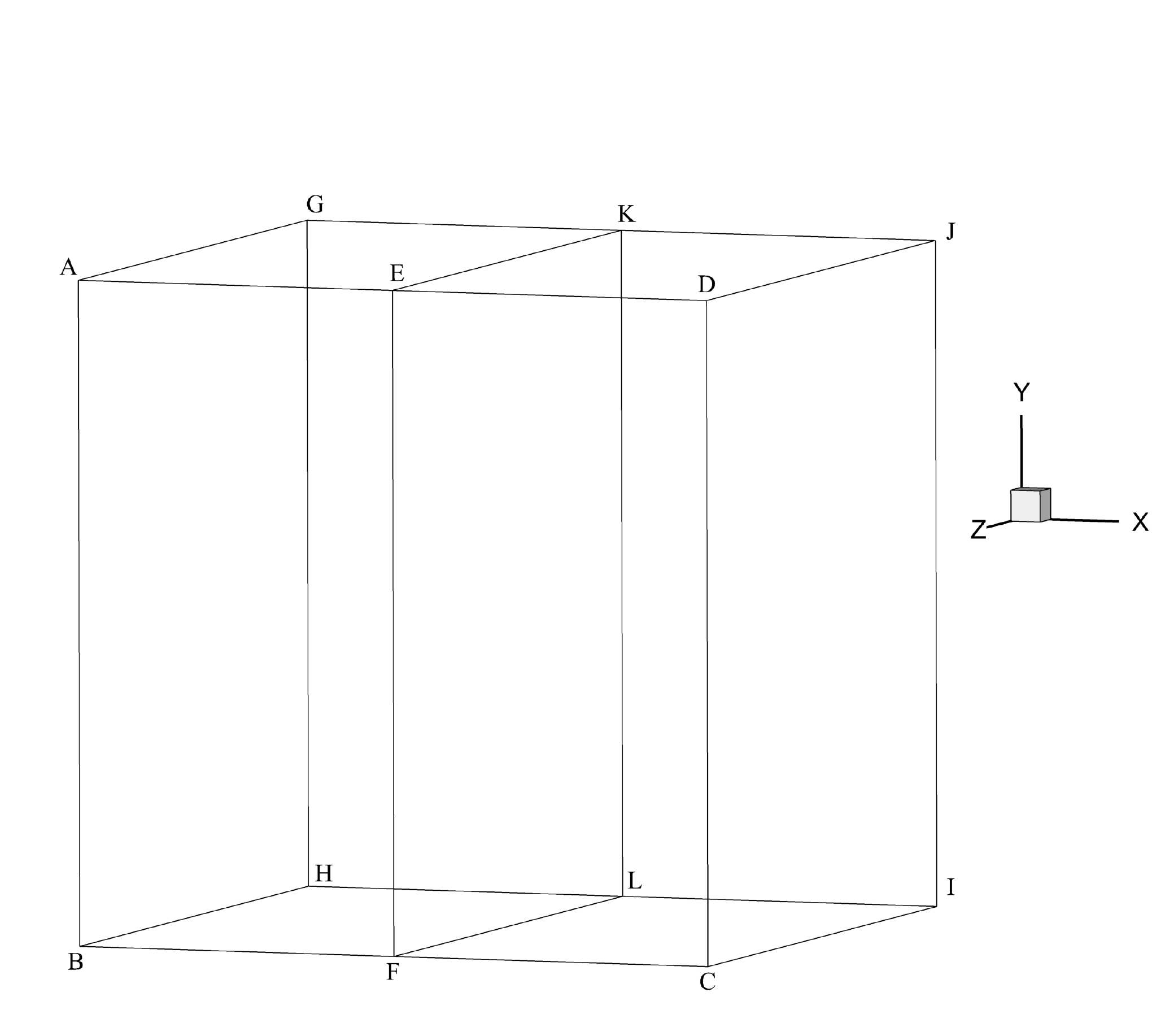}}
  }
  \centerline{
  \subfigure[Spatial Convergence]{\label{Spatial}\includegraphics[width=0.5\textwidth]{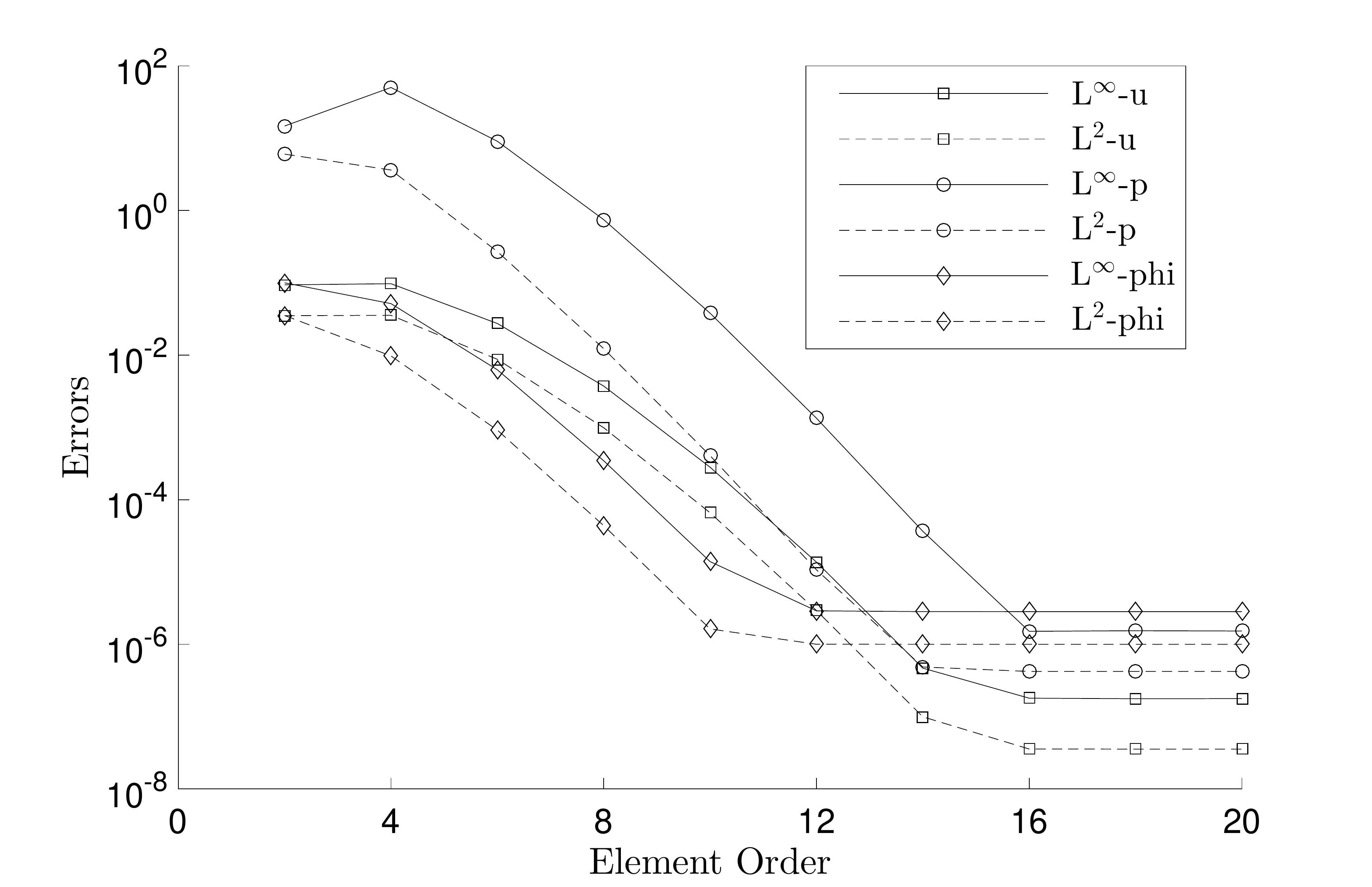}}%
  \subfigure[Temporal Convergence]{\label{Temporal}\includegraphics[width=0.5\textwidth]{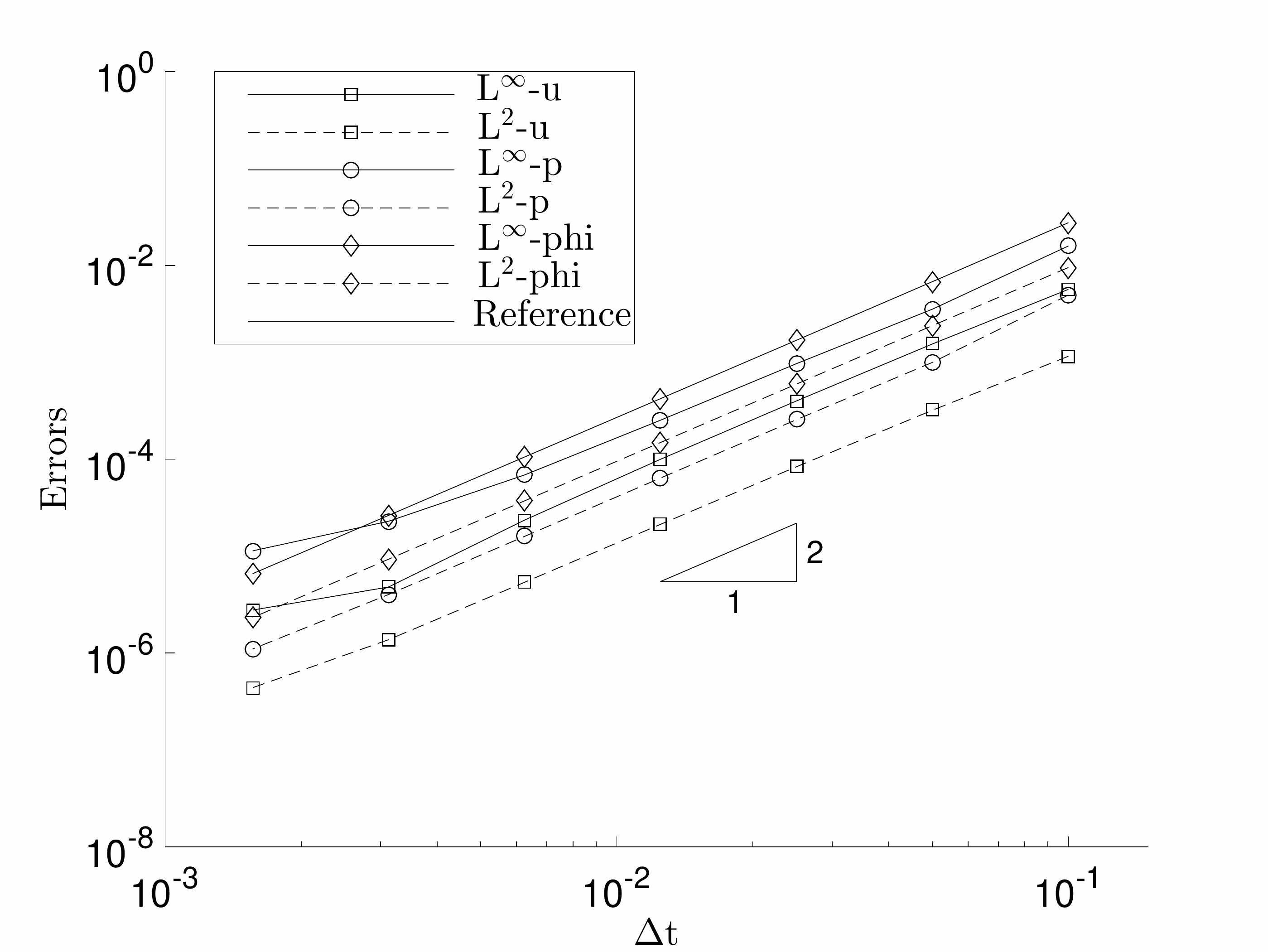}}
  }
  \caption{ Spatial/Temporal convergence rates: L$^\infty$ and L$^2$ errors as a function of the element order with fixed $\Delta$t=0.001 (a),
    and as a function of $\Delta$t with a fixed element order 16 (b).
  }
  \label{Fig.1}
\end{figure}

    Consider the flow domain sketched in Figure \ref{Fig.1}(a),
defined by $\Omega=[(x,y,z):0\leqslant x\leqslant 2,-1\leqslant y \leqslant 1,0\leqslant z\leqslant 2]$.  
Consider the following analytic solution to the 3D governing equations
\eqref{equ:nse_trans}, \eqref{e4.1b}--\eqref{e4.1c} on this
domain:
    \begin{equation}
    \begin{cases}
   u=A\cos(a x)\cos(b y)\cos(c z)\sin(\beta t)\\
   v=0\\
   w=\frac{A a}{c}\sin(a x)\cos(b y)\sin(c z)\sin(\beta t)\\
   P=A\sin(a x)\sin(b y)\sin(c z)\cos(\beta t)\\
   \phi=A_1\cos(a_1 x)\cos(b_1 y)\cos(c_1 z)\sin(\beta_1 t)
   \end{cases} 
   \label{e4.47}
    \end{equation}
where $A, A_1, b,c,a_1,b_1,c_1,\beta,\beta_1$ are prescribed
constants (given below), $(u,v,w)$ are the $(x,y,z)$ components of
the velocity \textbf{u}. $\phi$ is the phase field function and P is the effective pressure. The body force and the source field terms such as \textbf{f} in \eqref{equ:nse_trans}, $g$ in \eqref{e4.1c}, $g_c$ in \eqref{e4.4a}, $g_b$ in \eqref{e4.4b}, and the boundary velocity $\textbf{w}$ in \eqref{e4.3}, are chosen in a way such
that the analytical expressions in \eqref{e4.47} satisfy the governing equations \eqref{equ:nse_trans}, \eqref{e4.1b} and \eqref{e4.1c}, and also satisfy the boundary conditions \eqref{e4.3}, \eqref{e4.4a} and \eqref{e4.4b}.

We impose the Dirichlet boundary condition for the velocity based on \eqref{e4.47} and the contact angle boundary conditions \eqref{e4.4a} and \eqref{e4.4b} for the phase field function on the faces $\overline{ABHG}$, $\overline{DCIJ}$, $\overline{ADJG}$ and $\overline{BCIH}$.
We set the initial conditions for the velocity $\textbf{u}$ and the phase field function $\phi$ to be equal to the values obtained by the analytic expressions
in \eqref{e4.47} at time $t=0$.

The following parameter values are employed in the subsequent tests:
    \begin{equation}
    \begin{cases}
    A=1.0,\; a=1.0 \,\pi,\; b=1.5\, \pi,\; c=1.0\,\pi,\; a_1=b_1=c_1=1.0\, \pi,  \\
    A_1=1.0, \; \beta=1.0, \; \beta _1=1.0, \;\rho_1=1,\; \frac{\rho_2}{\rho_1}=3.0,\; \mu_1=0.01,\; \frac{\mu_2}{\mu_1}=2.0,\; J=2, \\
    \eta=0.1,\; \lambda=0.001,\; \gamma_1=0.001,\; \theta_s=60^0,\\
   \displaystyle \nu_m=\frac{1}{2}\left(\frac{\mu_1}{\rho_1}+\frac{\mu_2}{\rho_2} \right)=8.333 \times 10^{-3}, \; \rho_0=\min(\rho_1,\rho_2). \\
    \end{cases}
    \end{equation}
    To simulate the problem we discretize the domain using $8$ Fourier planes along the z-direction, and  further partition each plane along the x-direction using two quadrilateral spectral elements of equal sizes, as shown in Figure \ref{Fig.1}(a). The simulations are performed from time t=0 to a specified time t=t$_f$,
after which we calculate the errors by comparing the numerical results with the analytical solution at the final time t=t$_f$.
To study the spatial convergence rate of the method,
we fix the time step size to $\Delta t$=0.001 and also fix the number of time steps to be 100, and so the total integration time is $t_f$=0.1.
Then, we increase the element order sequentially from 2 to 20, and
carry out simulations for each element order. 
To study the temporal convergence rate, we fix the element order to a
large value $16$. Then, we reduce the time step size from $\Delta t$=0.1 to $\Delta t$=0.0015625 sequentially by a factor of two,
and also change the number of time steps  so that the total integration time is fixed at $t_f$=0.5. Simulations are performed for each $\Delta t$ and
the corresponding numerical errors are computed.

Figures \ref{Fig.1}(b) and \ref{Fig.1}(c) show the $L^{\infty}$ and $L^2$ errors of the velocity ($x$ component),
pressure and the phase field function for the spatial and temporal convergence tests respectively. From Figure \ref{Fig.1}(b), we observe that the errors  decrease exponentially as
the element order increases, until the element order
reaches around $10$ to $16$ (for different variables). The errors stagnate as the element order increases beyond $16$. This is because
at these high element orders the temporal truncation error becomes dominant and saturates the total error. From Figure \ref{Fig.1}(c), we observe a second-order convergence rate in time for the flow variables. 
        
The above results indicate that the current numerical
method  achieves a spatial exponential convergence rate and
a temporal second-order convergence rate for three-dimensional problems.

\subsection{Co-Current Flow of Two Immiscible Fluids in a Pipe}

\begin{figure}[tb]
\centering
  \includegraphics[width=0.5\textwidth]{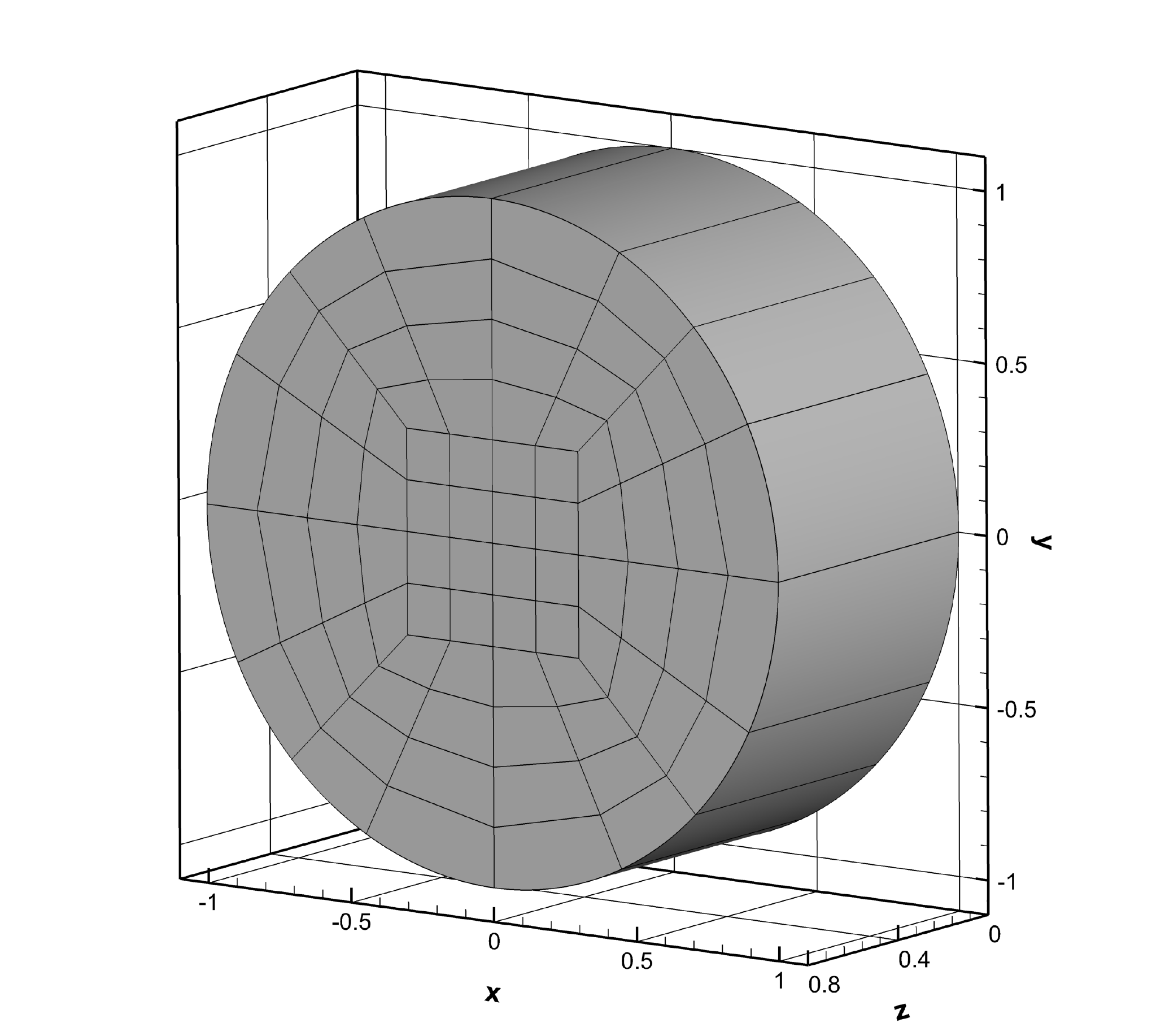}%
  \caption{ Geometry and mesh of co-current pipe flow.}
  \label{Fig.2}
\end{figure}

 In this subsection we simulate the co-current flow of two immiscible
incompressible fluids in a circular pipe, and compare simulation results
with the theoretical solution to this problem to test the accuracy
of the method developed herein.

Consider a fully-developed laminar co-current flow of two  immiscible incompressible
Newtonian fluids in a circular pipe. The radius of the  pipe is $r_2$. We assume that there is no interface instability between the two fluids and that the surface tension between them is zero. The first fluid occupies the inner core
region $0\leqslant r\leqslant r_1$,
where $0<r_1<r_2$,
and the second fluid occupies the outer region $r1\leqslant r \leqslant r_2$. 
The flow is driven by a pressure gradient $\bar{K}$ along
the axial direction.
Note that the flow is fully developed. There is no radial or azimuthal flow, and the only non-zero velocity component is in the axial direction.
We assume zero gravity in this problem.
Let $\mu_1$ and $\mu_2$ denote the dynamic viscosities of 
the two fluids. Note that the densities of the fluids have no effect on the fully-developed velocity profiles, and so
they are both taken as unit values in the simulation.

This problem has an exact physical solution.
The axial velocity profiles for the two fluids are 
given by \cite{NogueiraE.1990Htsi}:
  \begin{subequations}
 \begin{equation}
 \frac{w_1(r)}{\bar{w}}=\frac{2\left(1-\frac{r_1^2}{r_2^2}+\frac{\mu_2}{\mu_1}
 \frac{r_1^2-r^2}{r_2^2}\right)}{\frac{r_1^4}{r_2^4}
 \left(\frac{\mu_2}{\mu_1}-1\right)+1}, 
 \quad 0\leqslant r\leqslant r_1,
 \label{e49a}
 \end{equation}
 \begin{equation}
  \frac{w_2(r)}{\bar{w}}=\frac{ 2\left[1-\left(\frac{r}{r_2} \right)^2\right]} 
  { \frac{r_1^4}{r_2^4}
 \left(\frac{\mu_2}{\mu_1}-1\right)+1 }, 
 \quad r_1 \leqslant r\leqslant r_2,
 \label{e49b}
 \end{equation}
 \end{subequations}
where $w_1$ and $w_2$ denote the axial velocities of the first and second fluids respectively.  
$\bar{w}$ is the average velocity across the pipe, given by
 \begin{equation}
 \bar{w}=\frac{\bar{K}r_2^2}{8\mu_2}  
 \left[\frac{r_1^4}{r_2^4}
 \left(\frac{\mu_2}{\mu_1}-1\right)+1 \right].
 \end{equation}
We will simulate this problem and compare the results
with the analytic expressions given by 
equations \eqref{e49a}--\eqref{e49b}.


We consider a computational domain shown in
Figure \ref{Fig.2}. The axial dimension of the domain
is $\frac{4}{5}r_2$. All flow variables are 
assumed to be periodic along the axial direction.
When the surface tension is zero ($\sigma=0$),
$\lambda=\frac{3}{2\sqrt{2}}\sigma\eta=0$,
and the very nature of the partial differential equation \eqref{e4.1c}
for the phase field function changes, because of disappearance of
the terms containing the highest derivatives on
the right hand side.
Our current implementation of the algorithm cannot handle
this case for the convection equation. But we note that with
the convection equation, the phase field function is convected
along the pipe with the axial velocity,
and its equilibrium distribution will have isofaces that
are cylindrical surfaces around the pipe axis.
So for this problem we will employ the following distribution
for the phase field function,
$
\phi(\mathbf{x}) = -\tanh\frac{\sqrt{x^2+y^2}-r_1}{\sqrt{2}\eta}
$
($\eta$ is the interfacial thickness scale)
for solving the momentum equations
\eqref{e4.1a}--\eqref{e4.1b}. The
equation \eqref{e4.1c} with $\lambda=0$,
which reduces to the convection equation, will not be  solved.
No-slip condition for the velocity is imposed on the pipe wall.


We use the pipe radius $r_2$ as the length scale $L$, and the fluid
density (same density for the two fluids) as the density scale
$\varrho_d$. The velocity scale is chosen as
$U_0 = 10\sqrt{|\bar{K}|L}$, where $\bar{K}$ is the pressure drop
along the pipe axis over a unit length.
All the flow variables and parameters are then
normalized based on the constants given in
Table \ref{tab:general_normalization}.

  \begin{figure}[tb]
\centerline{
  \includegraphics[width=0.5\textwidth]{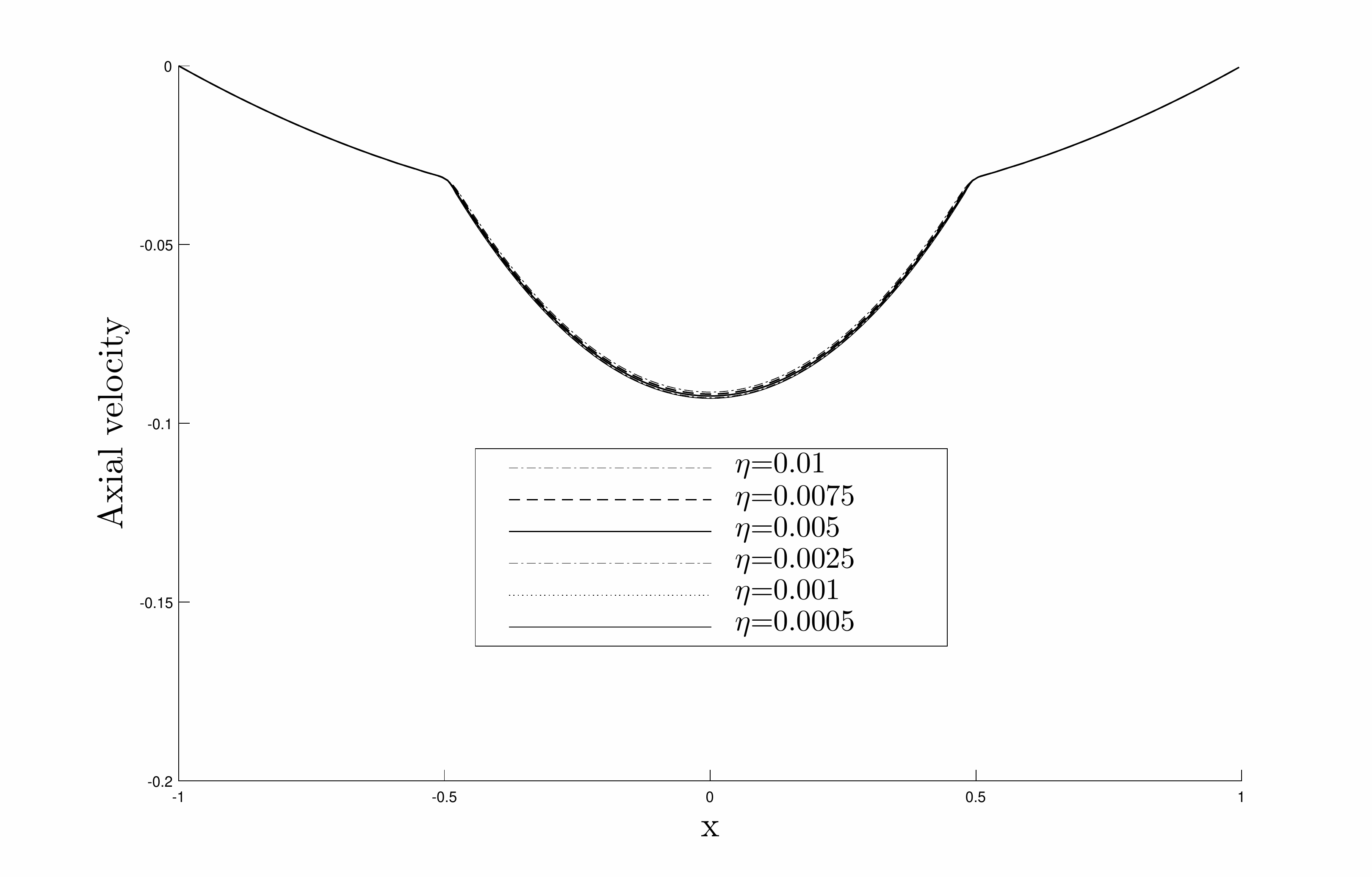}(a)
  \includegraphics[width=0.5\textwidth]{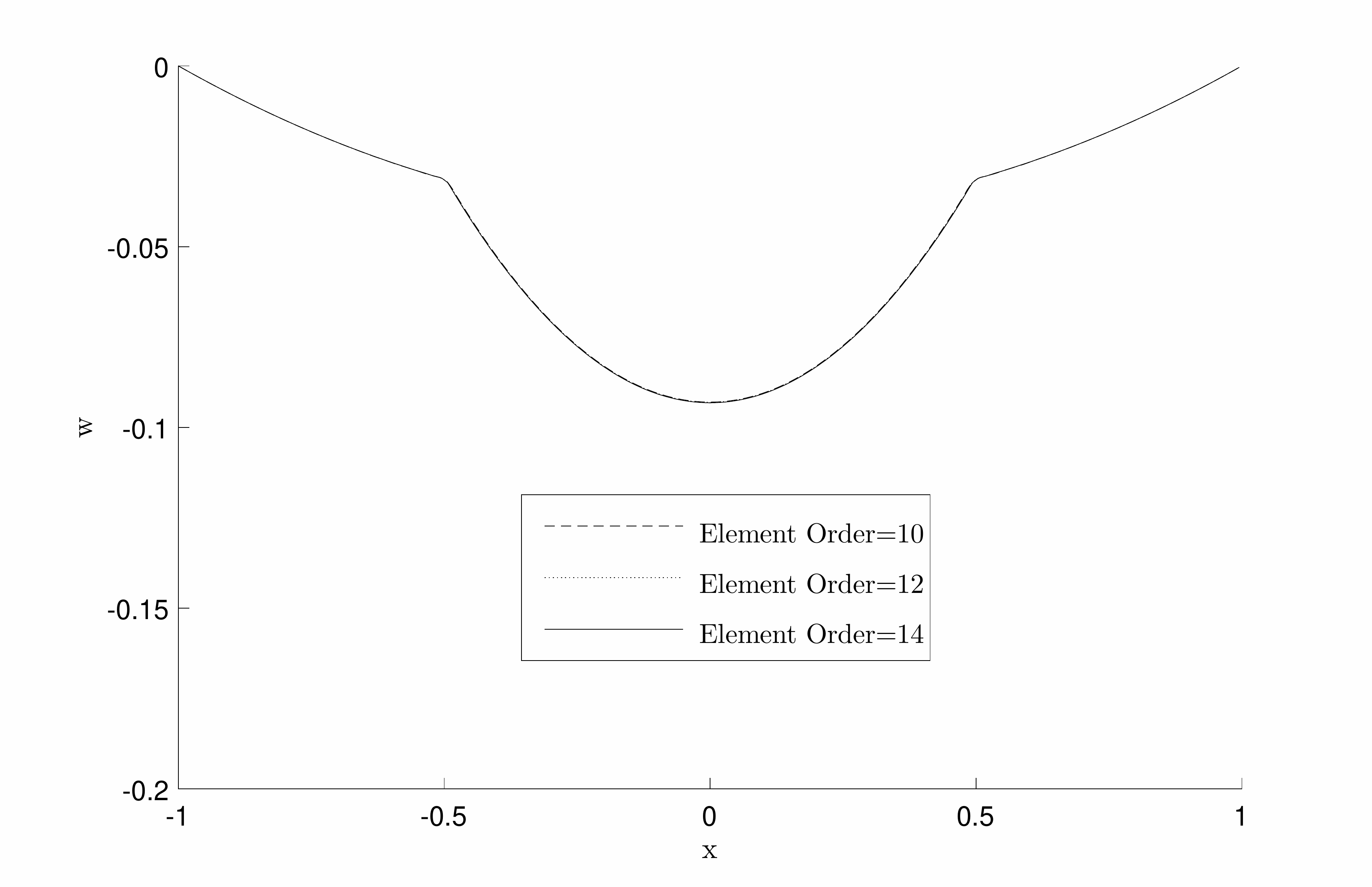}(b)
  }
  \caption{ 
  Steady-state axial velocity profiles obtained using
  different interfacial thickness scale $\eta$ (a),
  and different element orders (b).
  In (a) a fixed element order $10$ is employed,
  and in (b) a fixed $\eta=0.0005$ is employed.
  }
  \label{fig:pipe_eta}
\end{figure}


We discretize the computational domain using $16$ Fourier
planes along the axial ($z$) direction. Each plane is discretized
using a mesh of $80$ quadrilateral spectral elements
(see Figure \ref{Fig.2}), and the element order is varied
over a range of values in the tests.
Values for the non-dimensionalized 
simulation parameters are given below:
    \begin{equation}
    \begin{cases}
    r_1=0.5, \; r_2=1, \ 
    \mu_1=0.01, \ \frac{\mu_2}{\mu_1}=6, \ \rho_1=\rho_2=1, \\
    \bar{K}=-0.01, \
    \eta=0.0005 \ \text{(or varied)},\; J=2,\; \text{element order}=10\sim 14, \\
    \Delta t=5\times 10^{-5}. 
    \end{cases}
    \end{equation}




Starting with an initial zero velocity field,
we have performed a long-time simulation of this problem
using the method from  Section \ref{sec:method}, until
the flow has reached a steady state.
Figure \ref{fig:pipe_eta}(a) shows a study of the effect of
the interfacial thickness scale $\eta$ in the phase field
distribution
on the simulation results. Here we plot
the steady-state axial velocity profiles 
obtained with several $\eta$ values ranging
between $0.01$ and $0.0005$. A fixed element order $10$
is employed in these simulations.
In the region of the outer fluid, the velocity
profiles obtained with the various $\eta$ values
overlap with one another.
In the region of the inner fluid, on
the other hand, some
 differences can be observed. As $\eta$ decreases
from $0.01$, we initially observe an increase in 
the velocity magnitude in the inner fluid region.
When $\eta$ is decreased further (below about $\eta=0.001$),
no significant change in the velocity profiles
can be observed.

     \begin{figure}[tb]
\centering
  \includegraphics[width=0.6\textwidth]{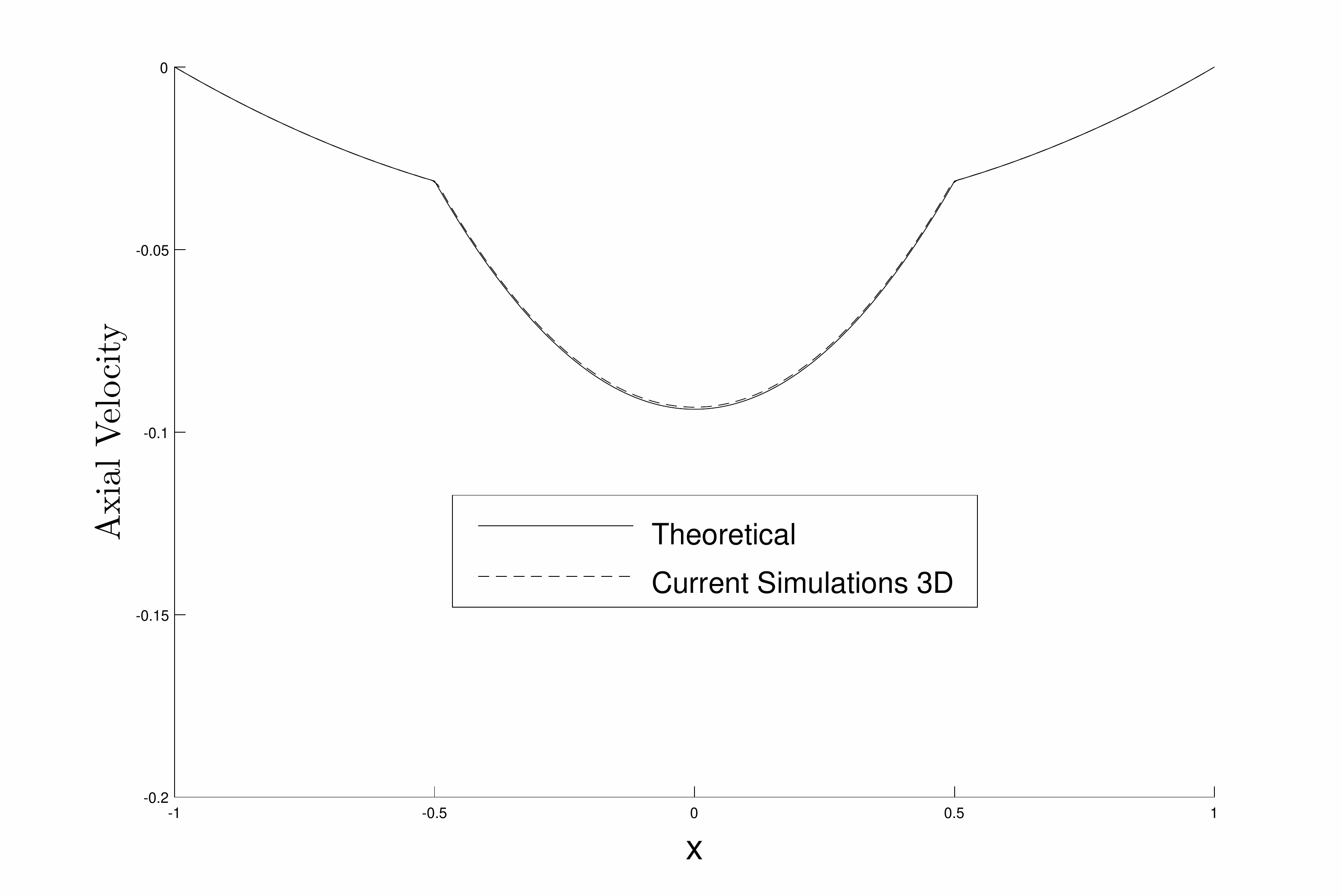}%
  \caption{ Comparison of steady-state axial velocity profiles
  from current simulation and the exact physical 
  solution of \cite{NogueiraE.1990Htsi}
  }
  \label{fig:pipe_compare}
\end{figure}

Fig. \ref{fig:pipe_eta}(b) illustrates the effect of
the spatial resolution in the simulations.
It is a plot of the 
profiles of the steady-state axial velocity across the pipe
obtained using several element orders ranging from
$10$ to $14$ in the simulations. The interfacial thickness
scale is fixed at $\eta=0.0005$ in these simulations.
The velocity profiles obtained with different element orders
essentially overlap with one another, with a negligible
difference. This suggests that these element orders
and the spatial resolutions are adequate for the simulations
of the current problem.



Fig. \ref{fig:pipe_compare} compares
the steady-state axial velocity profiles obtained 
from the current simulation and the exact physical solution
given by \cite{NogueiraE.1990Htsi}.
The simulation result corresponds to an element order $14$
and the interfacial thickness scale $\eta=0.0005$.
We observe that the simulation result is in good agreement
with the theoretical result, suggesting that
our method has captured the flow field accurately.

\subsection{Equilibrium Shape of a Liquid Drop on a Partially Wettable Wall}

        \begin{figure}[tb]
\centering
  \includegraphics[width=0.6\textwidth]{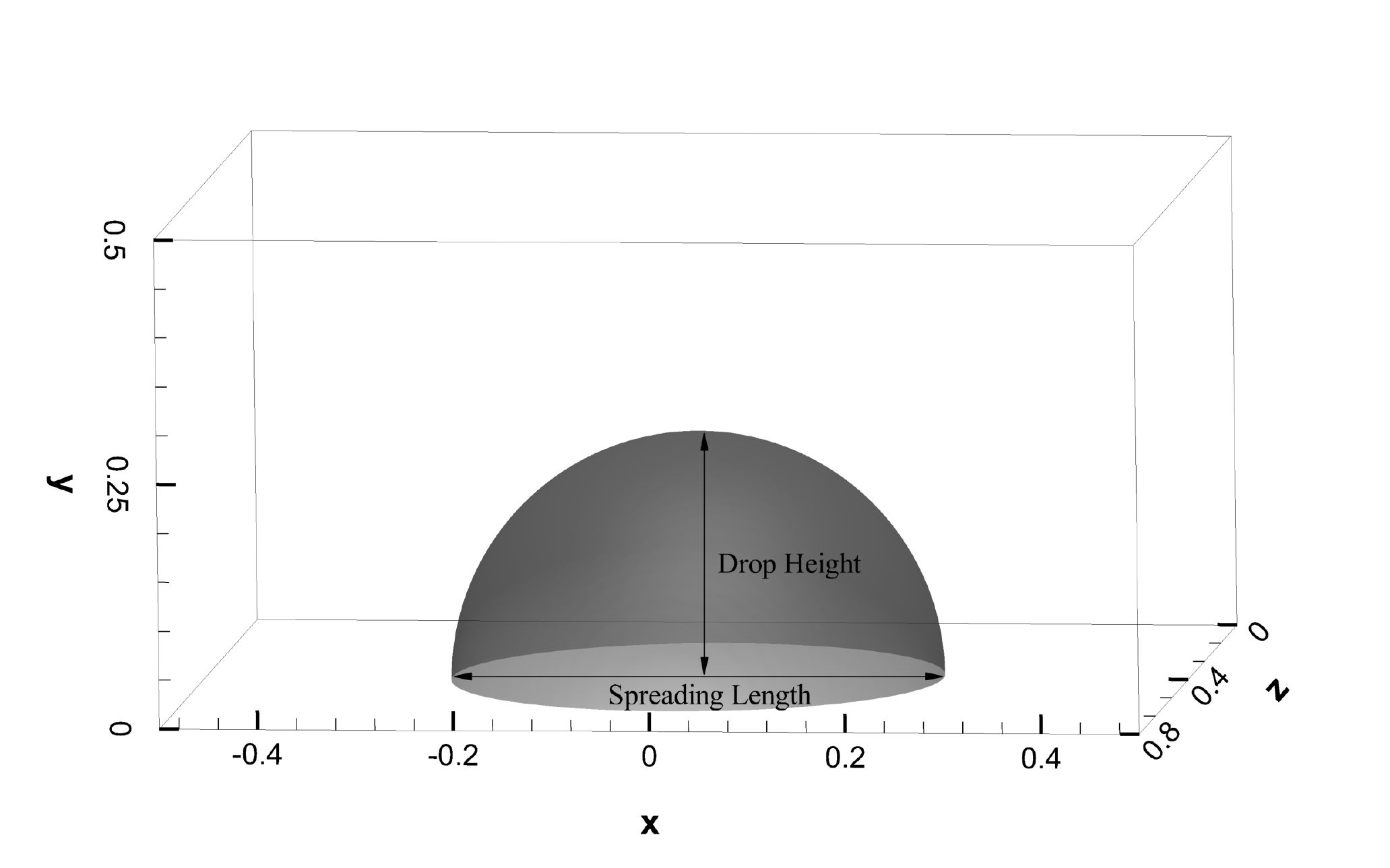}%
  \caption{Equilibrium liquid drop on the wall in an ambient fluid:
  sketch of the flow configuration and parameters.}  
  \label{WaterDropConfig}
\end{figure}

    In this subsection, we study the equilibrium configuration of
a liquid drop on a  horizontal wall that is partially wettable.
The effect of the contact angle on the equilibrium drop shape
will be investigated under zero gravity.

More specifically, we consider a liquid drop (the second fluid)
in an ambient lighter liquid (the first fluid); see
Figure \ref{WaterDropConfig} for a sketch of the problem.
The liquid drop (hemisphere) and the ambient fluid are initially at rest on
a horizontal wall with certain wetting properties.
The gravity is assumed to be zero for this problem.
The system is then released, and the drop starts to evolve
under the effect of surface tension between the two fluids 
and the imposed contact angle at the wall, eventually reaching
an equilibrium state. According to \cite{deGennesBQ2003},
the liquid drop  becomes a
spherical cap at equilibrium. We can define
the spreading length (diameter of the drop base) and
the drop height for the equilibrium shape; see
Figure \ref{WaterDropConfig} for a sketch of these parameters.
Our goal here is to study 
the equilibrium configuration of the drop, and compare
the simulation results with the theoretical results
from \cite{deGennesBQ2003}. 


To simulate the problem,
we consider the flow domain given by
$\Omega=\left\{(x,y,z):-\frac{L}{2}\leqslant x \leqslant \frac{L}{2}, 0\leqslant y\leqslant \frac{L}{2}, 0\leqslant z \leqslant \frac{4L}{5}\right\}$ as sketched in Figure \ref{WaterDropConfig},
where $L$ is a characteristic length scale. 
The initial drop shape is assumed to be a 
hemisphere with radius $R_0=\frac{L}{4}$.
We choose the density of the first fluid as
the characteristic density scale, i.e.~$\varrho_d=\rho_1$.
We choose the characteristic velocity scale as
$U_0 = \frac{1}{10}\sqrt{\frac{\sigma}{\varrho_dL}}$, where $\sigma$
is the surface tension.
All the physical variables and
parameters are then normalized based on
Table \ref{tab:general_normalization}.

Let R denote the radius of the spherical cap at equilibrium  and $\theta_e$ denote the contact angle measured on the side of
the heavier fluid (i.e.~the second fluid). Then,
based on the conservation of the volume of the liquid drop,
one can obtain the theoretical 
values of the spreading length and the drop height at
equilibrium as follows~\cite{deGennesBQ2003}:
\begin{equation}
\frac{R}{R_0}=\sqrt[\leftroot{-2}\uproot{15}3]{\frac{2}{(2+\cos\theta_e)(1-\cos\theta_e)^2}},\; L_s=2R\sin\theta_e,\; H=R(1-\cos\theta_e)
\label{equ:drop_theory}
\end{equation}
where $R_0$ is the radius of the initial hemisphere, 
$L_s$ is the spreading length, and $H$ is the drop height.

Along the $x$ and $z$ directions we impose periodic conditions for
all the flow variables. 
The top and bottom boundaries (i.e.~$y=0$ and $y=\frac{L}{2}$)
are assumed to be solid walls, and we impose the
no-slip condition 
for the velocity and the contact-angle boundary
conditions (with $g_a=0$ and $g_c=0$)
for the phase field function. To simulate the problem we discretize the
 flow domain using $96$ Fourier planes along the z direction, and each plane is further discretized using a mesh of $200$ quadrilateral spectral elements (with $20$ elements along the $x$ direction and $10$ elements
 along the $y$ direction).
The source term in the Cahn-Hilliard equation is set to $g=0$.
A number of contact angles ranging from $\theta_e=60^0$ to $120^0$ are
considered in the simulations.
%
The following non-dimensional parameter values have been used in the simulations:
           \begin{equation*}
        \begin{cases}
        \rho_1=1,\;\frac{\rho_2}{\rho_1}=5,\;\mu_1=0.05, \ \frac{\mu_2}{\mu_1}=2, \  \text{element order}=8\sim 12, \\
        \eta = 0.01, \ \sigma=100, \ \lambda = \frac{3}{2\sqrt{2}}\sigma\eta, \ 
        \gamma_1 = \frac{10^{-8}}{\lambda} \\
        \rho_0 = \min(\rho_1,\rho_2), \ 
        \nu_m = \frac{\max(\mu_1,\mu_2)}{\min(\rho_1,\rho_2)}, \ 
         \Delta t=1\times10^{-5}, \ J=2.
        \end{cases}
        \end{equation*}

\begin{figure}[tb]
  \subfigure[Contact angle $60^0$]{\label{Drop_120}\includegraphics[width=0.5\textwidth,scale=0.7]{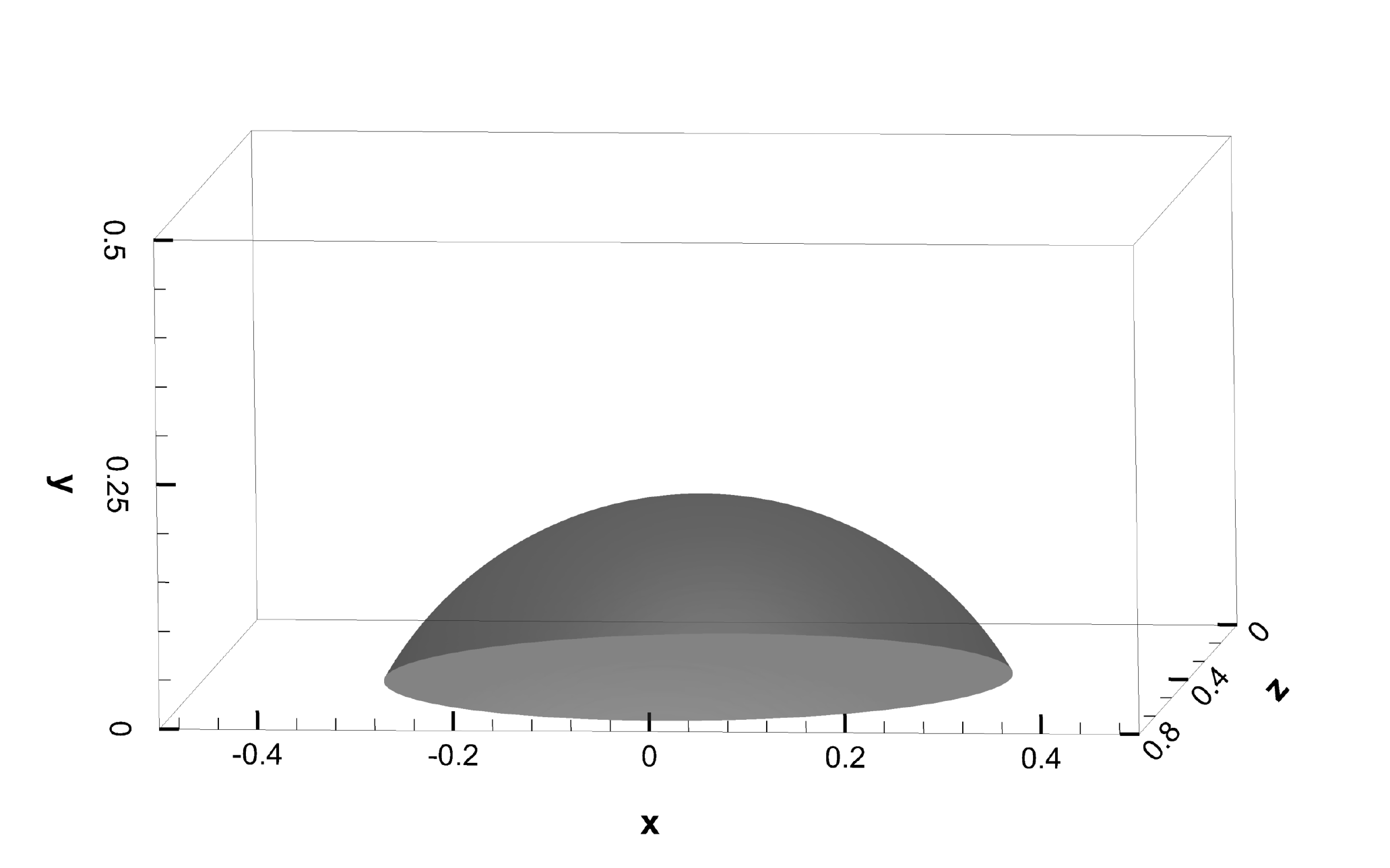}}%
  \hskip 0.2truein
  \subfigure[Contact angle $120^0$]{\label{Drop_60}\includegraphics[width=0.5\textwidth,scale=0.7]{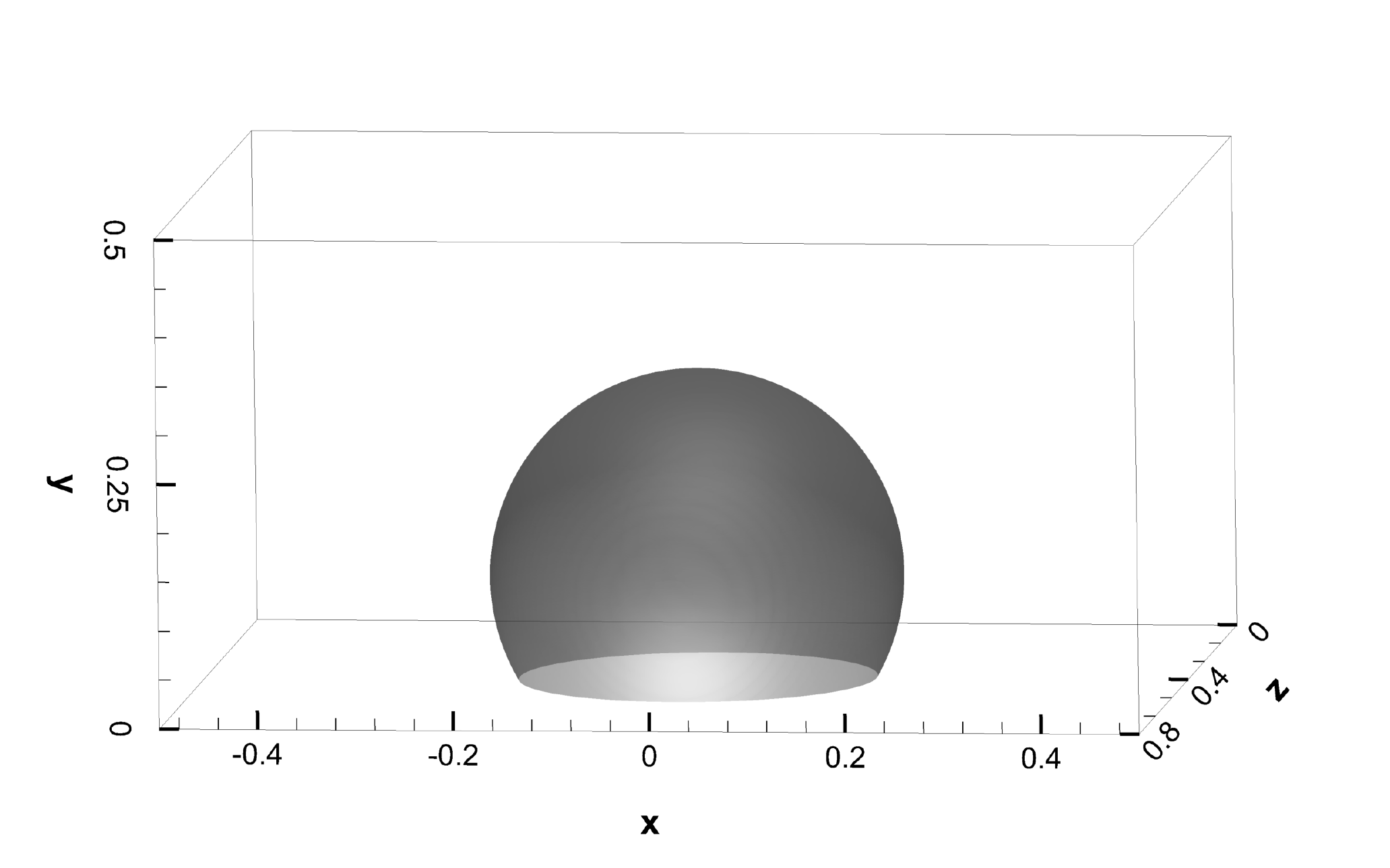}}
  \caption{ Liquid drop shapes at two contact angles:
  (a) $\theta_e =60^0$, (b) $\theta_e=120^0$. Plotted is the iso-surface of the
  phase field function $\phi=0$.
  }  
  \label{EqLiquidDrop}
\end{figure}

     Figures ~\ref{Drop_120} and ~\ref{Drop_60} show the equilibrium shapes of the liquid drop corresponding to the contact 
angles $\theta_e=60^0$ and $\theta_e=120^0$ respectively. 
Plotted here is the iso-surface $\phi=0$ for the phase field
function. It is evident that the drop resembles a spherical cap at
equilibrium,
qualitatively consistent with the theory~\cite{deGennesBQ2003}.

\begin{figure}[tb]
\centering
  \includegraphics[width=0.5\textwidth]{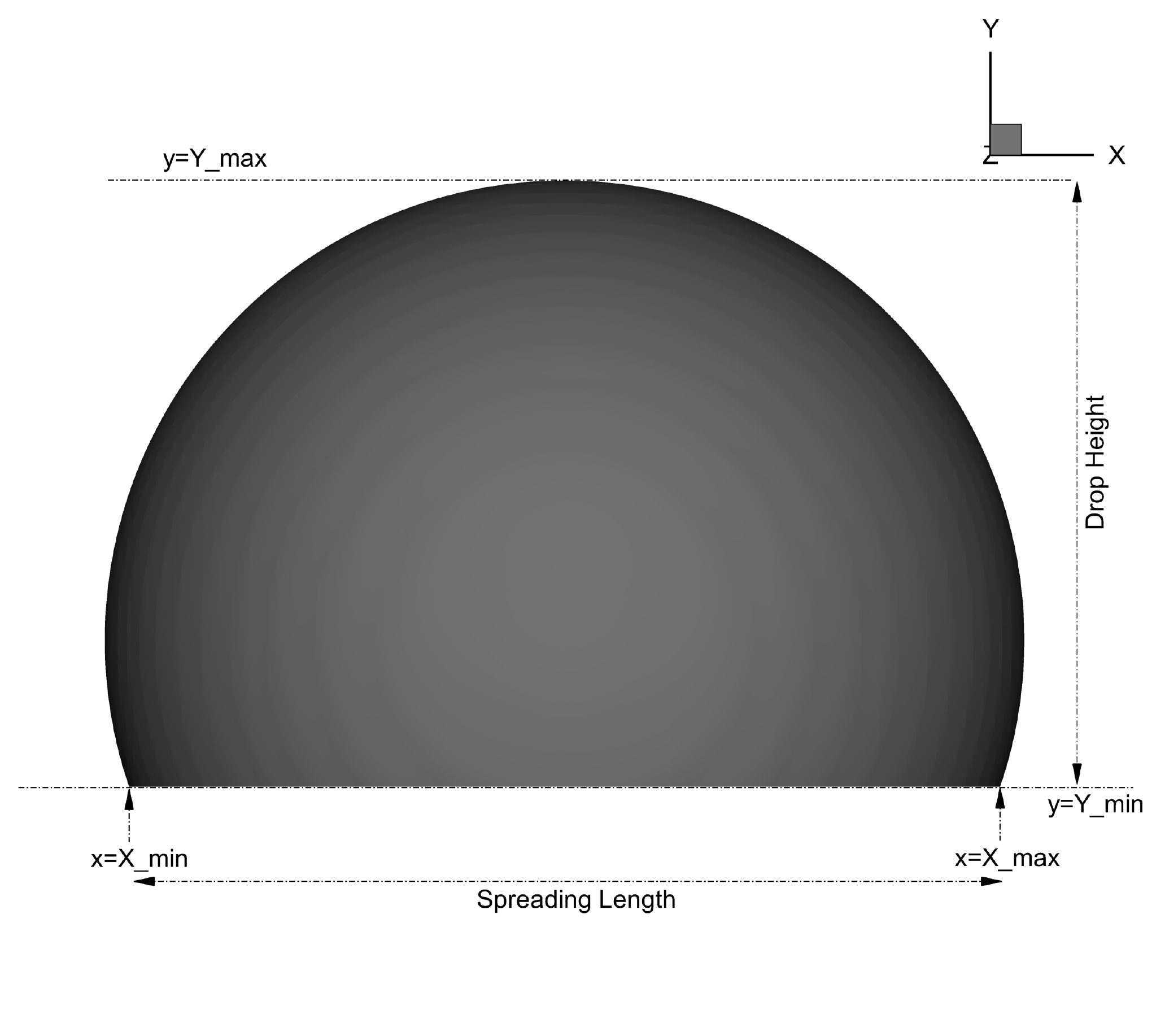}%
  \caption{Projection of the isosurface $\phi=0$ onto the $x-y$ plane,
  with $\theta_e=105^0$ }  
  \label{Isosurface}
\end{figure}

To make a quantitative comparison between the simulation and theory,
we need to extract and compute several parameters from the 
simulation data.
Figure \ref{Isosurface} illustrates how to extract the spreading
length and the height of the equilibrium drop. It
shows a projection of the iso-surface $\phi=0$ onto the $x-y$ plane using the visualization software 
Tecplot 360, which corresponds to a contact angle $\theta_e=105^0$. 
From the figure we can determine the minimum and the maximum coordinates
of the drop base along the $x$ direction, denoted by
$X_{\min}$ and $X_{\max}$ respectively. Then the spreading length
is $L_s = X_{\max} - X_{\min}$. 
Similarly, the minimum and the maximum coordinates of the drop profile in
the $y$ direction can also be determined, denoted by
$Y_{\min}$ ($Y_{\min}=0$) and $Y_{\max}$ respectively.
The drop height is given by $H = Y_{\max}-Y_{\min}$.
According to equation \eqref{equ:drop_theory},
\begin{equation}
\frac{L_s}{H} = \frac{2\sin\theta_e}{1-\cos\theta_e}.
\label{equ:simul_angle}
\end{equation}
Therefore, once the spreading length $L_s$ and the drop 
height $H$ are calculated, we can solve equation \eqref{equ:simul_angle} 
to obtain the actual contact angle based on the simulated
drop profile, which
can then be compared with the angle imposed through the contact-angle
boundary condition. 

\begin{table}[tb]
  \caption{Effect of spatial resolution (element order) on the
    spreading length ($L_s/R_0$) and the drop height ($H/R_0$).
    The contact angle is $\theta_e=105^0$}
\begin{center}
\label{table_ResolutionTest}
\begin{tabular}{l | l l l}
\hline
 & Element Order  & Spreading Length & Drop Height \\
 \hline
Simulation & 8&	0.432&	0.283\\
 & 10& 0.433&0.282  \\
 & 12& 0.433&0.282 \\ \hline
Theoretical & &0.434&0.283\\
\hline
\end{tabular}
\end{center}
\end{table}

Table \ref{table_ResolutionTest} shows results of 
the resolution test for this problem.
We list the spreading length and the drop height
obtained from the simulations using three element orders
($8$, $10$ and $12$) corresponding to an imposed contact angle
$\theta_e=105^0$. The theoretical values of these parameters,
computed based on equation \eqref{equ:drop_theory},
are also included for comparison. We observe that 
the change in these parameters values are negligible
with the increase of the element order, indicating 
the adequacy of the mesh resolutions for the current problem.
We can also observe that the values from the simulations
agree well with those of the theoretical values.
Based on the resolution test, we have employed an element order
$10$ in the majority of simulations.

      \begin{figure}[tb]
\centering
  \includegraphics[width=0.6\textwidth]{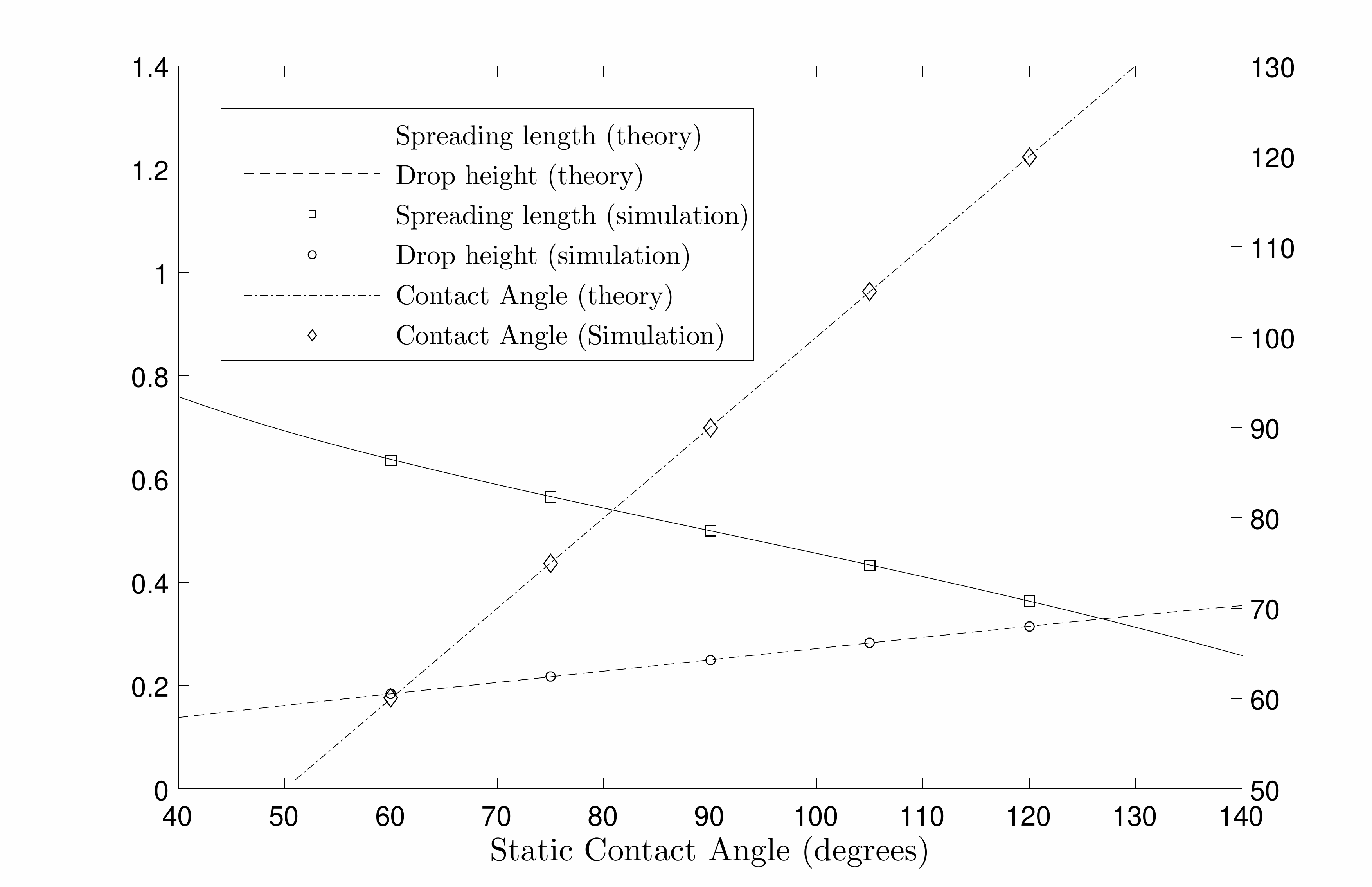}%
  \caption{ 
  Comparison of the drop height and spreading length 
  (normalized by $R_0$) as a function of the imposed contact 
  angle between the simulation and the theory. Comparison of
  the actual contact angle obtained from the simulated drop
  profiles with the imposed contact angle is also shown.
  The left vertical axis is for the drop height and spreading
  length, and the right axis is for the contact angle (in degrees).
  }  
  \label{fig:drop_compare}
\end{figure}

We have varied the imposed contact angle systematically between
$60^0$ and $120^0$, and computed the spreading length,
the drop height, and the actual contact angle obtained from
the simulated drop profiles at equilibrium.
Figure  \ref{fig:drop_compare} shows a comparison of these 
quantities as a function of the imposed contact angle between
the simulation and the theoretical results.
The theoretical values for the spreading length and
the drop height are computed based on equation \eqref{equ:drop_theory}. The actual contact angles
obtained from the drop profiles are computed based on
equation \eqref{equ:simul_angle}. It is observed that
the simulation results agree well with
the theoretical values. These results indicate that
our method has captured these flow parameters accurately.

\subsection{Dynamics of a Rising Air Bubble in Water}


  In this subsection, we use the  method
developed herein to study
the rise of an air bubble through water in a wall-bounded 
domain.
The goal is to investigate the effect of the contact
angle on the dynamics of the fluid interfaces,
and to demonstrate the potential of the method 
for  two-phase problems under  realistic
physical parameters such as
the density ratios and viscosity ratios.

\begin{figure}[tb]
\centerline{
    \subfigure[] {\includegraphics[width=0.25\textwidth]{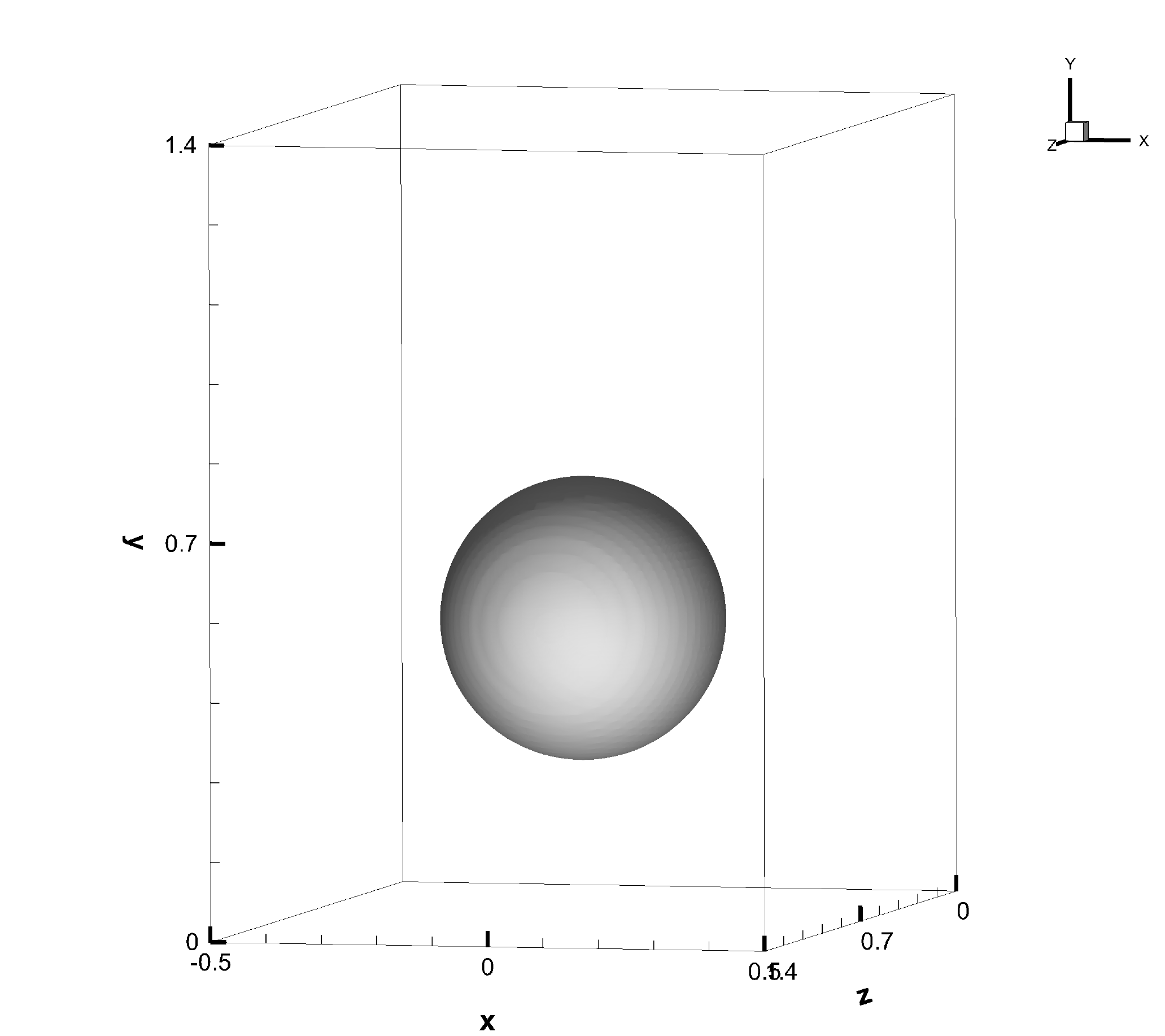}} %
    \subfigure[] {\includegraphics[width=0.25\textwidth]{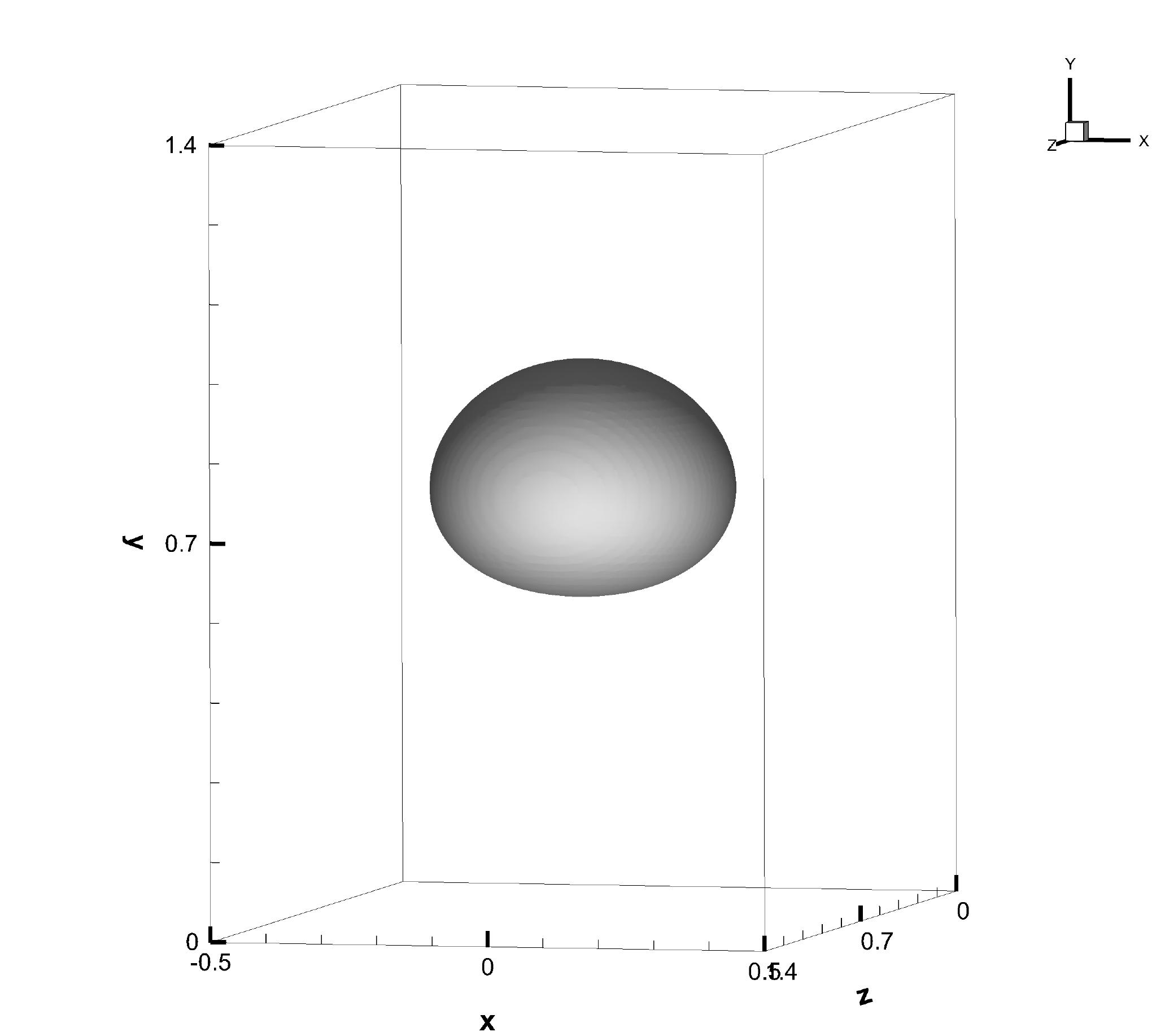}} %
    \subfigure[] {\includegraphics[width=0.25\textwidth]{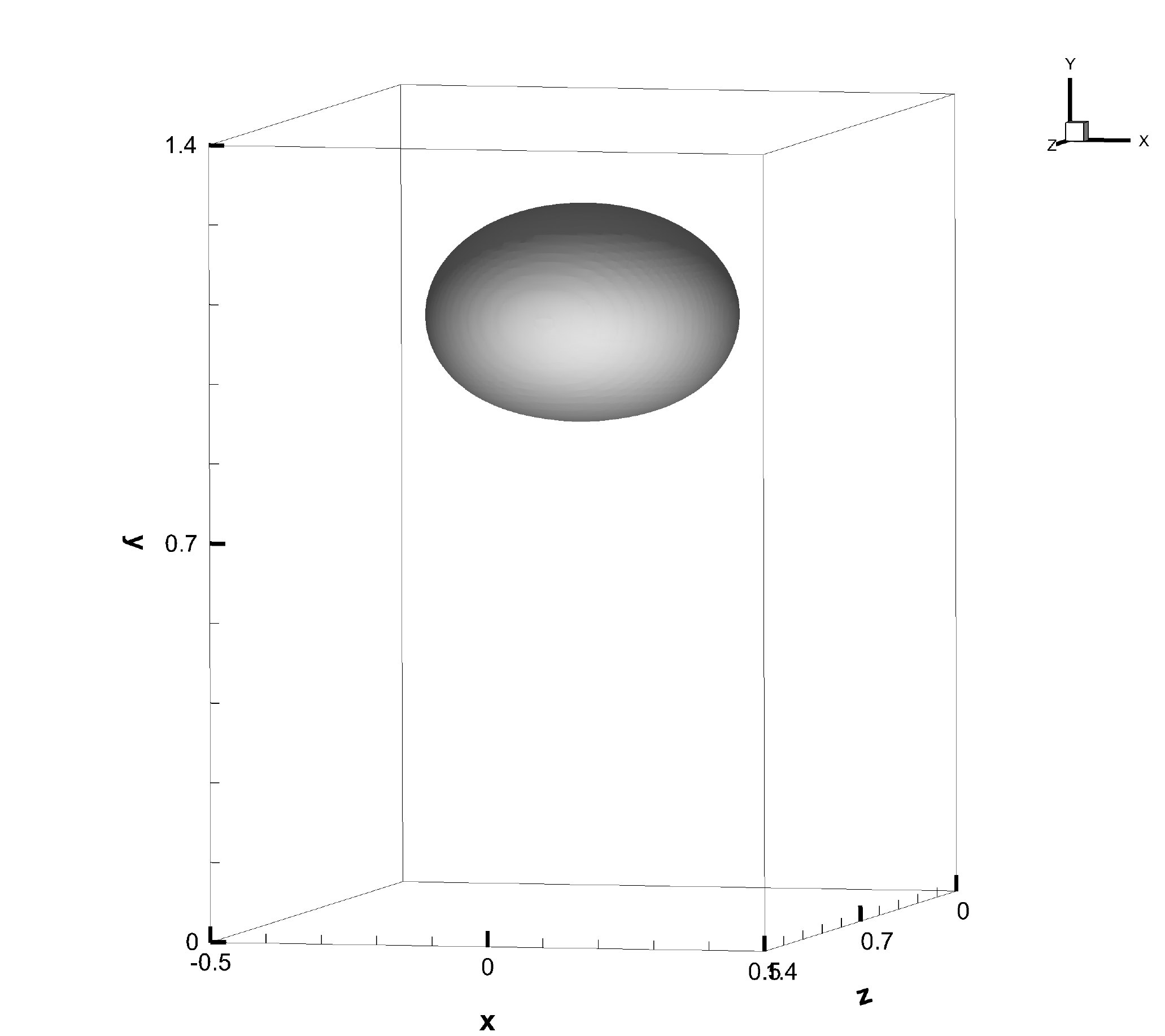}} \\
    \subfigure[] {\includegraphics[width=0.25\textwidth]{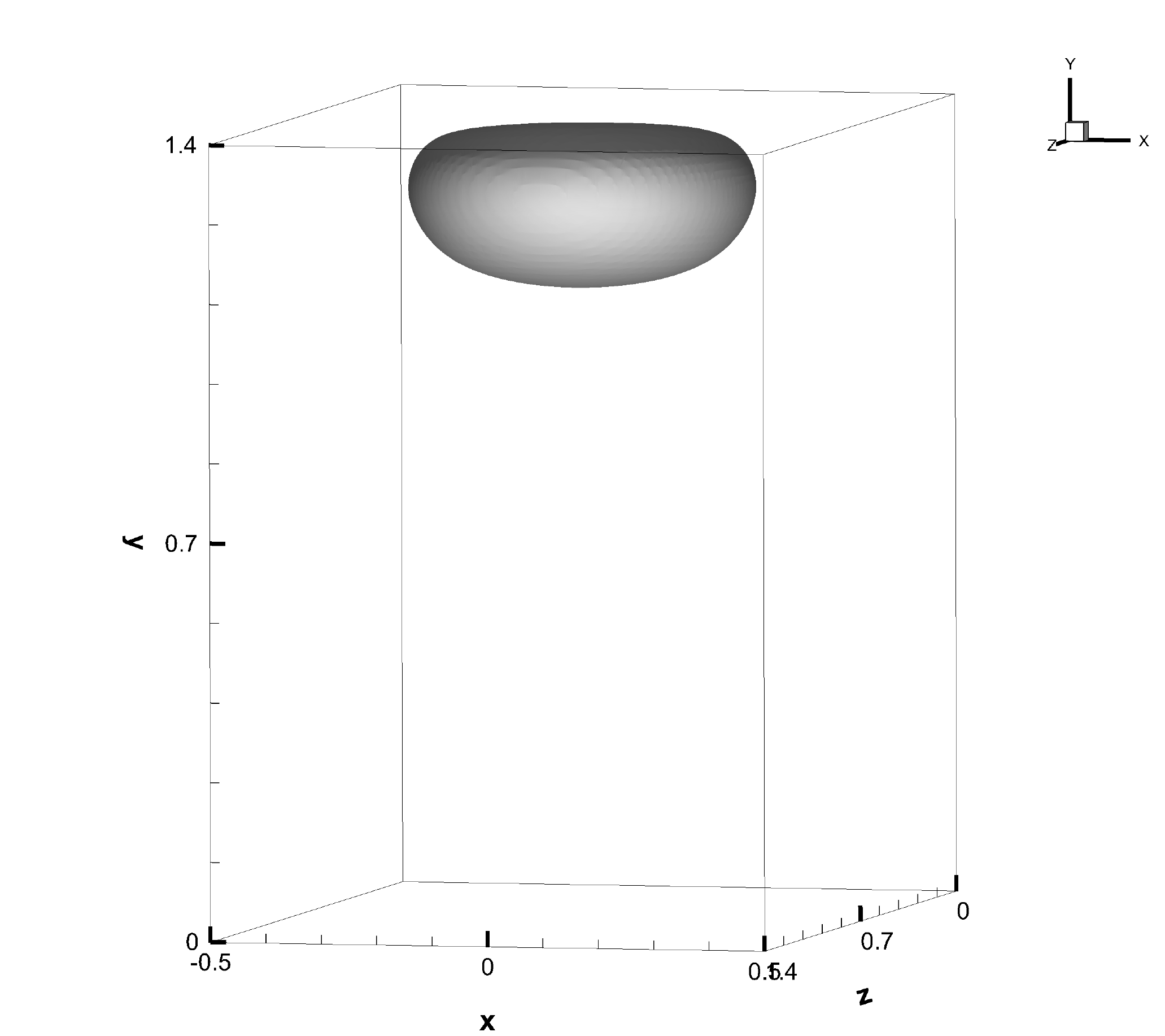}} %
    }
   \centerline{
    \subfigure[] {\includegraphics[width=0.25\textwidth]{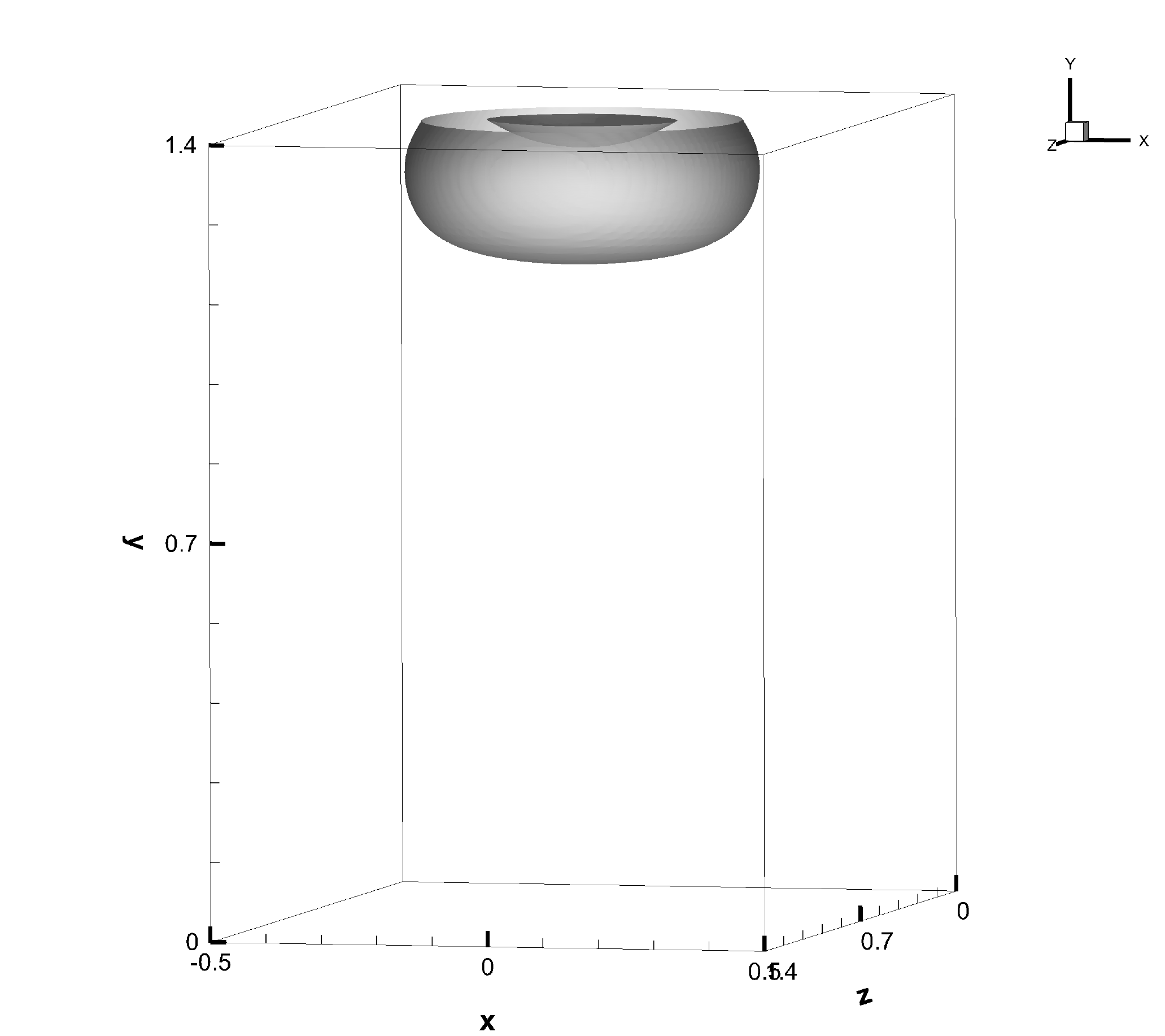}} \\
    \subfigure []{\includegraphics[width=0.25\textwidth]{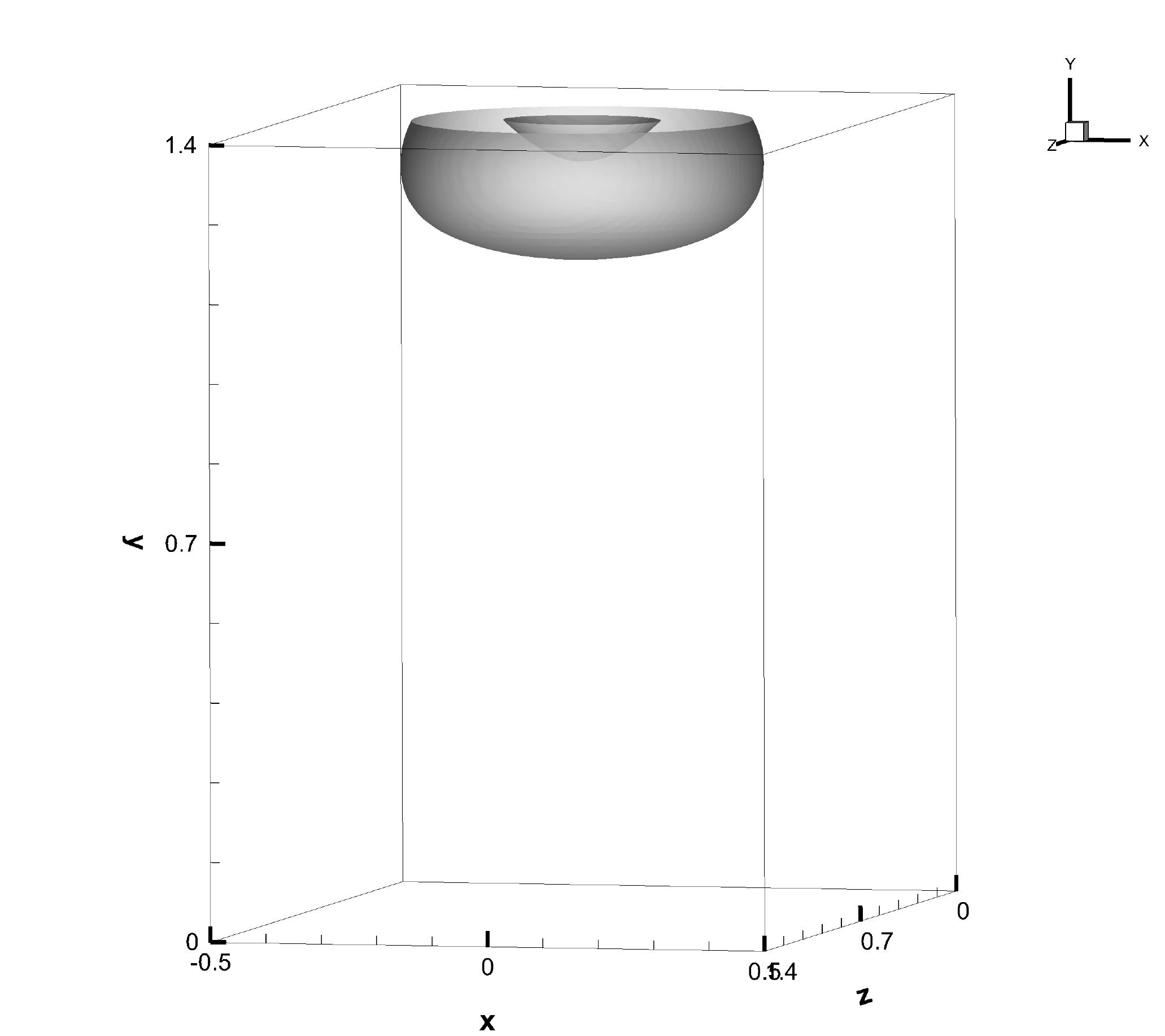}} %
    \subfigure []{\includegraphics[width=0.25\textwidth]{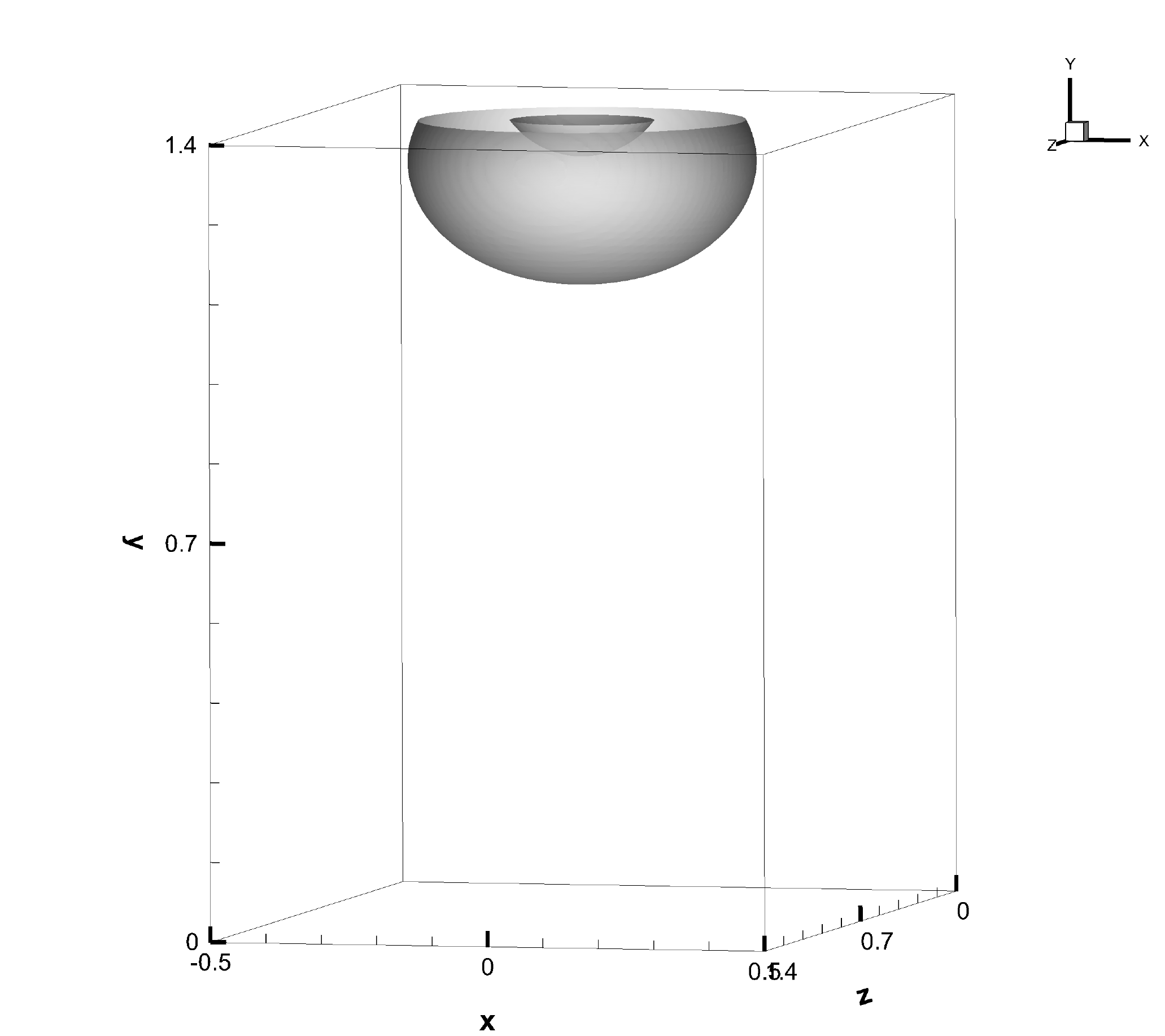}} %
    \subfigure []{\includegraphics[width=0.25\textwidth]{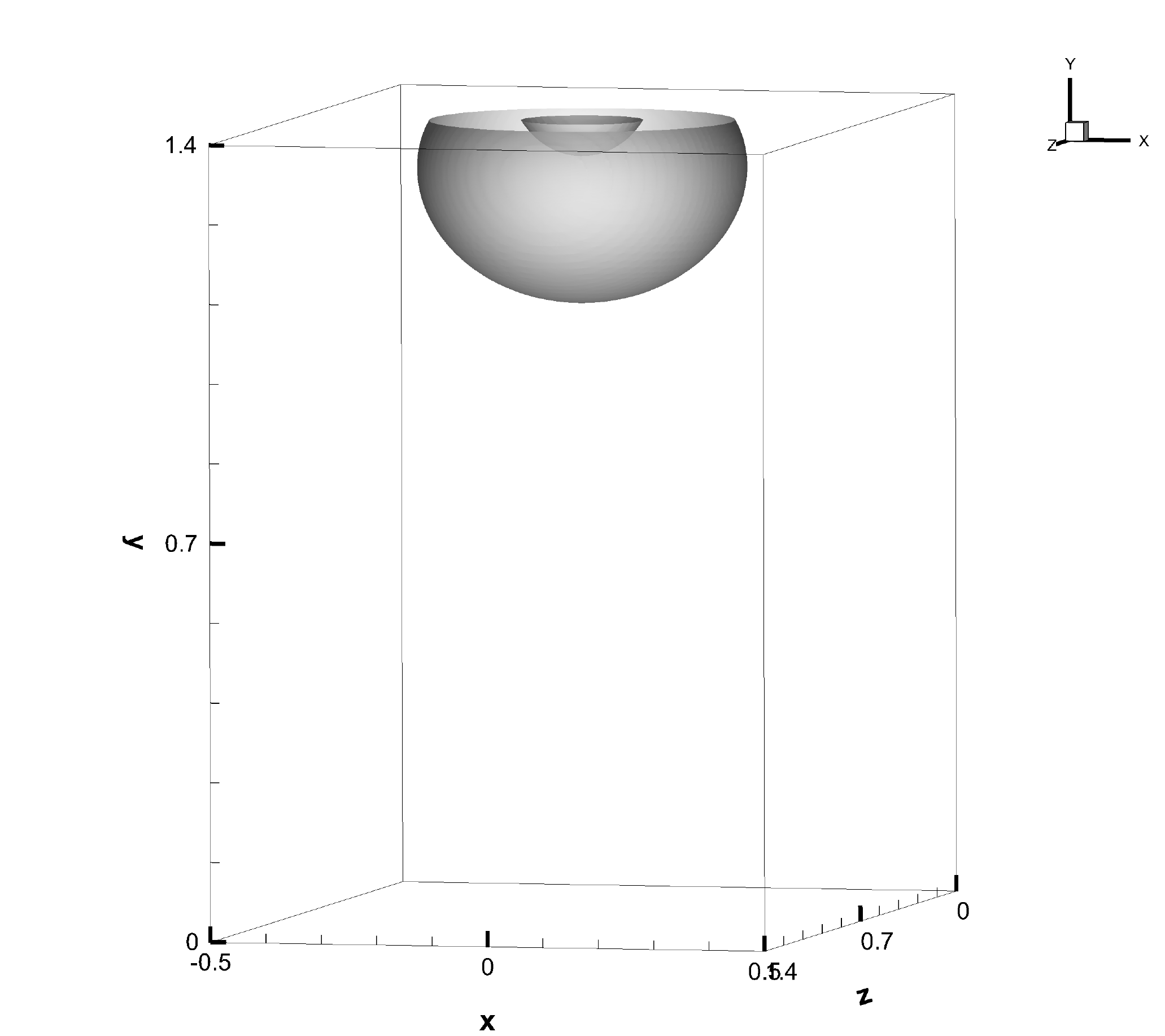}} 
    }
     \caption{Temporal sequence of snapshots of the rising air bubble in water with a contact angle $60^0$: (a)t=0.06, (b)t=0.26, 
     (c)t=0.44, (d)t=0.60, 
     (e)t=0.67, (f)t=0.70, (g)t=0.85, (h)t=1.95. }
      \label{F6}
\end{figure}

Specifically, we consider a three-dimensional
domain with dimensions $1cm\times 1.4cm\times 1.4cm$
along the three directions; see Figure \ref{F6}(a)
for illustration.
The top and the bottom sides of the domain are solid walls,
while in the $x$ and $z$ directions the domain
is  periodic. The domain
is filled with water.  An air bubble,
initially spherical with a radius $2.5mm$,
is trapped in the water (at rest),
with its center located $5mm$ above the bottom wall.
The gravity is assumed to point downward
($-y$ direction). At $t=0$, the system is released.
The air bubble rises through the water due to buoyancy,
and touches the upper boundary to form an
air dome attached to the upper wall.
We will employ the method developed in Section \ref{sec:method} to
investigate this dynamic process
and the topological changes of the air bubble.
The physical parameter values of
the air and water employed in the current problem
are listed in Table \ref{tab:air_water_param}.

\begin{table}[tb]
\begin{center}
\begin{tabular} {l | l  l}
\hline
 & Air &  Water \\ \hline
 Density [$kg/m^3$] & $1.204$ &  $998.207$ \\
 Dynamic viscosity [$kg/(m\cdot s)$] & $1.78\times 10^{-5}$ 
 & $1.002\times 10^{-3}$ \\
 Surface tension [$kg/s^2$] &  $0.0728$ \\
 Gravitational acceleration [$m/s^2$] & $9.8$ \\
\hline
\end{tabular}
\caption{Physical properties for the air and water.}
\label{tab:air_water_param}
\end{center}
\end{table}

We set up a coordinate system, in which the domain is given by
$-\frac{L}{2} \leqslant x \leqslant \frac{L}{2}$,
$0\leqslant y \leqslant \frac{7L}{5}$,
and $0\leqslant z \leqslant \frac{7L}{5}L$,
where $L=1cm$ is the characteristic length scale.
We use the air density as the characteristic density scale,
$\varrho_d=1.204 kg/m^3$.
We choose the characteristic velocity scale as 
$U_0=\sqrt[]{g_0 L}$, where $g_0=1 m/s^2$. 
The normalization of the physical variables and parameters
is then performed according to Table \ref{tab:general_normalization}.


In the simulations the air and the water are chosen as
the first and the second fluids, respectively.
The flow domain is discretized using $96$ Fourier planes
along the $z$ direction, and each $x$-$y$ plane is 
discretized with a mesh of $200$ quadrilateral elements
(with an element order $12$).
Periodic boundary conditions are imposed for all
flow variables along the $x$ direction (at $x=\pm\frac{L}{2}$).
Periodic conditions are also enforced in the $z$ direction
because of the Fourier spectral expansion in this direction.
On the top and bottom walls, a no-slip condition (zero velocity)
is imposed for the velocity, and
the contact-angle conditions \eqref{e4.4a}--\eqref{e4.4b} 
with $g_c=0$ and $g_b=0$ are imposed for the phase field 
function. The initial velocity is set to zero, and
the initial distribution of the phase field function
is given by
   \begin{equation}
   \phi_{in}(\textbf{x})=-\tanh\frac{\sqrt{(x-x_0)^2+(y-y_0)^2 + (z-z_0)^2}-R_0}{\sqrt[]{2}\eta},
   \end{equation}
where $R_0=\frac{L}{4}$ is the initial radius of the air bubble,
and $(x_0,y_0,z_0)=(0, 0.5L, 0.7L)$ is the initial coordinate
of the center of bubble.
A number of contact angles ranging from $60^0$ to $105^0$
have been considered in the simulations,
which correspond to hydrophilic and hydrophobic walls.
Here the contact angle refers to the angle formed between
the air/water interface and the wall measured on the side of the water.
The values of the non-dimensionalized simulation
parameters for this problem are given below:
   \begin{equation*}
   \begin{cases}
   \rho_1 =1,\; \frac{\rho_2}{\rho_1}=829,\; \mu_1=0.0147828,\; \frac{\mu_2}{\mu_1}=56.29, \\
   \sigma = 604.601, \ \eta = 0.02, \ 
   \lambda = \frac{3}{2\sqrt{2}}\sigma\eta, \ 
   \gamma_1 = \frac{10^{-6}}{\lambda}, \\
   \rho_0 = \min(\rho_1,\rho_2), \ 
   \nu_m = \max\left(\frac{\mu_1}{\rho_1}, \frac{\mu_2}{\rho_2} \right), \\
   96\ \text{planes in z}, \ 200\ \text{elements per plane}, \  \text{element order}=12,\;\Delta t=2.5\times 10^{-5}. \\
   \text{contact angle:} \  60^0 \sim 105^0
   \end{cases}
   \end{equation*}

We have conducted simulations of this air/water two-phase
problem under several contact angles, and let us 
now look into the dynamics of the system. 
Figure \ref{F6} shows a temporal sequence of snapshots
of the air/water interface in the flow, which correspond to
a contact angle $60^0$ (hydrophilic wall).
The fluid interface is visualized by the iso-surface
$\phi=0$ of the phase field function.
It is observed that as the system is released the air bubble
rises through the water due to buoyancy, and that the bubble
experiences a notable deformation (Figures \ref{F6}(a)-(c)).
Then as the bubble touches the upper wall,
a topological change in the air-water interface 
occurs (Figures \ref{F6}(d)-(e)). We observe that
the air-water interface, which was originally a single piece separating
the bulk of air and water, splits into two pieces 
(Figures \ref{F6}(e)-(f)) on the upper wall.
The inner piece of interface traps 
a small pocket of water on the wall.
It evolves into
a small water drop that is suspended from the upper wall
and trapped inside the bulk of air (Figures \ref{F6}(f)-(h)).
The outer piece of interface separates the air from the bulk of water,
and it evolves into a spherical cap
attached to the upper wall over time (Figure \ref{F6}(h)).

\begin{figure}[tb]  
\centerline{
    \subfigure[] {\includegraphics[width=0.25\textwidth]{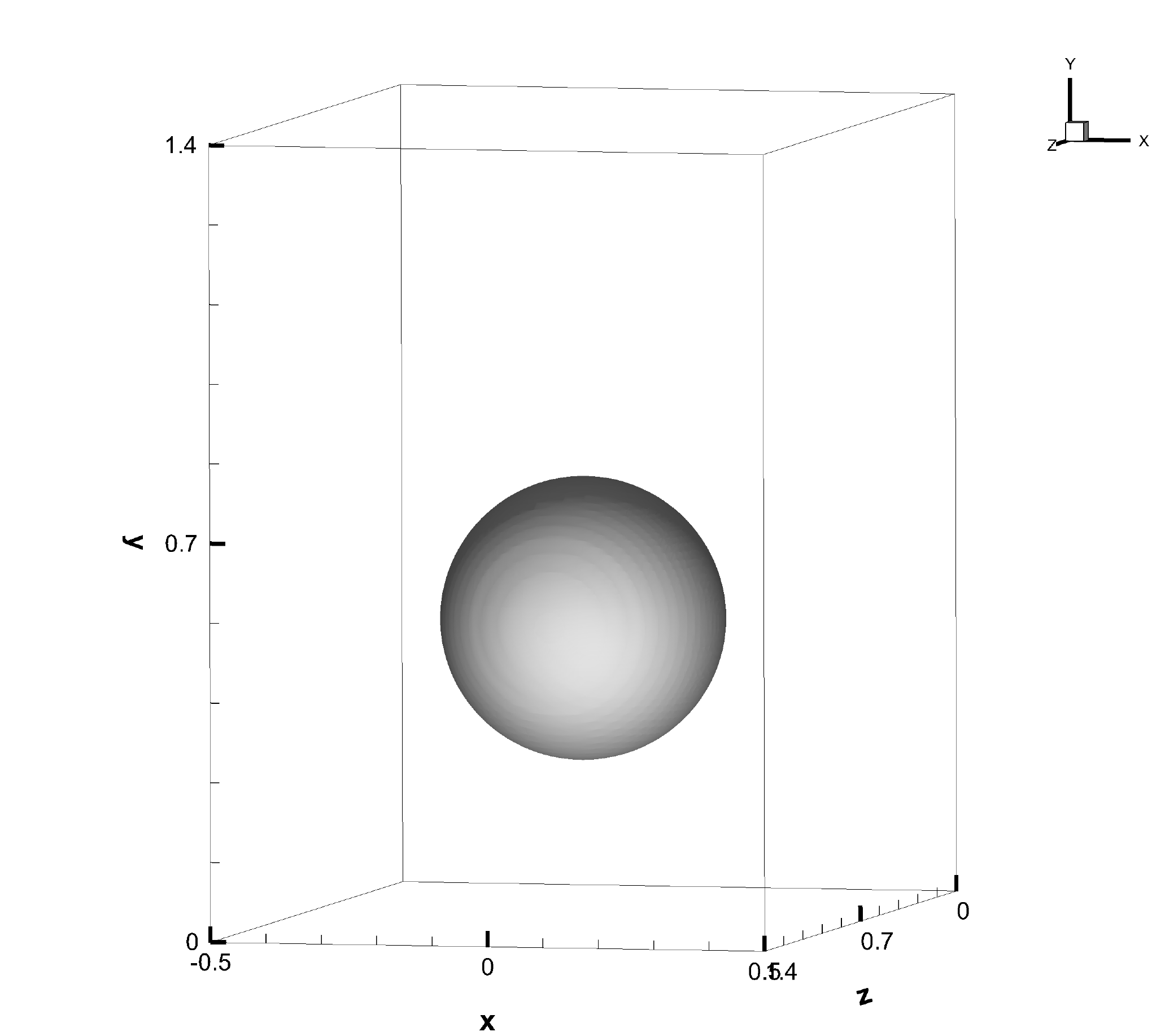}} %
    \subfigure[] {\includegraphics[width=0.25\textwidth]{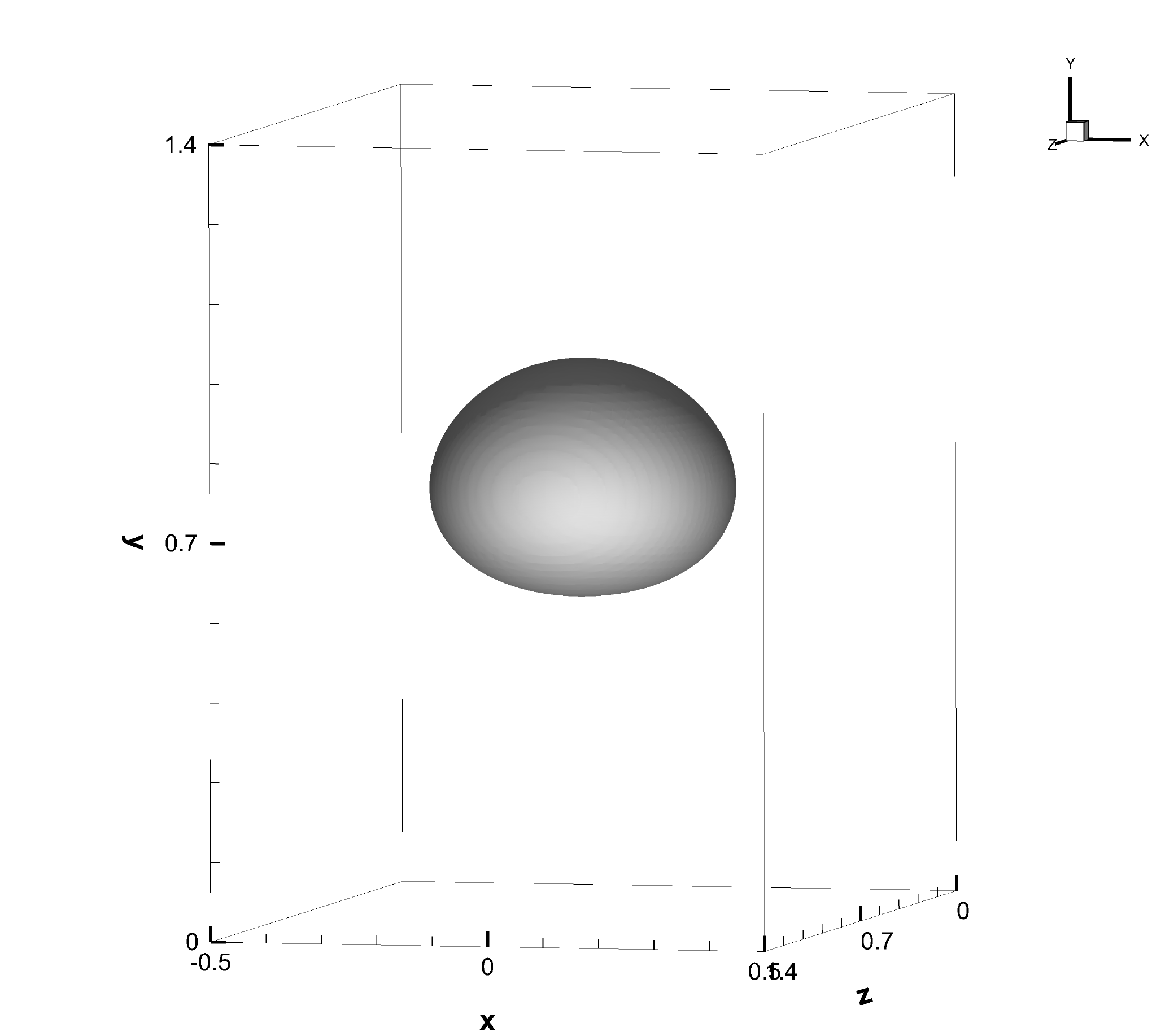}} %
    \subfigure[] {\includegraphics[width=0.25\textwidth]{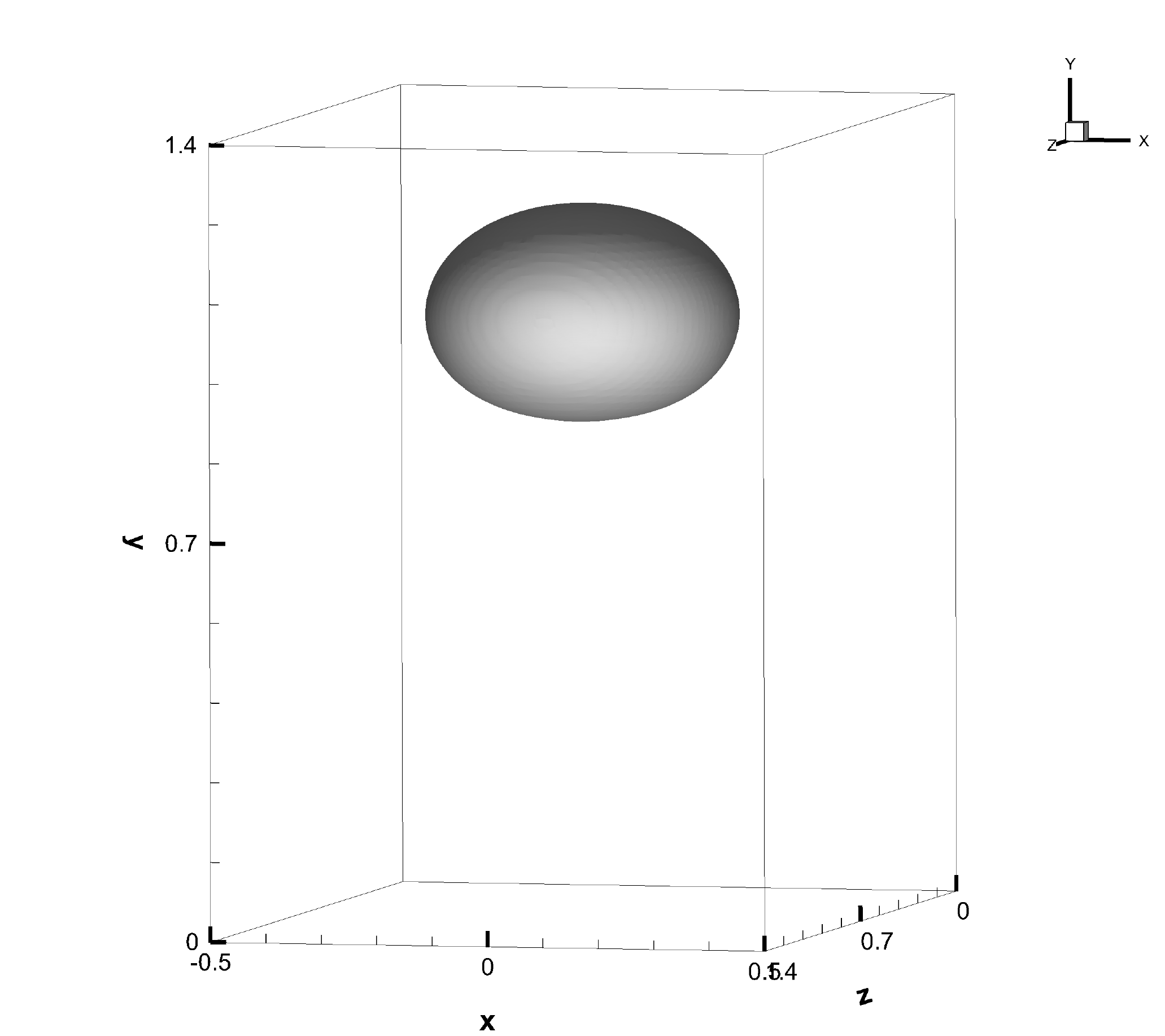}} \\
    \subfigure[] {\includegraphics[width=0.25\textwidth]{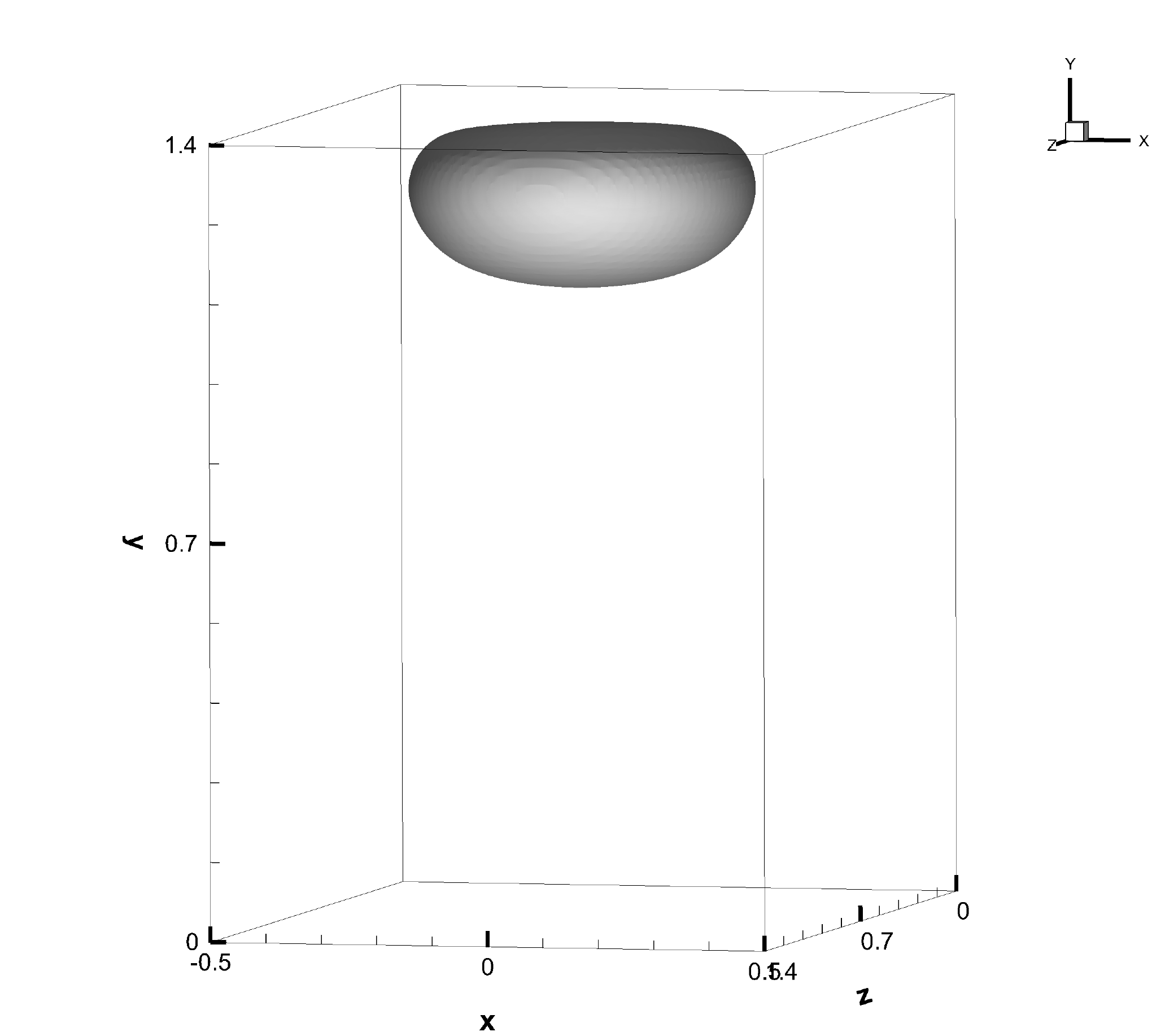}} %
}
\centerline{
    \subfigure[] {\includegraphics[width=0.25\textwidth]{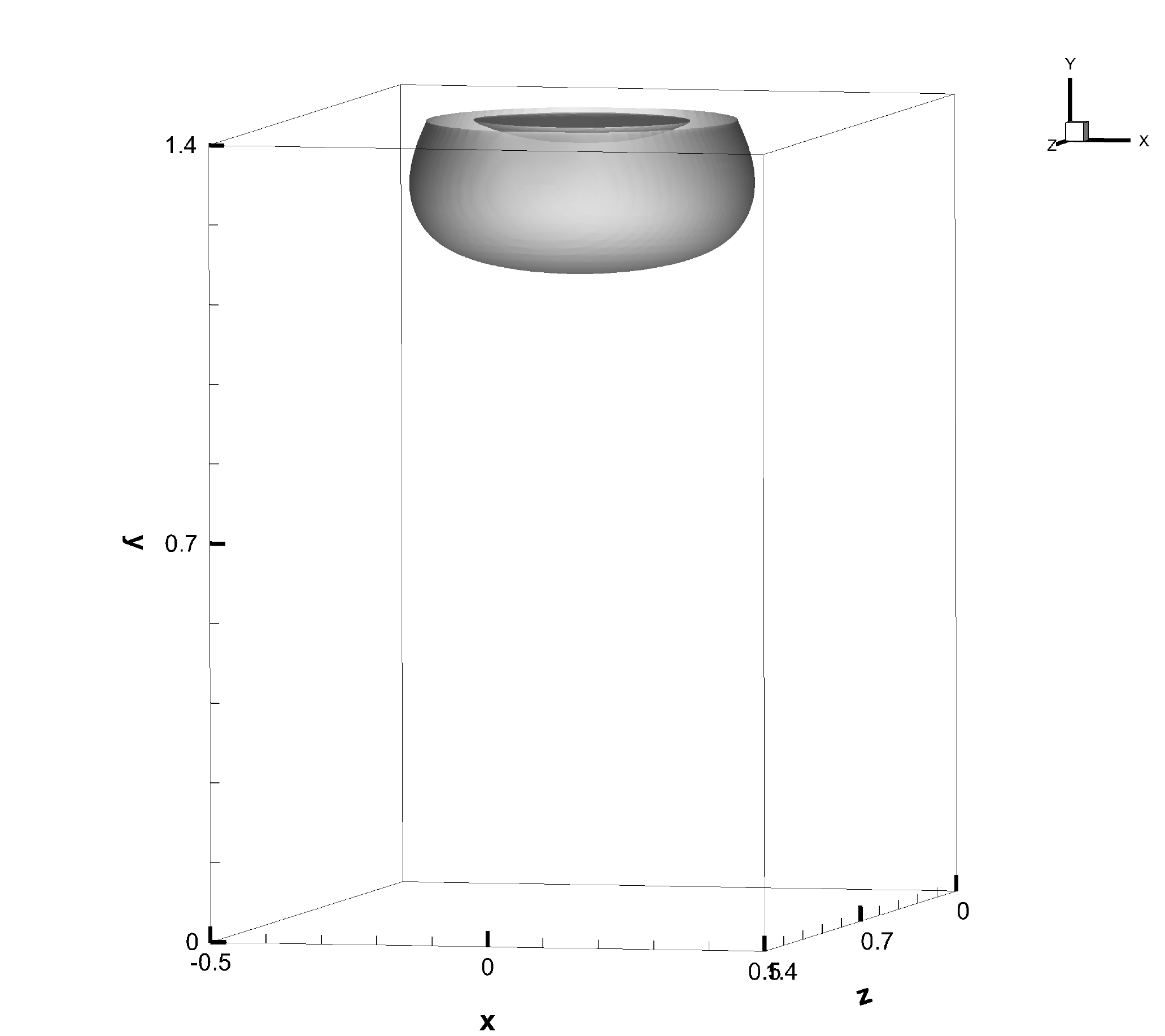}} %
    \subfigure[] {\includegraphics[width=0.25\textwidth]{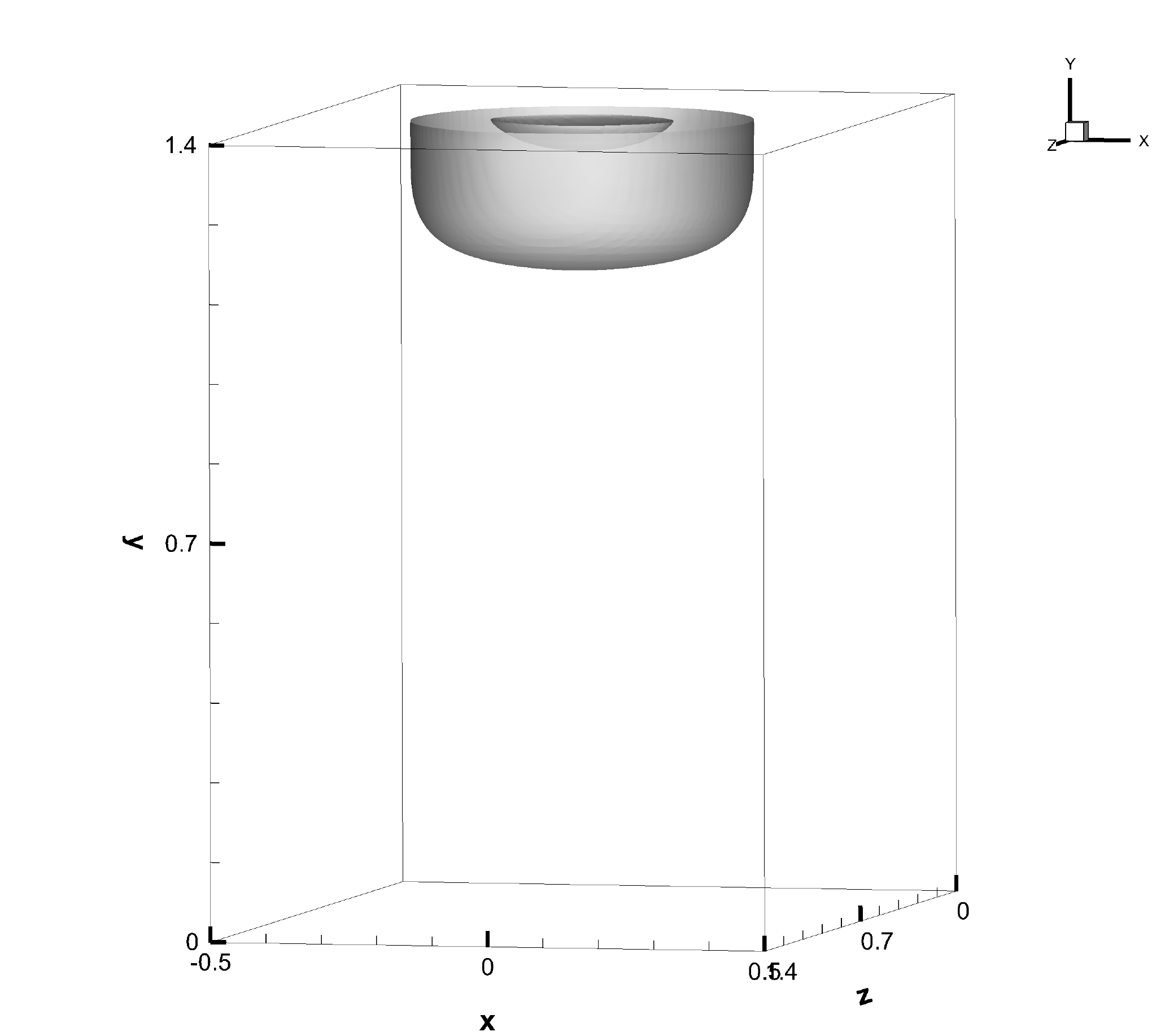}} \\
    \subfigure[] {\includegraphics[width=0.25\textwidth]{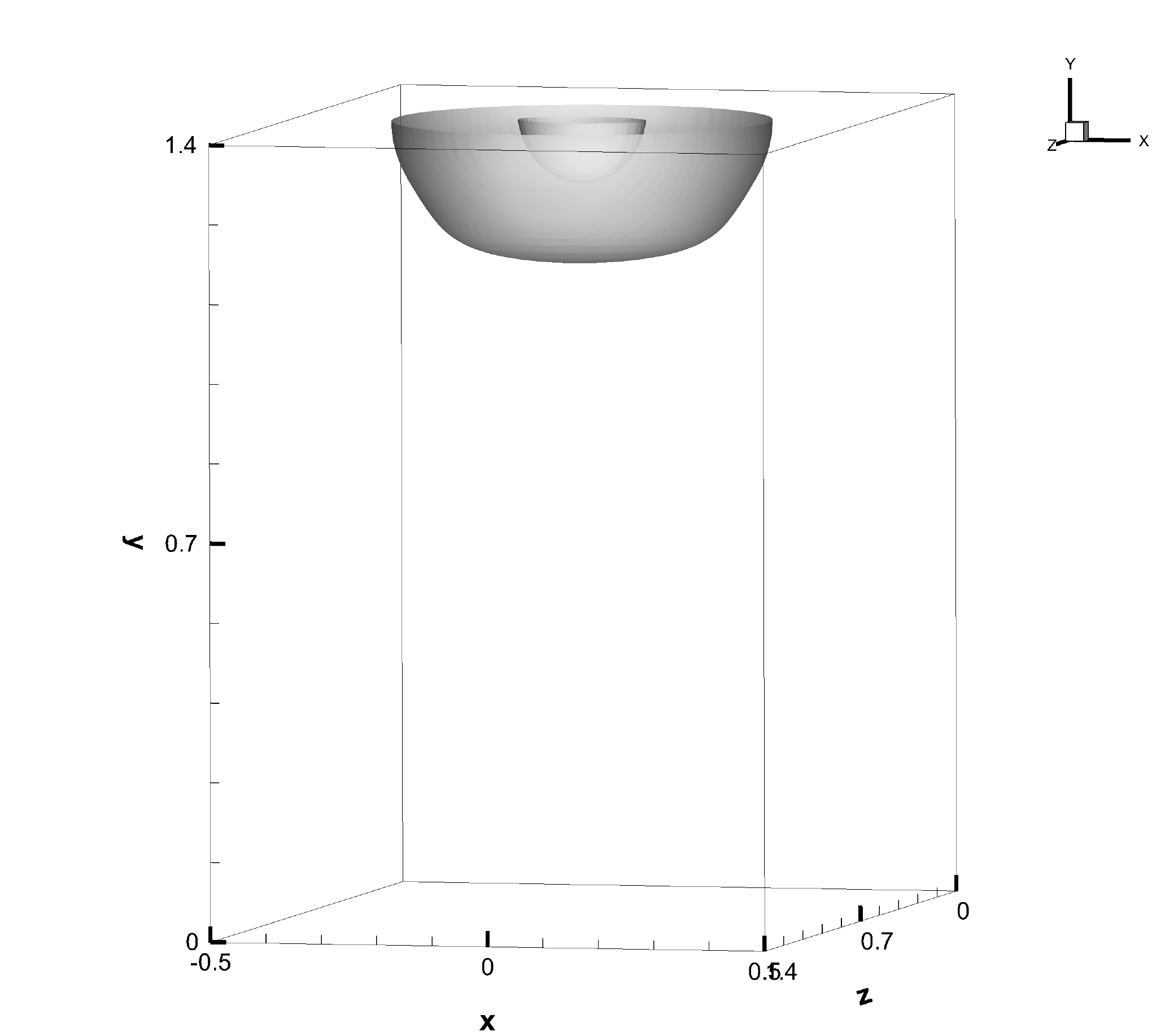}} %
    \subfigure[] {\includegraphics[width=0.25\textwidth]{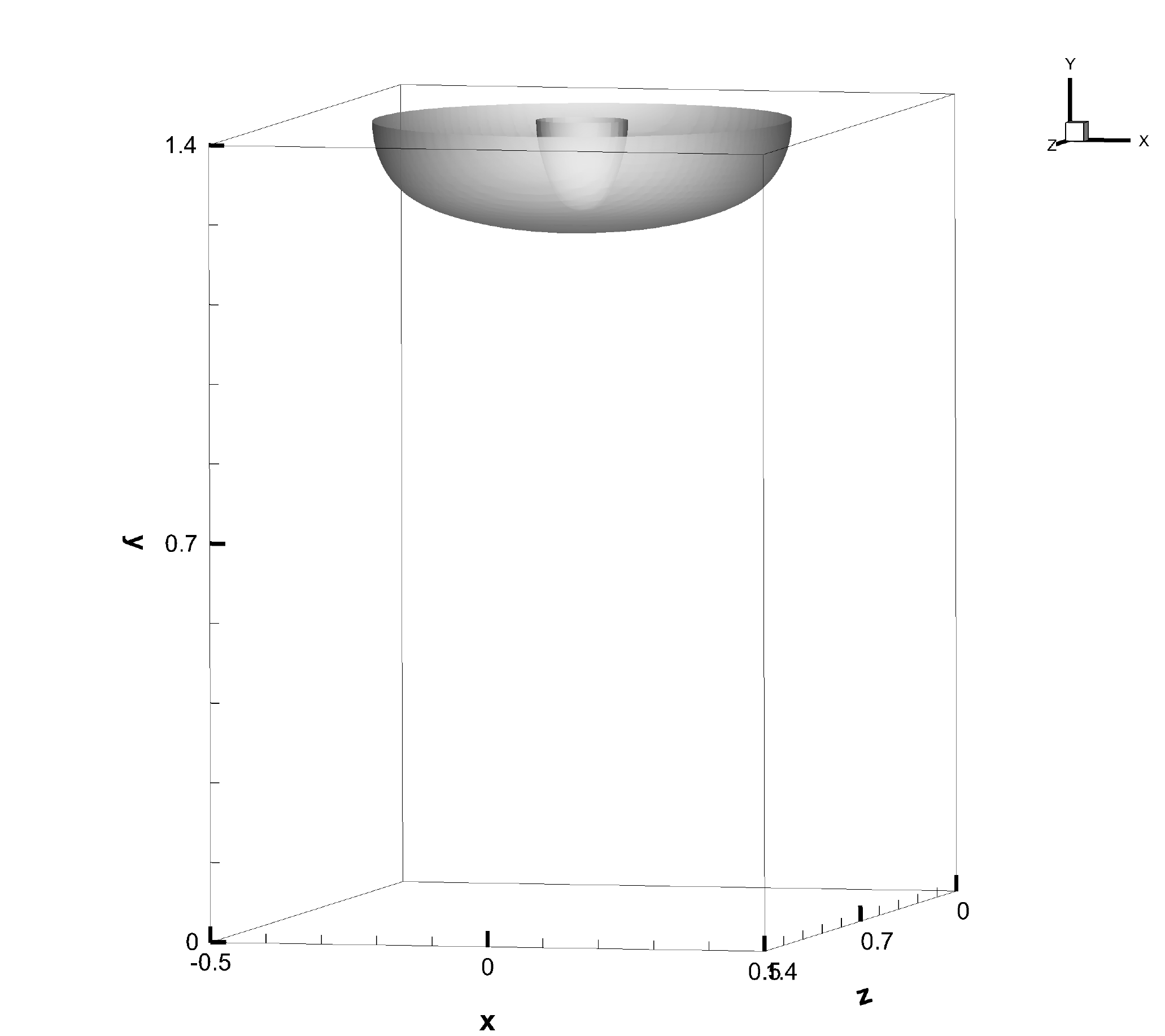}} %
}
\centerline{
    \subfigure[] {\includegraphics[width=0.25\textwidth]{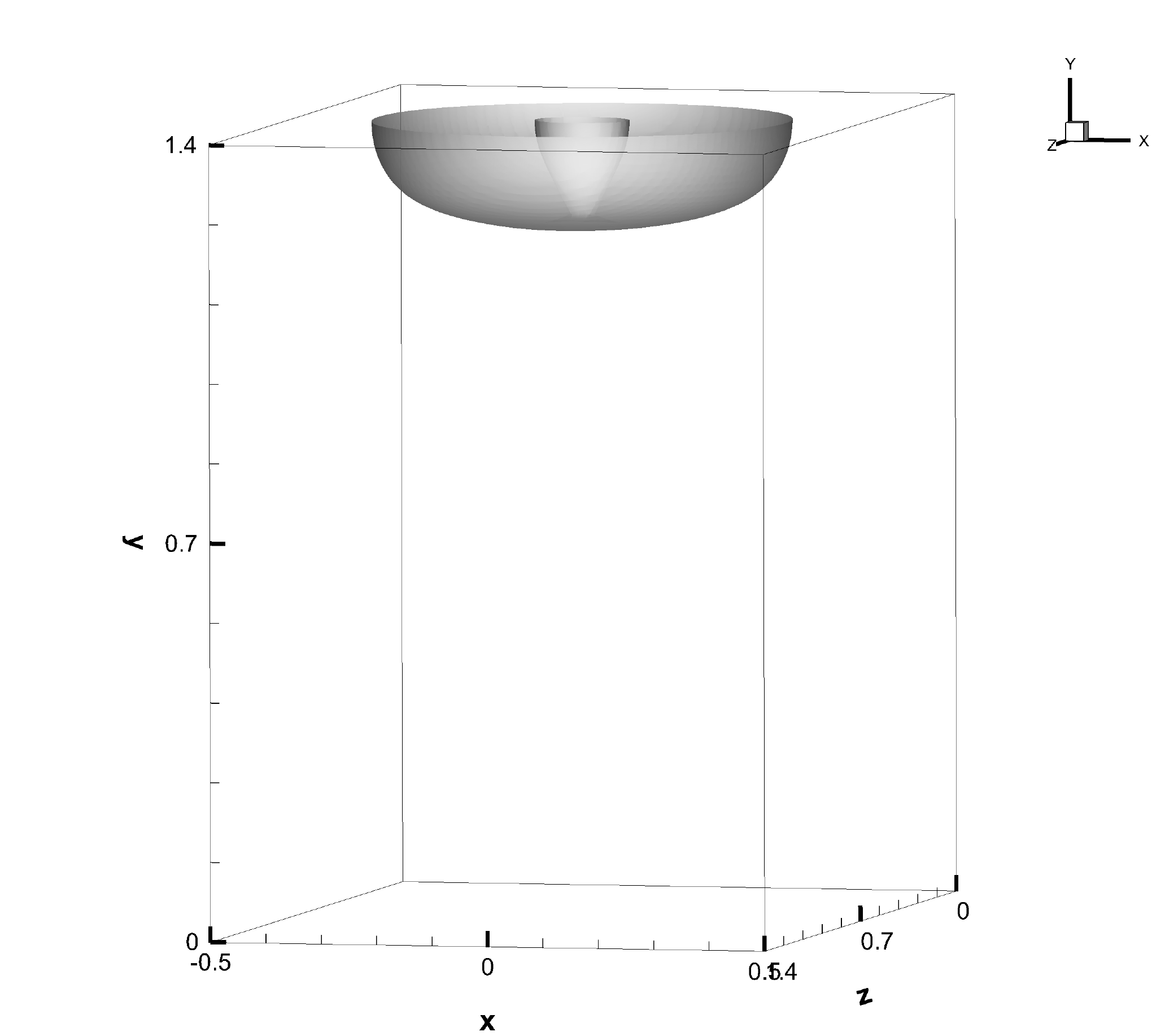}} \\
    \subfigure[] {\includegraphics[width=0.25\textwidth]{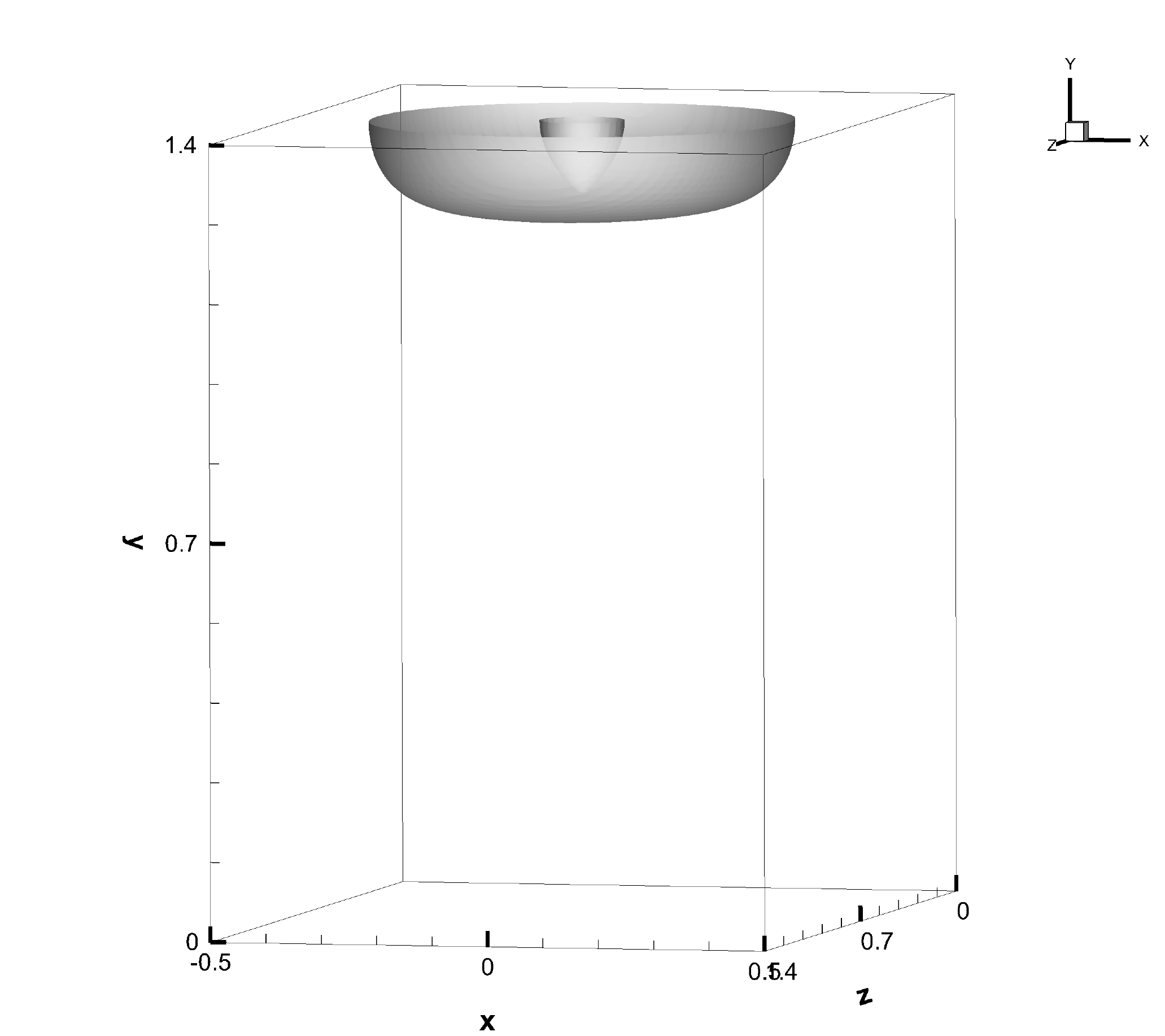}} %
    \subfigure[] {\includegraphics[width=0.25\textwidth]{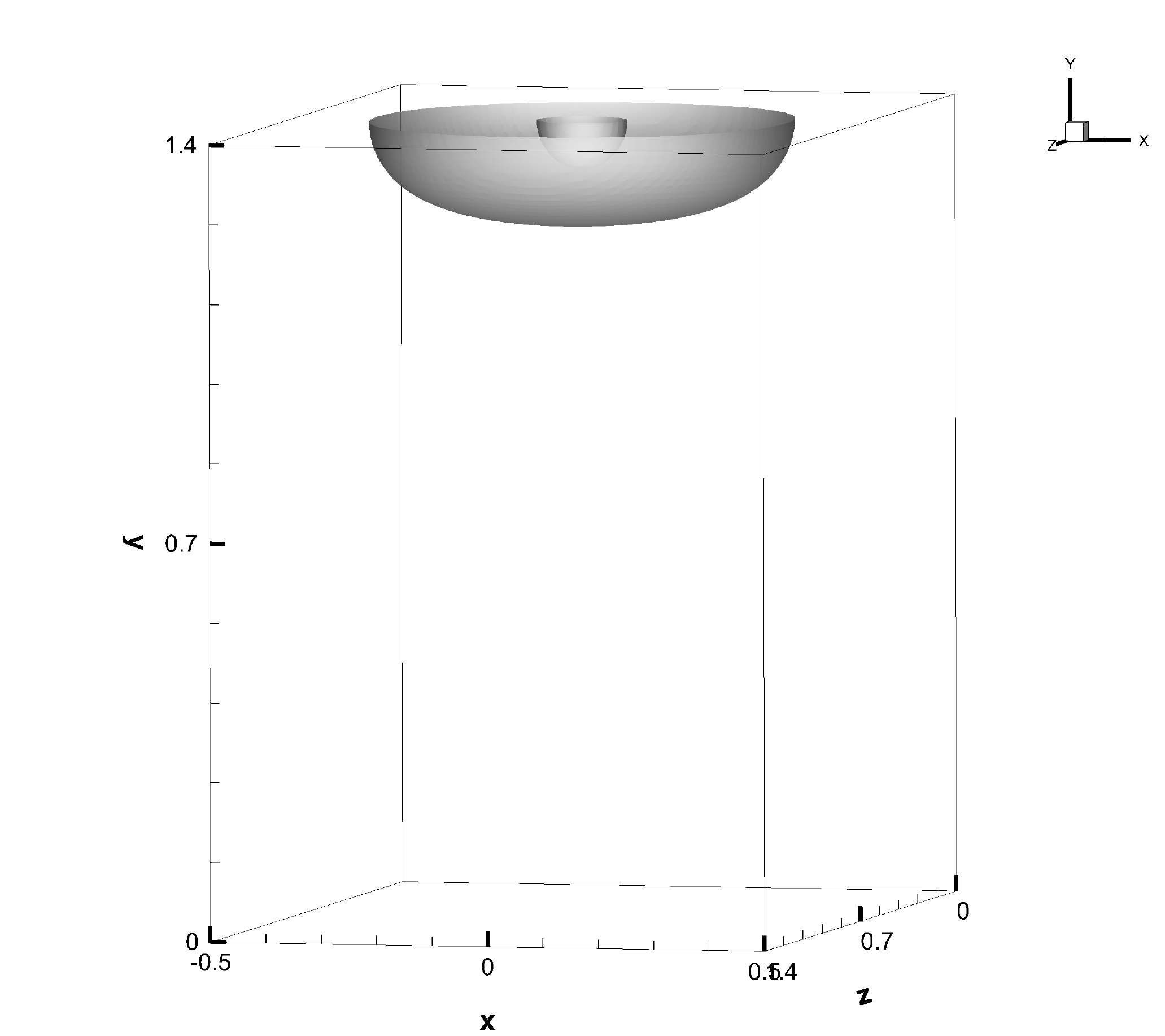}} %
    \subfigure[] {\includegraphics[width=0.25\textwidth]{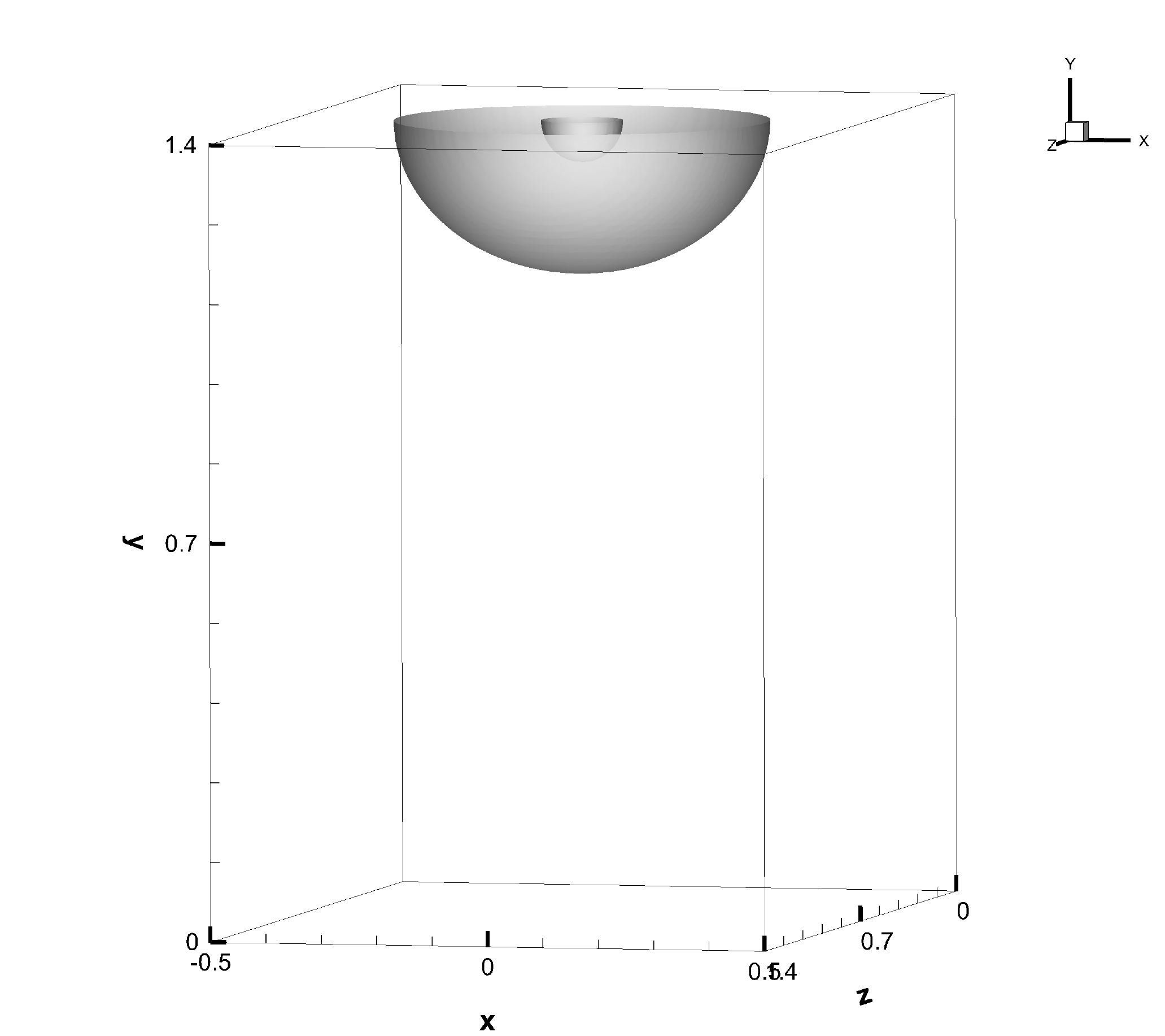}}
  }
     \caption{Temporal sequence of snapshots of the rising air bubble in water with contact angle $90^0$: (a)t=0.06, 
     (b)t=0.26, (c)t=0.44, (d)t=0.60, (e)t=0.63, 
     (f)t=0.64, (g)t=0.66, (h)t=0.73, (i)t=0.74, 
     (j)t=0.78, (k)t=0.85, (l)t=1.45.}
      \label{F8}
\end{figure}

Figure \ref{F8} illustrates the dynamics of the air bubble
corresponding to a contact angle of $90^0$ (neutral wettability).
We observe  a behavior similar to that 
with the contact angle of $60^0$ at the early stage, before the bubble
reaches the wall.
But when the bubble touches the upper wall, 
the system exhibits  some notable difference in the dynamics,
and in this case it is notably more complicated. 
With the $90^0$ contact angle,
after the air/water interface breaks up into 
two individual pieces, the pocket of water trapped by
the inner piece of interface appears more
mobile on the upper wall (Figures \ref{F8}(e)-(i)).
As the surface area of the water pocket contracts due to
surface tension,
the contact line formed between this interface and the wall
moves inward and the base of the water pocket shrinks
on the wall (Figures \ref{F8}(e)-(g)).
As a result, the water  deforms into a long slender
column at some point (Figure \ref{F8}(h)).
Simultaneously, the contact line formed between the outer
piece of the air/water interface and the wall moves
outward, as the bulk of air  
spreads outward on the upper wall (Figures \ref{F8}(f)-(h)).
The pillar-like inner interface and the dome-like outer
interface touches each other and re-connects near the center, 
and a certain amount of water
escapes from the drop on the upper wall into
the bulk of water (Figures \ref{F8}(h)-(i)).
Afterward, the two pieces of interface pinches off
and separates again.
The inner interface shrinks and traps the residual water
to form a smaller drop on the upper wall
(Figures \ref{F8}(k)-(l)), while the outer interface
evolves into a hemisphere eventually (Figure \ref{F8}(l)).

    \begin{figure}[tb]
    \centerline{
  \subfigure[] {\includegraphics[width=0.25\textwidth]{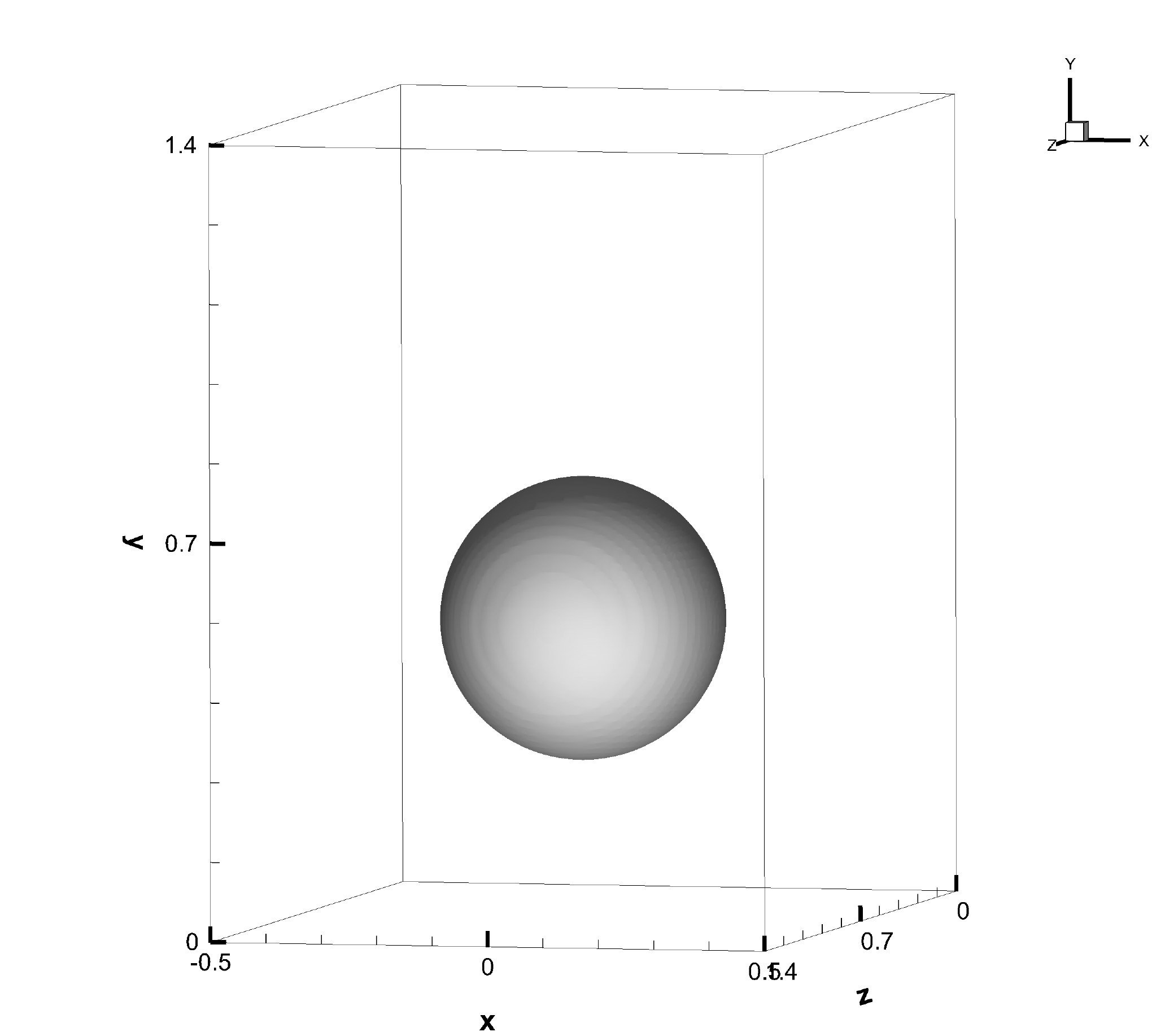}} %
    \subfigure[] {\includegraphics[width=0.25\textwidth]{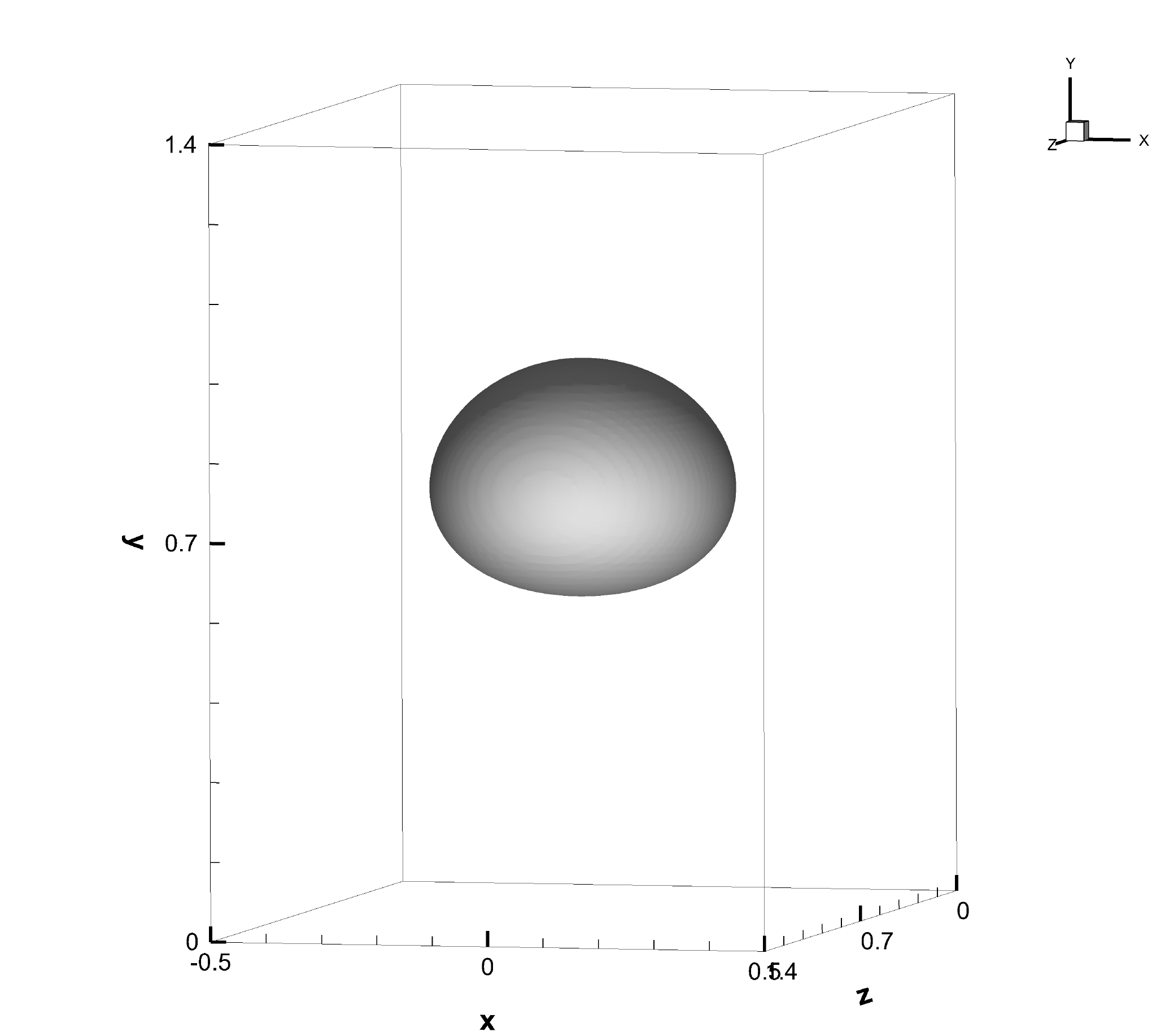}} %
    \subfigure[] {\includegraphics[width=0.25\textwidth]{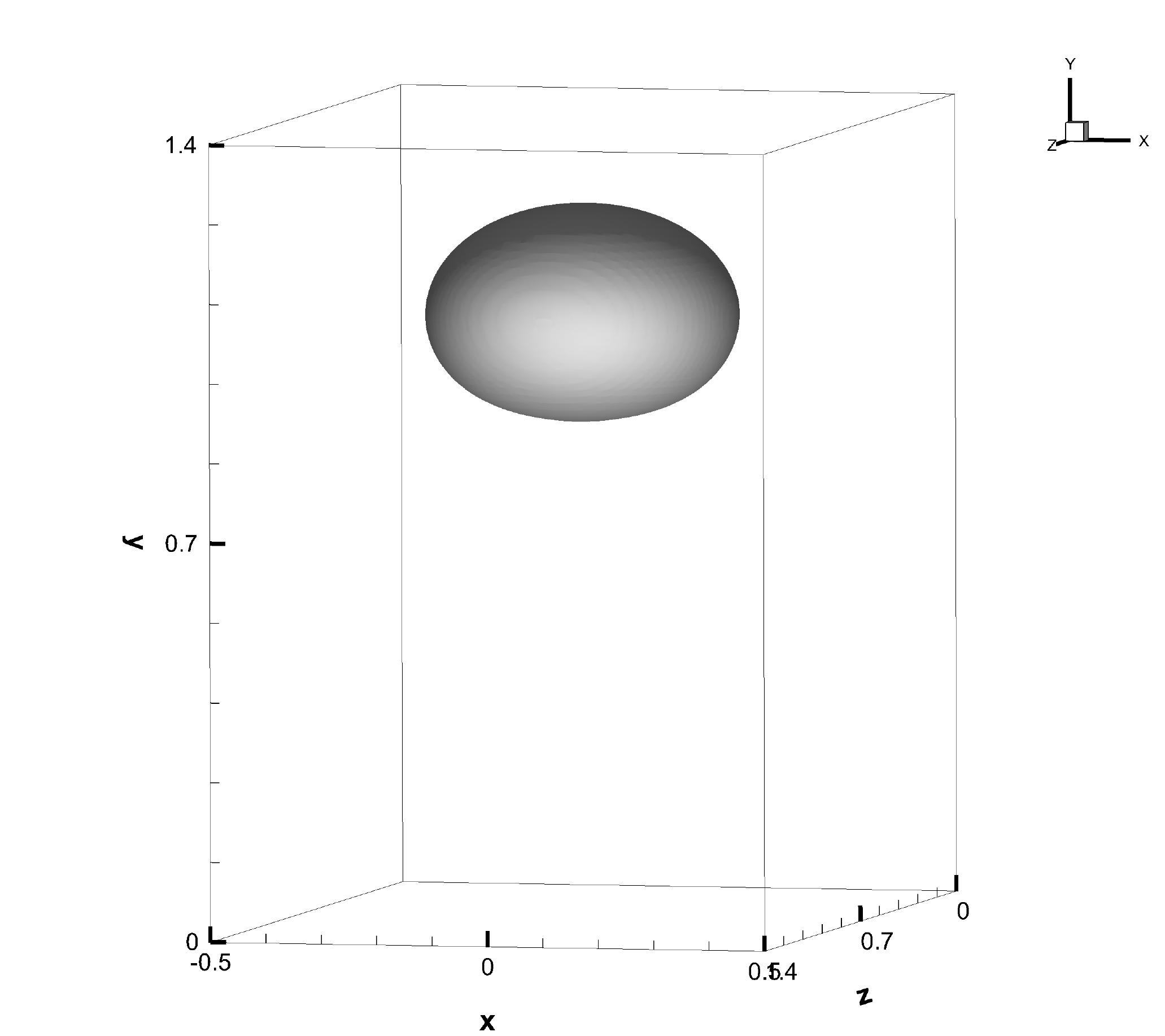}} \\
    \subfigure[] {\includegraphics[width=0.25\textwidth]{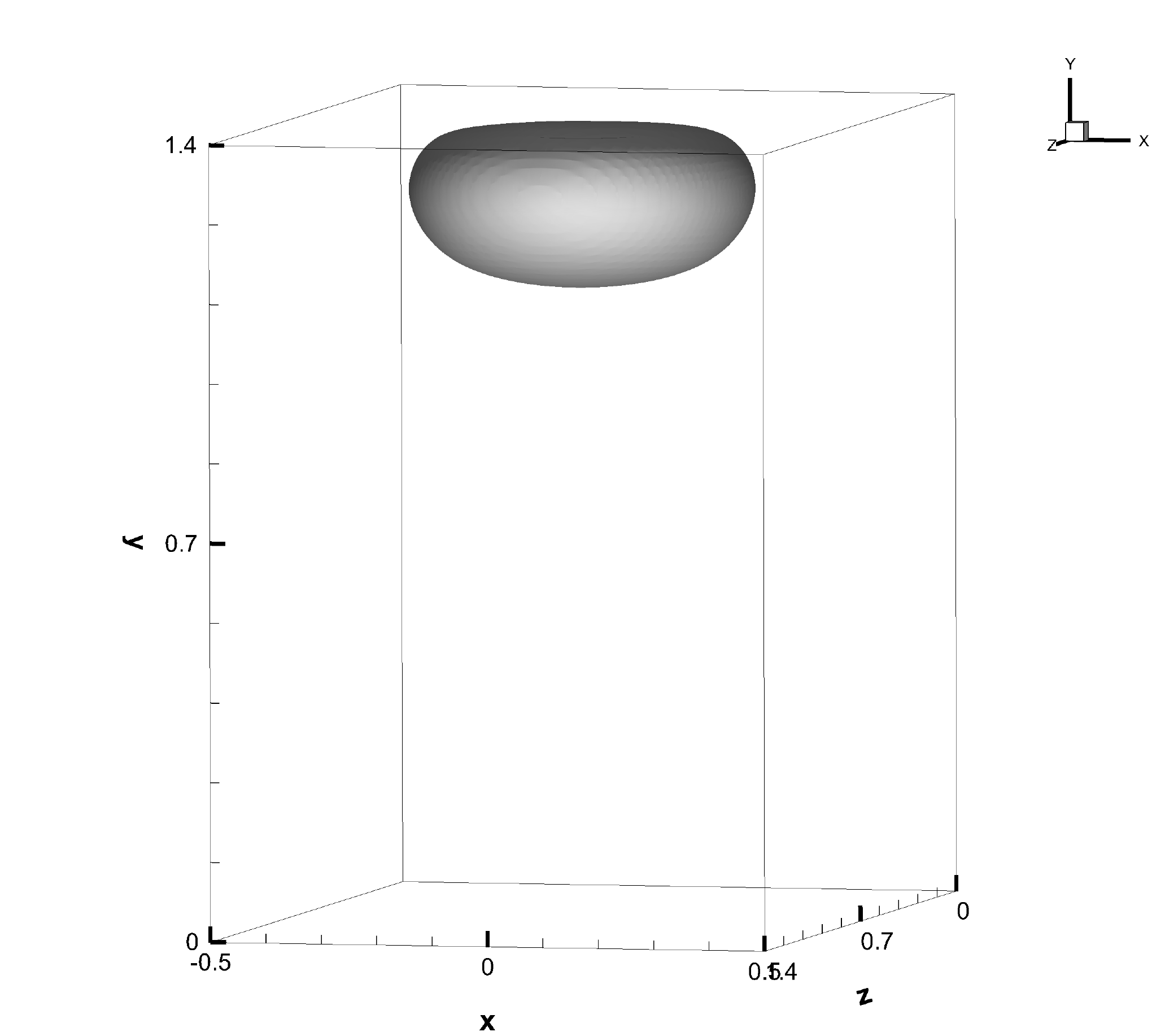}} %
    }
    \centerline{
    \subfigure[] {\includegraphics[width=0.25\textwidth]{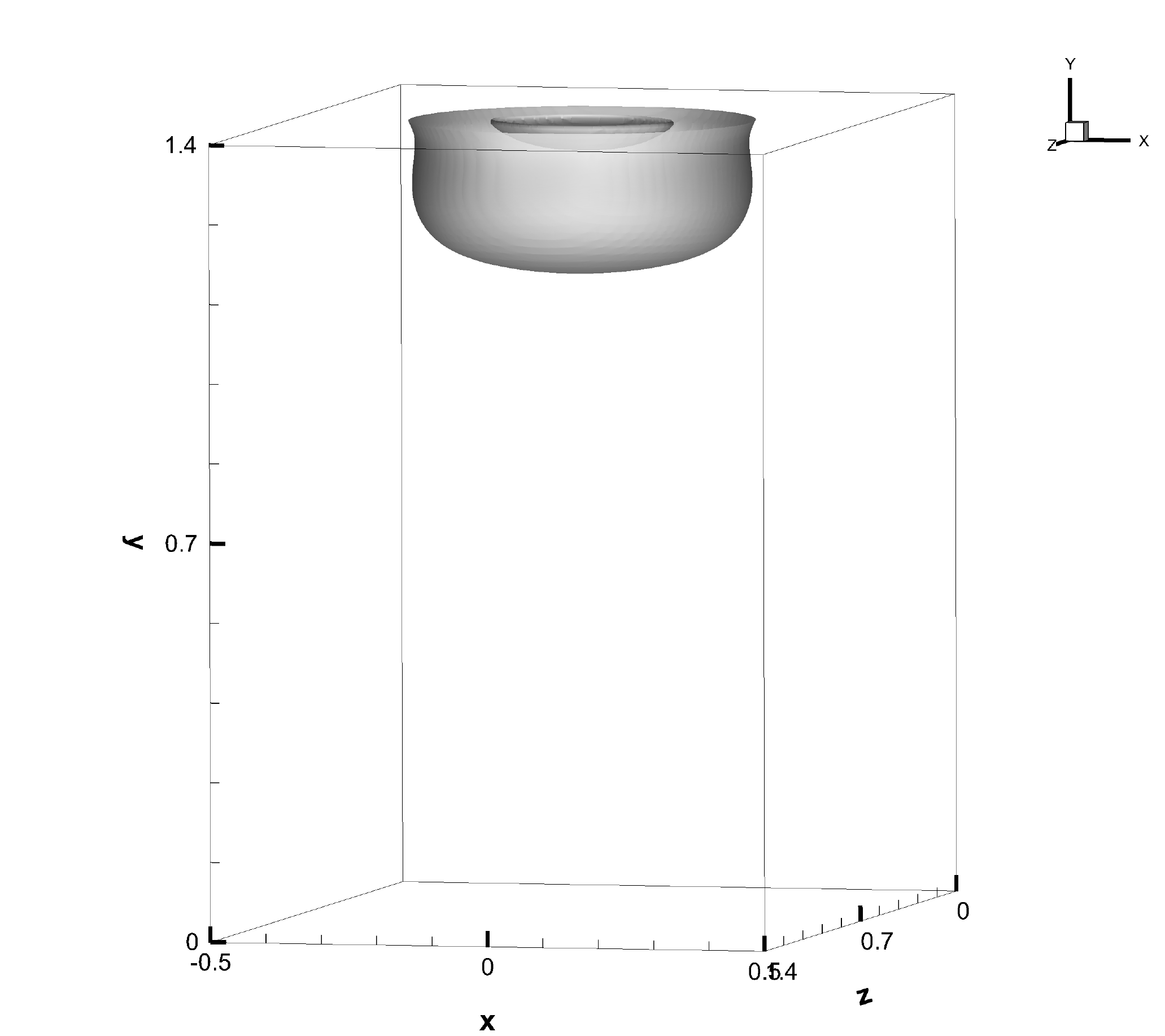}} %
    \subfigure[] {\includegraphics[width=0.25\textwidth]{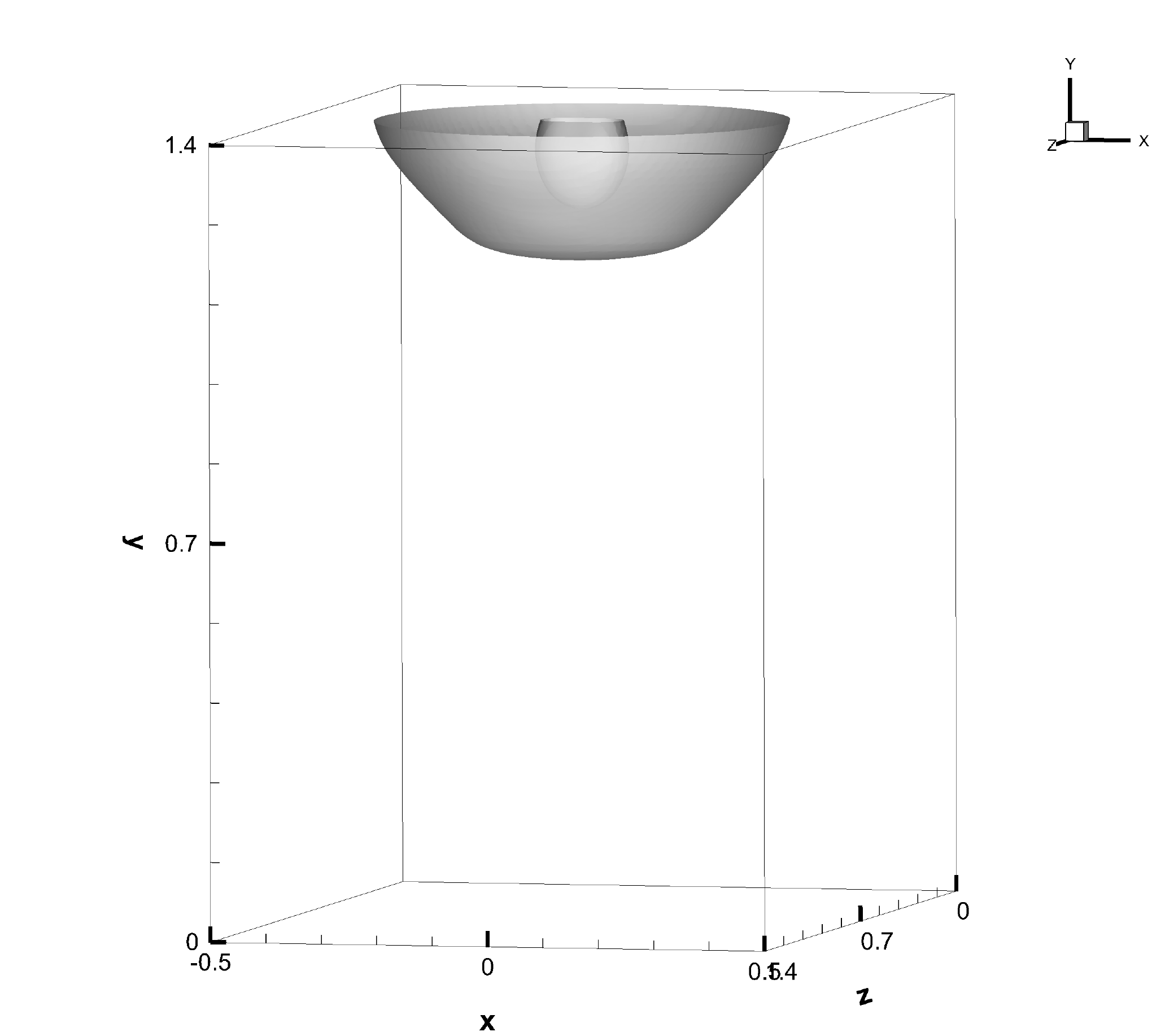}} \\
    \subfigure[] {\includegraphics[width=0.25\textwidth]{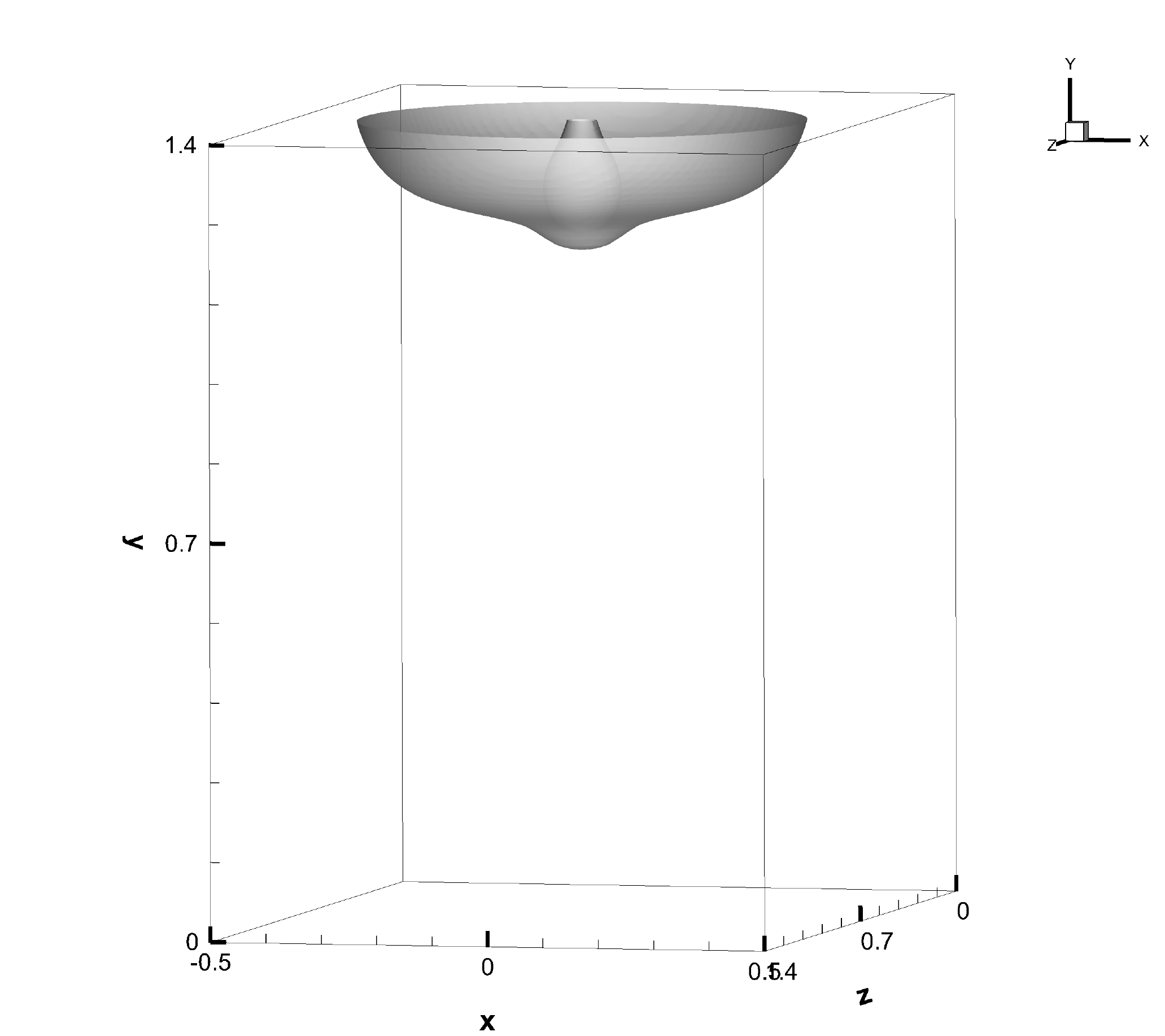}} %
    \subfigure[] {\includegraphics[width=0.25\textwidth]{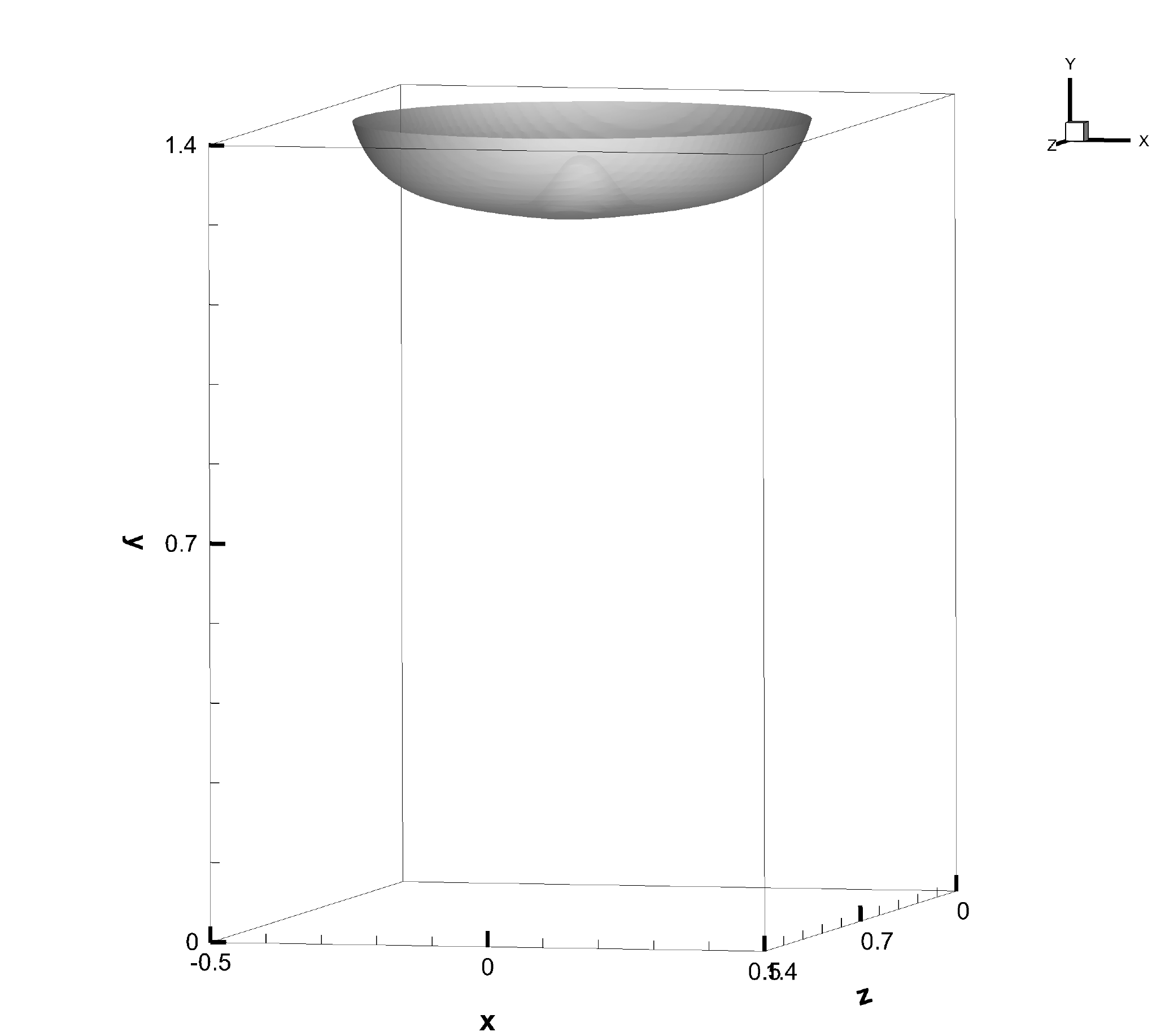}} %
    }
    \centerline{
    \subfigure[] {\includegraphics[width=0.25\textwidth]{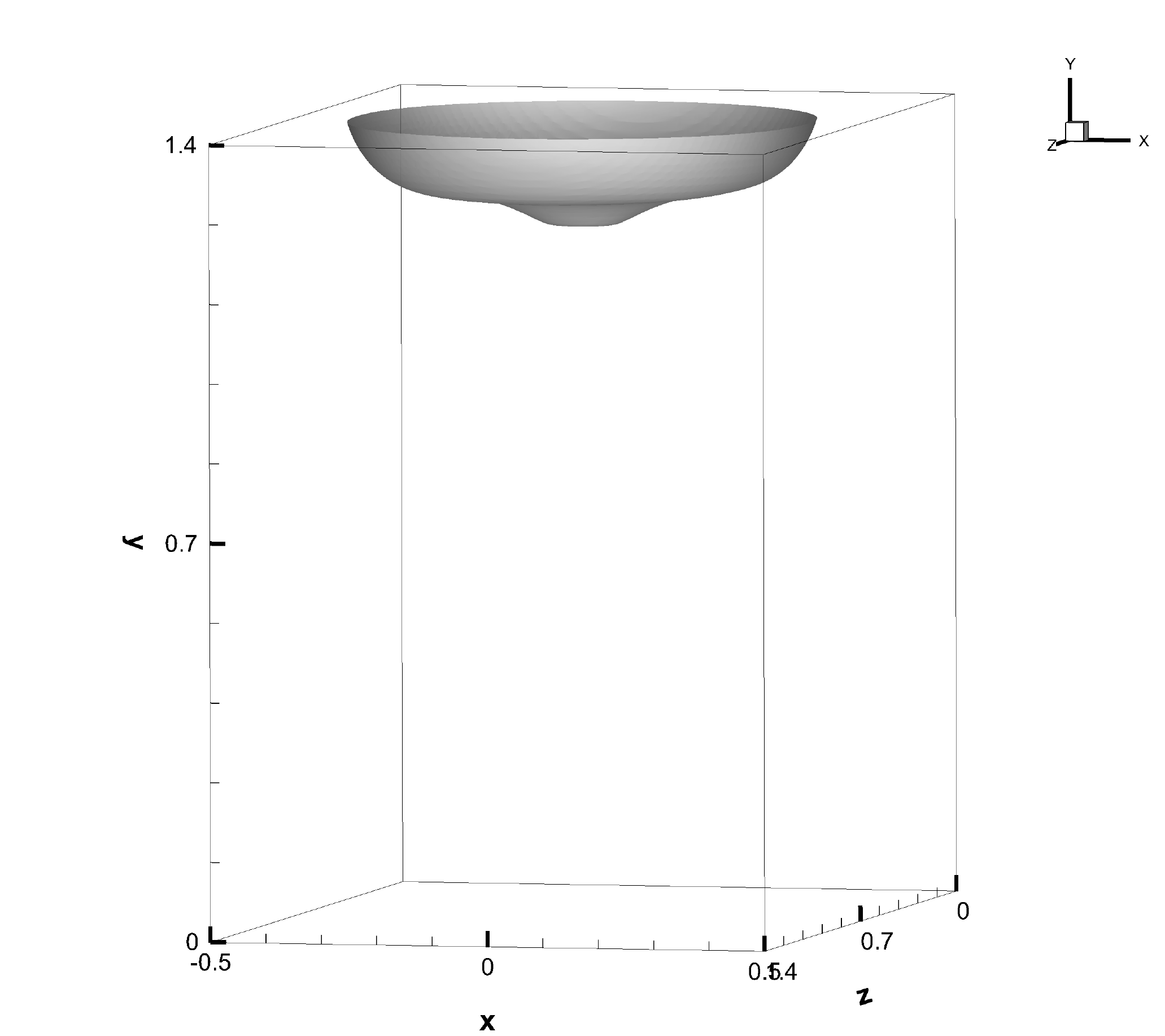}} \\
    \subfigure[] {\includegraphics[width=0.25\textwidth]{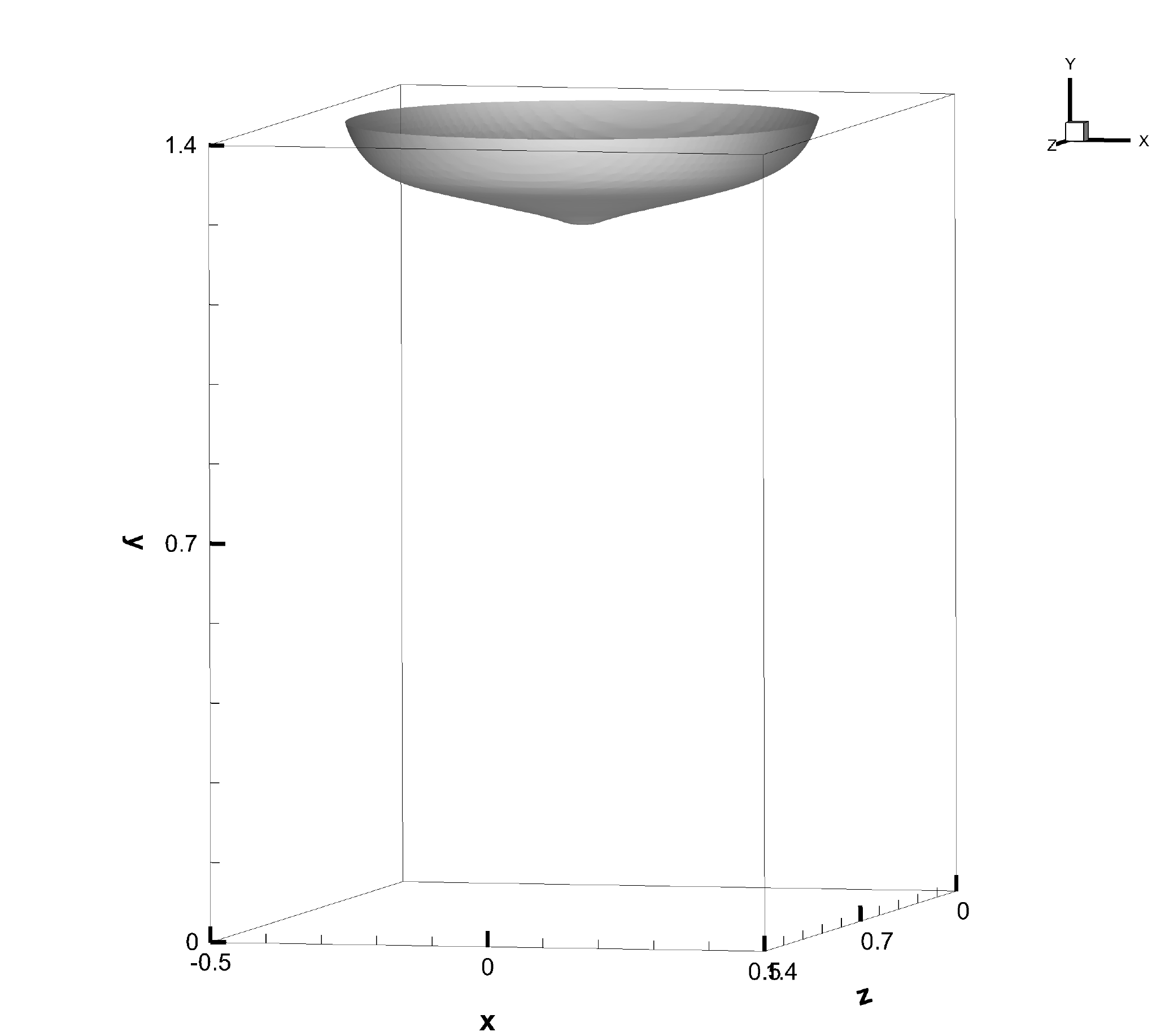}} %
    \subfigure[] {\includegraphics[width=0.25\textwidth]{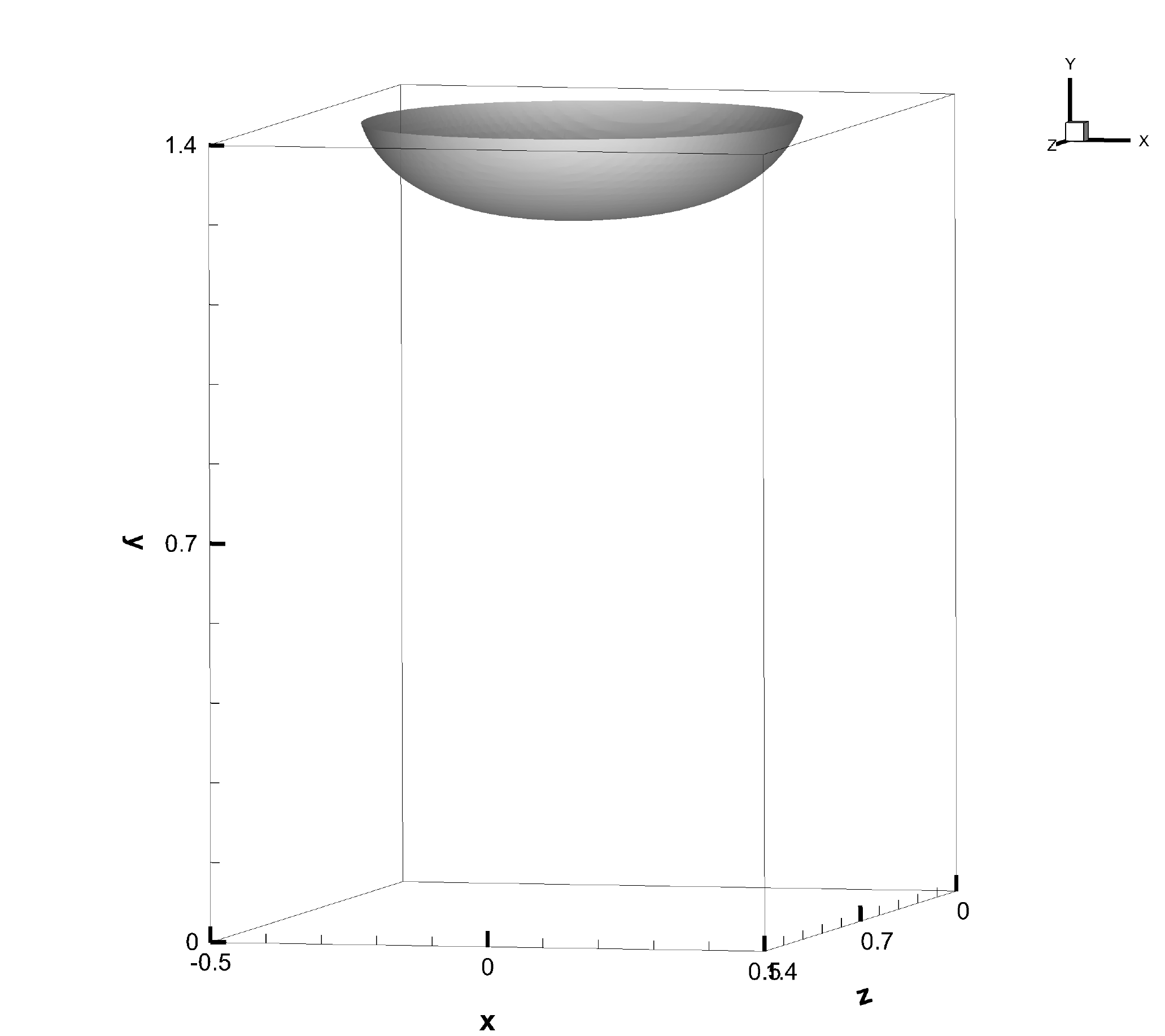}} %
    \subfigure[] {\includegraphics[width=0.25\textwidth]{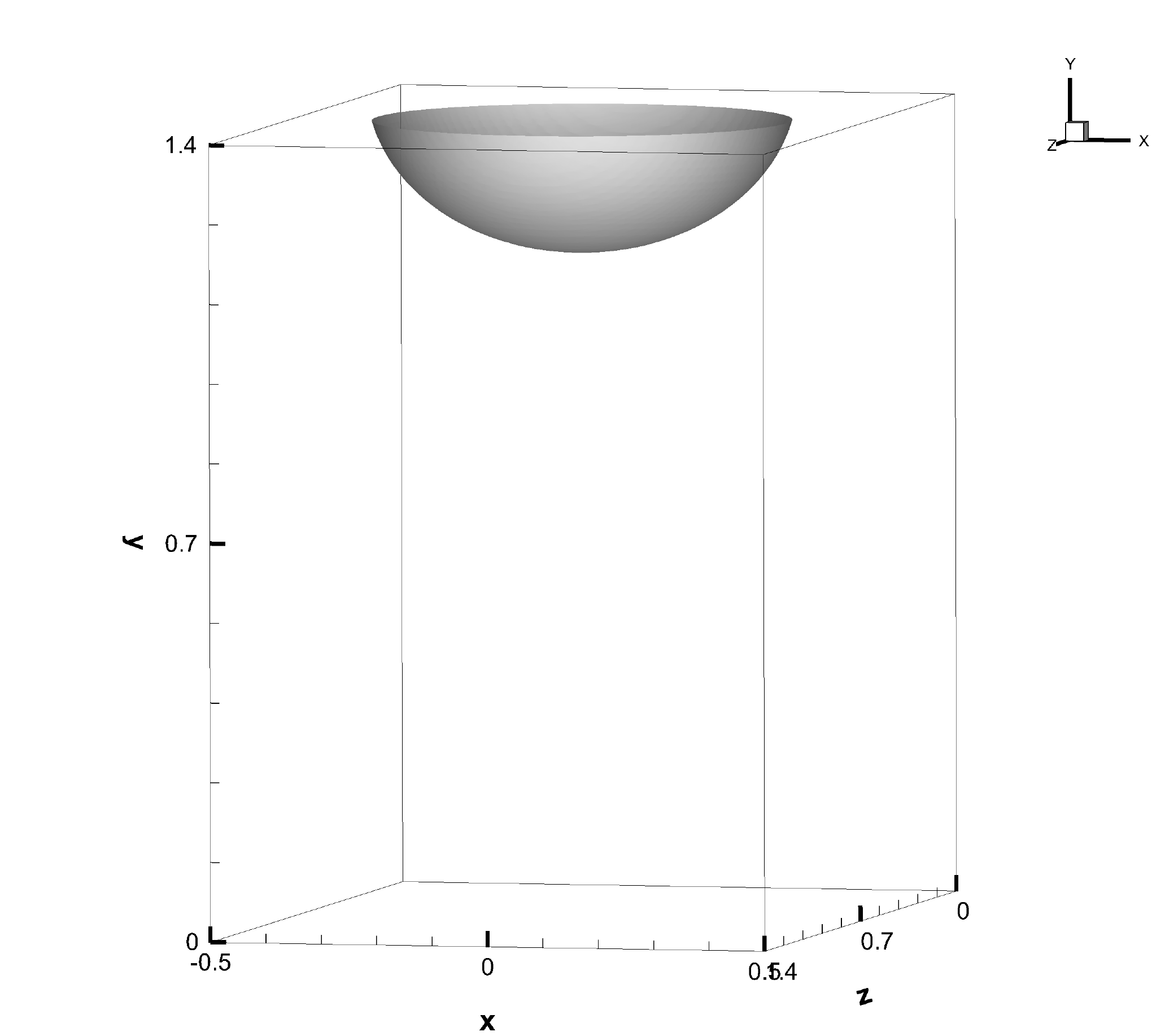}} 
    }
     \caption{Temporal sequence of snapshots of the rising air bubble in water with a contact angle $105^0$: (a)t=0.06, 
     (b)t=0.26, (c)t=0.44, (d)t=0.60, (e)t=0.63, (f)t=0.66,
     (g)t=0.70, (h)t=0.72, (i)t=0.75, (j)t=0.77, (k)t=0.90,
     (l)t=1.45.
     }
      \label{F9}
\end{figure}

Figure \ref{F9} shows the motion of the air bubble
with a contact angle $105^0$, which corresponds to
a hydrophobic wall. The same dynamic behavior can be
observed of the bubble at the initial stage, before
it touches the upper wall (Figures \ref{F9}(a)-(c)). 
Compared with
the other cases, the configuration of the system
is significantly different after the bubble breaks up
on the upper wall. At this contact angle, 
as the base of the water drop trapped by the inner
interface shrinks on the upper wall due to the contact line 
motion,
the surface tension apparently can no longer sustain 
the weight of the water on the wall.
We observe that the small water drop completely detaches
from the wall. It drips onto the outer 
piece of interface, and 
merges into the 
bulk of water (Figures \ref{F9}(f)-(j)).
The deformation caused to the outer air/water
interface due to this process is clearly visible
from Figures \ref{F9}(g) and \ref{F9}(i)-(j).
Only the air dome (without the small water drop inside)
is left on the upper wall for this 
case (Figure \ref{F9}(l)).

With the $60^0$ contact angle, the pocket of water
trapped by the inner interface becomes a
water drop suspended on the upper wall. As the contact angle
increases, the trapped water drop can become 
unstable and some water may escape from the wall due to
the interactions among the inertia (contact
line motion), contact angle, gravity and the surface tension.
For the $90^0$ contact angle, a portion of
the trapped water  escapes from the air bubble, 
 and some remains at  the wall. 
 This seems to be because the surface tension
 is not strong enough to carry the massive water drop and overcome
 the inertia due to the contact line motion. 
 But once some of the water escapes, it is sufficient
  to stabilize the lighter drop. 
 For the air-water interface with the $105^{0}$  
 contact angle, the trapped water drop completely escapes
 from the air-bubble into the bulk of water, and there is once more only a single air-water interface. 
The trapped water overall tends to break free from
the wall more likely with a larger contact angle.

The results of this section show that
the method developed herein can effectively 
capture the three-dimensional two-phase flow dynamics under
large and realistic density ratios and viscosity
ratios. The method will be useful for studying the
interactions between fluid interfaces and solid walls,
large interfacial deformations and topological changes
of the interfaces.

\section{Concluding Remarks}
\label{sec:summary}


The current work is focused on three-dimensional two-phase
flows on domains with at least one homogeneous direction.
We have presented a hybrid spectral element-Fourier spectral
method for efficient simulations of this class of
3D two-phase problems. 
A critical component of this method is a strategy we
developed in a previous work for dealing with
the variable density/viscosity of the two-phase mixture,
which makes the efficient use of FFTs in the current work possible for
these flows with different densities and viscosities
for the two fluids.
Our method transforms the computation of two-phase
flows in the 3D space into a set of computations
in the 2D planes of the non-homogeneous directions,
which are completely de-coupled from one another.
Therefore, only a series of 2D problems need to be
solved, which can be performed in parallel. 

Extensive numerical tests with several wall-bounded
two-phase flows
in 3D space have been presented to assess the accuracy
and performance of the method. 
Comparisons with theoretical results demonstrate
that the presented method can capture the 
physical effects such as those of the surface tension
and the contact angles accurately in 3D.
Simulation results show that our method 
can effectively deal with the topological changes
of the fluid interface, the moving contact lines,
and the interaction between fluid interfaces and
solid walls in the 3D space.


The presented method can be an efficient technique
for investigating a wide class of two-phase problems
such as the oil transport in the pipe, wind/ocean-wave
interactions, and the dynamics of contact lines.
We anticipate that it will be
useful and instrumental in a number of areas
such as microfluidics, surface ship hydrodynamics, 
and the study of functional surfaces.

\section*{Acknowledgment}

The work of S.H.C. and S.D. was partially
supported by the NSF DMS-1318820 and DMS-1522537.
L.D.Z. acknowledges the support of NSF DMS-1522554.


\bibliographystyle{plain} 
\bibliography{twohalf,multiphase,nphase,sem,mypub,contact_line,interface,cyl}

\end{document}